\documentclass{iopart}

\usepackage{amssymb,graphicx} 
\usepackage{subfigure}
\usepackage{hyperref}

\begin{document}

\title{Advanced Virgo: a second-generation interferometric gravitational wave detector}

\author{F.~Acernese$^{1,2}$, 
M.~Agathos$^{3}$, 
K.~Agatsuma$^{3}$, 
D.~Aisa$^{4,5}$, 
N.~Allemandou$^{6}$, 
A.~Allocca$^{7,8}$, 
J.~Amarni$^{9}$, 
P.~Astone$^{10}$, 
G.~Balestri$^{8}$, 
G.~Ballardin$^{11}$
% ...
}
\address{$^{1}$Universit\`a di Salerno, Fisciano, I-84084 Salerno, Italy}
\address{$^{2}$INFN, Sezione di Napoli, Complesso Universitario di Monte S.Angelo, I-80126 Napoli, Italy}
\address{$^{3}$Nikhef, Science Park, 1098 XG Amsterdam, The Netherlands}
\address{$^{4}$Universit\`a di Perugia, I-06123 Perugia, Italy}
\address{$^{5}$INFN, Sezione di Perugia, I-06123 Perugia, Italy}
\address{$^{6}$Laboratoire d'Annecy-le-Vieux de Physique des Particules (LAPP), Universit\'e de Savoie, CNRS/IN2P3, F-74941 Annecy-le-Vieux, France}
\address{$^{7}$Universit\`a di Siena, I-53100 Siena, Italy}
\address{$^{8}$INFN, Sezione di Pisa, I-56127 Pisa, Italy}
\address{$^{9}$APC, AstroParticule et Cosmologie, Universit\'e Paris Diderot, CNRS/IN2P3, CEA/Irfu, Observatoire de Paris, Sorbonne Paris Cit\'e, 10, rue Alice Domon et L\'eonie Duquet, F-75205 Paris Cedex 13, France}
\address{$^{10}$INFN, Sezione di Roma, I-00185 Roma, Italy}
\address{$^{11}$European Gravitational Observatory (EGO), I-56021 Cascina, Pisa, Italy}
% ...

\ead{losurdo@fi.infn.it}

\begin{abstract}

Advanced Virgo is the project to upgrade the Virgo interferometric detector of gravitational waves, with the aim of increasing the number of observable galaxies (and thus the detection rate) by three orders of magnitude. The project is now in an advanced construction phase and the  assembly and integration will be completed by the end of 2015. Advanced Virgo will be part of a network, alongside the two Advanced LIGO detectors in the US and GEO HF in Germany, with the goal of contributing to the early detections of gravitational waves and to the opening a new window of observation on the universe. In this paper we describe the main features of the Advanced Virgo detector and outline the status of the construction.

\end{abstract}

\pacs{04.80.Nn, 95.55.Ym}

\maketitle

\section{Introduction: scope of the Advanced Virgo upgrade}
Advanced Virgo (AdV) is the project to upgrade the Virgo detector~\cite{VirgoP} to a second generation instrument. It is designed to achieve a sensitivity of about one order of magnitude better than that of Virgo, which corresponds to an increase in the detection rate by about three orders of magnitude. Advanced Virgo will be part of the international network of detectors aiming to open the way to gravitational-wave (GW) astronomy \cite{rev1,rev2,rev3}. With respect to Virgo, most of the detector subsystems have to deliver a significantly improved performance to be compatible with the design sensitivity. The AdV design choices were made on the basis of the outcome of the different R\&D investigations carried out within the gravitational wave community and the experience gained with Virgo, while also taking into account budget and schedule constraints. The AdV upgrade was funded in December 2009 and is currently in an advanced phase of installation. In March 2014 a Memorandum of Understanding  for full data exchange, joint data analysis and publication policy with the LIGO Scientific Collaboration was renewed, thus strengthening the world-wide network of second generation detectors (including Advanced Virgo, the two Advanced LIGO~\cite{ligo_CQG} and GEO HF\cite{geo_CQG}).

In this introduction we briefly describe the main upgrades foreseen in the AdV design. Comprehensive and more detailed
descriptions can be found in the following sections and in the Technical Design Report \cite{ADV:TDR}.

\paragraph{Interferometer optical configuration} 

AdV will be a dual-recycled interferometer (ITF). Besides the standard power recycling, a signal-recycling (SR)
cavity will also be present. The tuning of the signal-recycling parameter allows for the changing of the shape of the sensitivity curve and the optimizing of the detector for different astrophysical sources. To reduce the impact of the thermal noise of the mirror coatings in the mid-frequency range, the beam spot size on the test masses has been enlarged. Therefore, unlike Virgo, the beam waist will be placed close to the center  of the 3\,km Fabry-Perot (FP) cavities. The cavity finesse will be higher than that of Virgo: a reference value of 443 has been chosen. 
Having a larger beam requires the installation of larger vacuum links in the central area and new mode matching telescopes at the interferometer input/output.
Locking all of the cavities at the same time might be difficult. To ease the lock of the full interferometer, a system of auxiliary green lasers is being developed.

\paragraph{Increased laser power} 

Improving the sensitivity at high frequency requires high laser power\footnote{Though squeezing is not part of the AdV baseline, the infrastructure has been prepared to host a squeezer in subsequent years.}. The AdV reference sensitivity is computed assuming 125 W entering the interferometer, after the Input Mode Cleaner (IMC). Therefore, considering the losses of the injection system, the laser must provide a power of at least 175 W. A 200 W laser, based on fiber amplifiers, will be installed after 2018, while during the first years of operation (at a lower power) AdV will use the Virgo laser,  capable of providing up to 60 W. The input optics for AdV must be compliant with the ten-fold increase in optical power. Specifically designed electro-optic modulators and Faraday isolators able to withstand high power have been developed and a DC readout scheme will be used, requiring a new design of the output mode cleaner.
A sophisticated thermal compensation system has been designed to cope with thermally-induced aberrations (but also with losses induced by intrinsic defects of the optics). The sensing is based on Hartmann sensors and phase cameras, while the ring heaters around several suspended optics will be used as actuators to change the radius of curvature. CO$_2$ laser projectors, which shine on dedicated compensation plates, allow the compensation of thermal lensing and optical defects. 

\paragraph{Mirrors}

To cope with the increased impact of radiation pressure fluctuations the AdV test masses will be twice as heavy (42 kg) as those of Virgo.
Fused silica grades with ultra low absorption and high homogeneity have been chosen for the most critical optics. State of the art polishing technology is used to reach a flatness better than 0.5 nm rms in the central area of the test masses. Low-loss and low-absorption coatings are used to limit as far as possible the level of thermal noise and the optical losses in the cavities, which eventually determine the sensitivity in the high frequency range.

\paragraph{Stray light control} 

Scattered light could be a significant limitation on detector sensitivity. In order to limit phase noise caused by part of the light being back-scattered into the interferometer, new diaphragm baffles will be installed. These will be either  suspended around the mirrors, or ground-connected inside the vacuum links. All photodiodes to be used in science mode will be seismically isolated and in vacuum. To this end, a compact vibration isolation system, accommodated inside a vacuum chamber, has been built. Five of these {\it minitowers} will be installed.   

\paragraph{Payloads and vibration isolation} 

A new design of the payloads has been developed. This was triggered mainly by the need to suspend heavier mirrors, baffles and compensation plates, while controllability and mechanical losses have also been improved. The Virgo Super Attenuators (SA) provide already provide a seismic isolation compliant with AdV requirements. However, some upgrades are planned, to take the new payload design into account and to further improve robustness in high seismicity conditions: the possibility to actively control the ground tilt will be implemented. 
 
\paragraph{General detector infrastructure}

Important modifications have been undertaken in the main experimental hall, in order to be able to host the minitowers and upgrade the laboratories where the laser, the input optics and the detection system are located, turning them into acoustically-isolated clean rooms. The vacuum has been upgraded by installing large cryotraps at the ends of the 3 km pipes, in order to lower the residual pressure by a factor of about 100. Several upgrades of the data acquisition and general purpose electronics are foreseen for AdV, in order to keep up with the increasing number of channels and the more demanding control system for the signal-recycling configuration, and to cope with the obsolescence of several boards.

\section{Sensitivity goals} \label{sec:sens}
\subsection{Target sensitivity for AdV} \label{sec:agencies}

\begin{figure}\begin{center} 
\centerline{\includegraphics[viewport=30 220 600 620,width=\textwidth,clip]{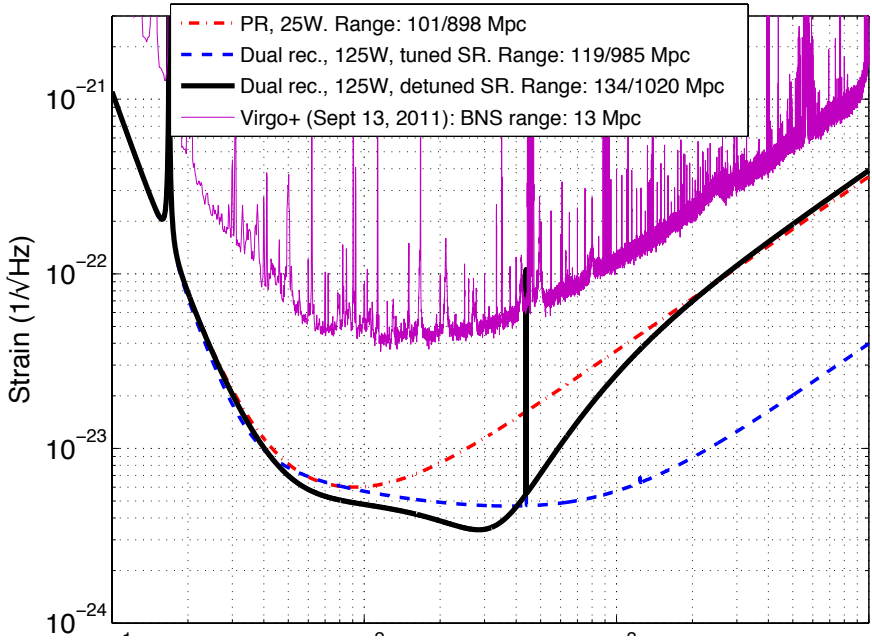}}
\caption{\sf AdV sensitivity for the three benchmark configurations as defined in the TDR: early operation (dash-dotted line), 25 W input power, no SR; mid-term operation, wideband tuning (dashed line), 125 W input power, tuned SR; late operation, optimized for BNS (black solid line), 125 W input power, detuned SR (0.35 rad). In the legend, the inspiral ranges for BNS and BBH (each BH of  30 $M_\odot$) in Mpc are reported. The best sensitivity obtained with Virgo+ is shown for comparison.}
\label{commsteps}
\end{center}\end{figure}

A target sensitivity curve for the new interferometer was defined, based on the detector design and noise modeling at that time (solid black line in Figure \ref{commsteps}) in the Advanced Virgo Technical Design Report (TDR)~\cite{ADV:TDR}, which was approved by the funding agencies in 2012.  The curve corresponds to a detector configuration with 125 W at the interferometer input and SR parameters chosen in order to maximize the sight distance for coalescing binary neutron stars (BNS). The corresponding inspiral ranges\footnote{The {\it inspiral range} is defined as the volume- and orientation-averaged distance at which a compact binary coalescence gives a matched filter signal-to-noise ratio of 8 in a single detector \cite{finn}.} are $\sim 140$ Mpc for BNS (1.4 M$_\odot$ each) and $\sim 1$ Gpc for 30 $M_\odot$ coalescing binary black holes (BBH). 

Some basic assumptions are made in the sensitivity computation, which are to be considered as system requirements. Some subsystem design choices were made to comply with these requirements:
\begin{itemize}
\item {\bf input power:} it is assumed that a laser power of 125 W will be available in the TEM$_{00}$ mode after the IMC;
\item {\bf round trip losses:} the power lost in a round trip inside a Fabry-Perot cavity must not exceed 75 ppm;
\item {\bf technical noises:} all technical noises\footnote{The distinction between {\it fundamental} and {\it technical} noise is common in the GW community, but not always clear. We define fundamental noises as those that require a large investment in infrastructure, a deep redesign of the detector, or new technological developments in order to be suppressed (e.g. thermal noise, shot noise, seismic noise). We define technical noises as those that require a relatively limited investment in commissioning or upgrade to be suppressed (e.g. the noise generated by scattered light or electronics).} must be reduced to a level such that the corresponding strain noise in amplitude is $<$0.1 of the design sensitivity in the 10 Hz-10 kHz frequency range.
\end{itemize}
However, the AdV detector is tunable in three ways: by changing the laser power, by changing the transmissivity of the SR mirror and by tuning the position of the SR mirror. The SR mirror transmittance influences the detector bandwidth, while the position of the SR mirror at a microscopic scale changes  the frequency of the maximal sensitivity. Thus, the presence of the SR cavity allows to think of AdV as a tunable detector (see Section \ref{sec:osd}): the sensitivity curve can be shaped in order to perform data-taking periods optimized to target different astrophysical sources. For the sake of simplicity, we refer to three different operation modes:
\begin{itemize}
\item {\bf power recycled, 25 W};
\item {\bf dual recycled, 125 W, tuned signal recycling};
\item {\bf dual recycled, 125 W, detuned signal recycling} (SR tuning chosen to optimize BNS inspiral range).
\end{itemize}

AdV will not initially be operated in the final configuration. The new features will create new problems, which must be faced with a step-by-step approach. The above-mentioned modes of operation correspond to commissioning steps of increasing complexity. They should be considered as {\it benchmark configurations}, while the commissioning will progress through many intermediate steps. Periods of commissioning will be alternated with periods of data-taking and such plans will need to be coordinated with Advanced LIGO in order to maximize the capabilities of the network. 

Figure \ref{commsteps} compares the target sensitivity curves for the reference scenarios described, as defined in the TDR\footnote{The sensitivity curves shown in this section have been plotted using {\it GWINC}, a MATLAB code developed within the LIGO Scientific Collaboration (LSC) \cite{bench} and adapted to AdV.}. 
\enlargethispage*{2.5\baselineskip}

\subsection{Further progress}

\begin{figure}[h]
\centerline{\includegraphics[viewport=30 200 600 600,width=\textwidth,clip]{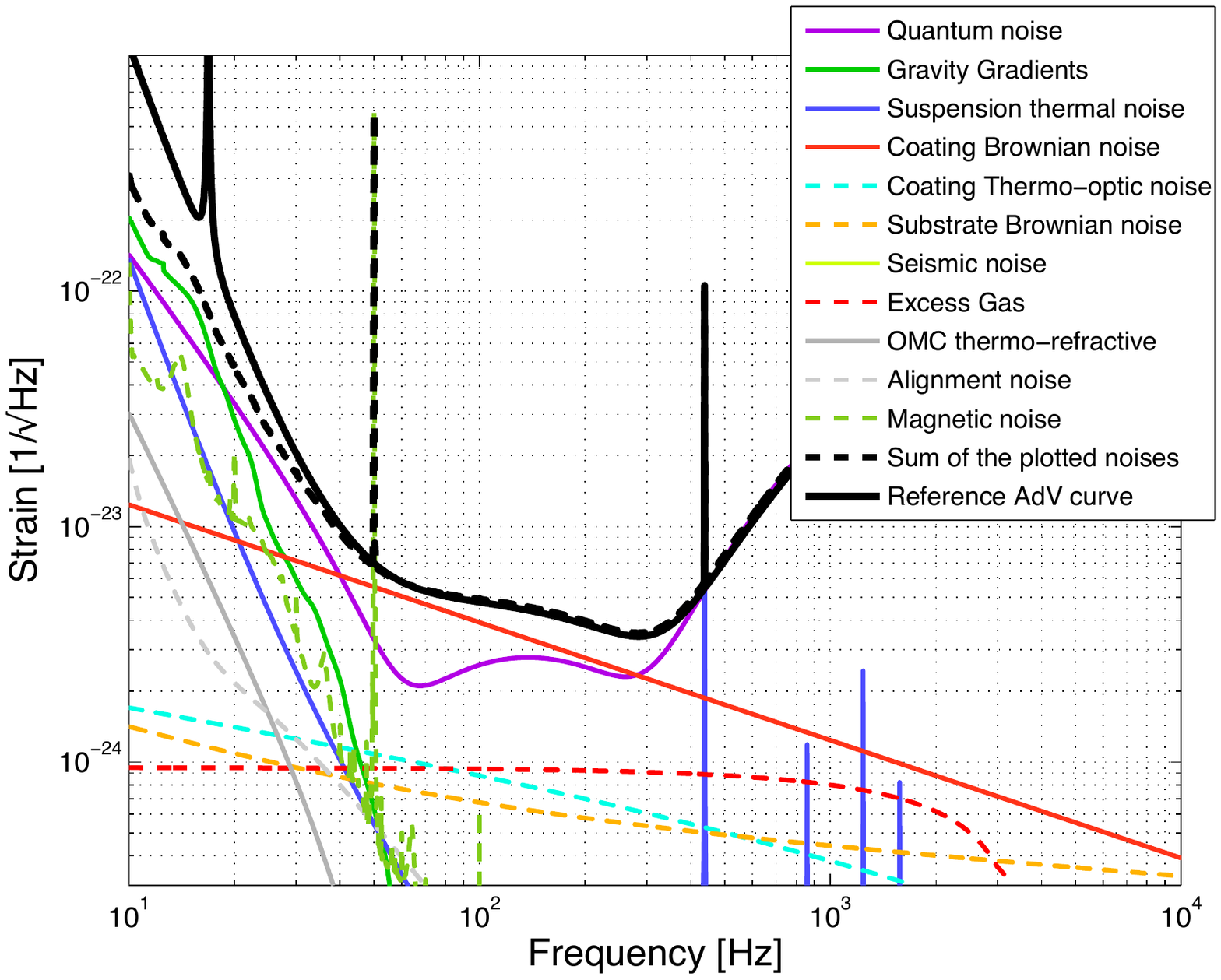}} 
\caption{\sf The AdV reference sensitivity (solid black) compared to an AdV noise budget (dashed black) using the new models for suspension thermal noise, gravity gradient noise and some modeled technical noises, computed in the configuration optimized for BNS detection with 125W of input laser power. See Section \ref{sec:agencies} for further details}
\label{newsens}
\end{figure}

Since the release of the TDR some progress has been made both in the design of the detector and the modeling of the noise:
\begin{itemize}
\item The design of the payload has been finalized and the recoil mass, present in the Virgo payload, has been removed. Furthermore, experimental tests have shown that mechanical dissipation at the level of the marionette is lower than previously assumed. This has led to a new modeling of the suspension thermal noise, which no longer limits the sensitivity at low frequency (see Section \ref{sec:pay});
\item The gravity gradient noise model has also evolved. It now uses the typical seismic noise spectrum measured on the Virgo site as input.
\item We have started to add some of the technical noises that we know how to model to the noise budget. 
\end{itemize}
In Figure \ref{newsens} we show the result of this work (which must be considered as work in progress).

\enlargethispage*{2.5\baselineskip}

\subsection{Timeline}

One of the main goals of the AdV project is to start taking data in 2016. AdV will be operated in two main phases:
\begin{itemize}
\item Early operation: the interferometer configuration will be a power-recycled Fabry-Perot Michelson and the power injected will not exceed 40 W. This configuration is, for many reasons, similar to that of Virgo+. Thus, we expect a shorter commissiong period and faster progress in the sensitivity improvement. The achievable BNS inspiral range is larger than 100 Mpc.
\item Late operation: the interferometer will subsequently be upgraded by installing the signal recycling mirror and the high-power laser, fulfilling the full specifications. The tentative date for this upgrade is 2018, though this will depend on the joint plans for science runs agreed with the partners in the network.
\end{itemize}

The installation and integration of the upgrades needed for the first phase and their acceptance are scheduled to be completed by fall 2015. The commissioning of parts of the detector will start prior to then:
\begin{itemize}
\item the commissioning of the input-mode-cleaner cavity started in June 2014, following its first lock;
\item in spring 2015, the beam will be available at the dark port, making it possible to commission the detection system;
\item a single 3-km arm  will be available shortly after.
\end{itemize}

\section{Optical design} \label{sec:osd}
The optical configuration of Advanced Virgo was designed to maximize improvements in the detector sensitivity, while inducing only minor changes to the infrastructure. The vacuum enclosure which housed the Virgo interferometer continues to constrain the cavity lengths for Advanced Virgo. As a consequence, the arm cavity length is 3\,km, while the recycling cavities are $\sim$12\,m long.

\begin{figure}
\begin{center}
\includegraphics[width = 0.9\textwidth]{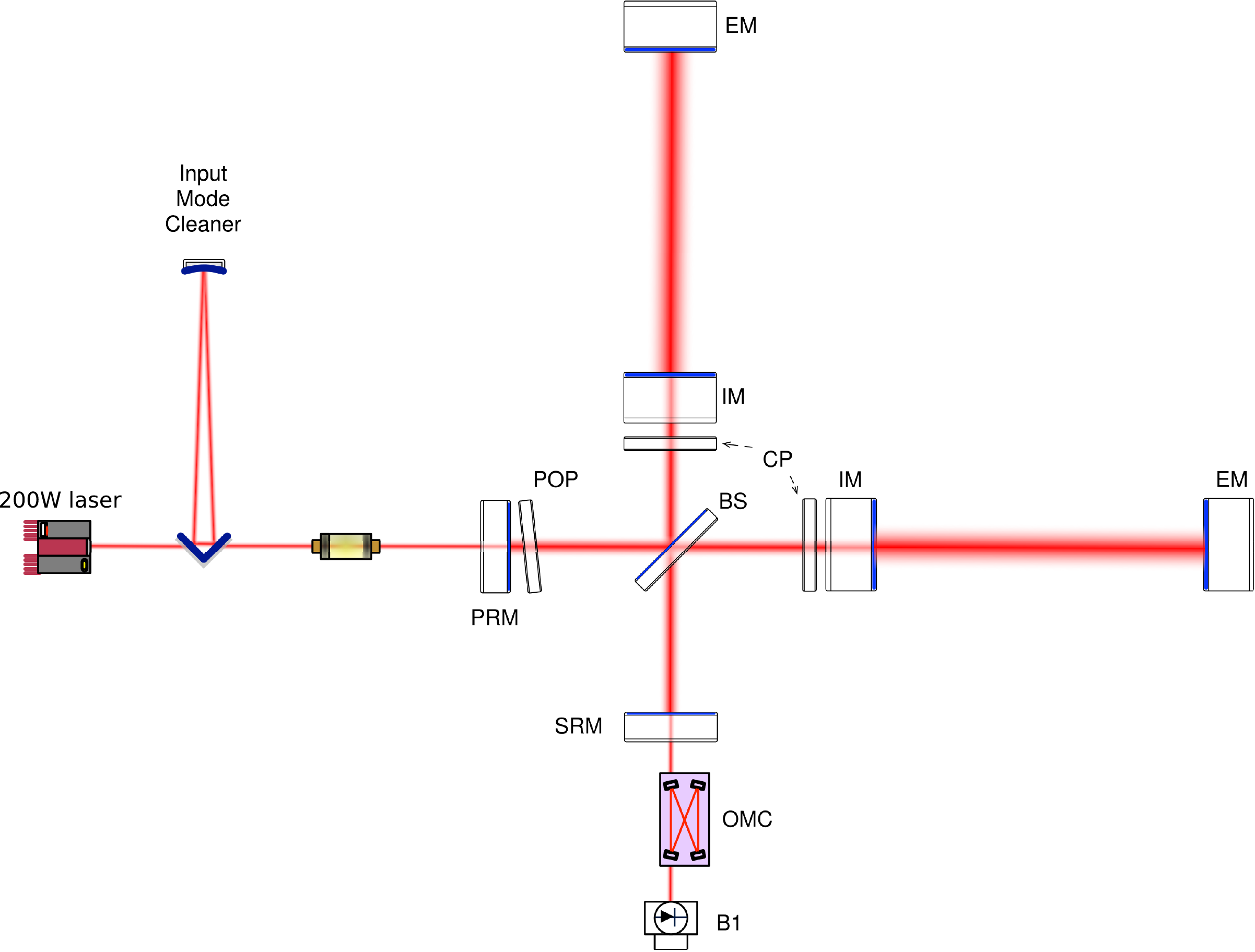}
\end{center}
\caption{Simplified optical layout of the Advanced Virgo interferometer. Each 3~km-long arm-cavity is formed by an Input Mirror (IM) and an End Mirror (EM). The recycling cavities at the center of the interferometer are 12 meters long and are formed by the Power Recycling Mirror (PRM), the Signal Recycling Mirror (SRM) and the two IM.}
\label{fig:OSD:config}
\end{figure}

A sketch of the optical configuration is presented in Figure \ref{fig:OSD:config}. The core of Advanced Virgo is composed of a dual recycled Michelson interferometer with Fabry-Perot arm cavities.

\subsection{Design of the arm cavities}

The arm cavities have a bi-concave geometry, with each mirror, also known as test mass, having a concave radius of curvature slightly larger than half the arm-cavity length to ensure the stability of the cavity. This near-concentric resonator configuration has been chosen for two main reasons: (1) to increase the beam size at the mirrors, thus averaging over a larger area of the mirror surface and reducing the relative contribution of mirror-coating thermal noise and thermal gradient in the substrate; and (2) to limit the effect of radiation-pressure-induced alignment instabilities.

With this type of topology the cavity waist is near the middle of the arm. In addition, having mirrors in the cavity with two different radii of curvature (RoC) displaces the cavity waist towards the mirror with the smaller radius of curvature, so that the beam size on that mirror is also reduced. We decided to have a smaller beam on the IM to reduce the clipping loss in the recycling cavities. This did not cause problems in terms of coating thermal noise since the IM have fewer coating layers compared to the End Mirror (EM), mmeaning less thermal noise if the beam sizes are identical on both mirrors. We have thus chosen to have the EM with a larger radius, and therefore larger beam size, than the IM.

The final decision on arm mirror radii of curvature was guided by different competing factors.  These included: mirror thermal noise; cavity stability; clipping losses as beams become larger; and tolerance to manufacturer errors in mirror production, which can limit the cavity stability in the cold (uncorrected) state.  The RoC should also be chosen in such a way as to minimize the accidental degeneracy of the higher-order modes in the arm cavities (i.e. we do not want a higher order optical mode to resonate at the same time as the fundamental one).

As such, we have chosen an average radius of 1551\,m for the IM and EM, which means the cavity fundamental mode resonance is between the resonances of higher-order mode of orders 8 and 9. As described above, the IM RoC is reduced and the EM RoC is increased to minimize the impact of mirror thermal noise, while also keeping clipping losses low by limiting the final beam size on the EM.  This procedure yields mirror radii of 1420\,m for the IM and 1683\,m for the EM, resulting in beam radii of respectively 48.7\,mm and 58\,mm on these mirrors.

Since Advanced Virgo will use the signal-recycling technique, the detector sensitivity will vary with the arm-cavity input-mirror transmissivity, the signal-recycling mirror transmissivity, the signal-recycling cavity length, and the circulating power.  Considering the combination of all these factors, the detector bandwidth depends only weakly upon the specific choice of arm cavity finesse (changes in the arm-cavity input-mirror transmissivity can be compensated by changing one of the other parameters).

Factors other than the ideal sensitivity curve thus drive the choice of arm cavity finesse. These factors include thermal loading in the central interferometer, length noise coupling from auxiliary degrees of freedom, and the relative impact of losses in the arm and signal-recycling cavities.

As a trade-off, we have chosen a value of finesse of 440. The resulting arm input mirror transmission is set to 1.4\% while the EM is almost perfectly reflective (transmission of few ppm).

\subsection{Choice of the recycling cavities}

The geometry of the recycling cavities is referred to as {\it marginally stable}. The cavities are formed by using a radius of curvature of 1420\,m  for the input test masses and a radius of curvature of 1430\,m for the power- and signal-recycling mirrors. The recycling-cavity length is 11.952\,m. This yields a beam diameter of about 5\,cm (the beam size is constant in the cavity, as there is no focusing element). In this configuration some high-order modes can resonate at the same time as the fundamental mode, since the Gouy phase accumulated during the free-space propagation inside the recycling cavity is not sufficient to move all high-order modes out of resonance. This configuration is conceptually similar to the one used in initial Virgo. However, in Advanced Virgo the increase in beam size from 2 to 5\,cm further reduces the round-trip Gouy phase and hence increases the degeneracy. This degeneracy results in this type of recycling cavity being extremely sensitive to optical aberrations or to thermal effects in the working interferometer. The carrier field is largely unaffected by these effects as it is stabilized by the arm cavities. The Radio-Frequency (RF) sidebands, however, do not resonate in the arm cavities and therefore excessive aberrations in the recycling cavities may result in noisy and unstable control signals.

The recycling cavity design differs from that of other Advanced GW interferometers, such as LIGO, which have opted for stable cavities, which will be less sensitive to aberrations. The decision to use marginally stable cavities was mainly driven by the construction schedule, budget and increased suspension complexity required for a stable-cavity solution. However, a thermal-compensation system in initial Virgo was successfully used  to reduce aberrations in the recycling cavity and this work will continue with an upgraded system in Advanced Virgo. Optical simulations have indicated that an acceptable RF sideband signal may be obtained if the total round-trip recycling cavity optical path distortions are reduced to less than 2\,nm.

The choice of PR mirror transmission was a trade-off between maximizing the circulating power in the arms and reducing the sensitivity of the power-recycling cavity to aberrations. The former requires the matching of the reflectivity of the power-recycling mirror with the effective reflectivity of the arm cavities (PR transmission of 2.8\,\%). The latter requires the reduction of the PR cavity finesse to a minimum. A PR transmission of 5\,\% was chosen as the best compromise. The choice of SR mirror transmission was a trade-off between optimizing the sensitivity to BNS inspirals, to BBH inspirals and in a broadband (zero detuning) configuration. A SR transmission of 20\,\% was chosen as the best compromise.

\section{Mirror technology} \label{sec:mir}
\subsection{Substrates}

A new type of fused silica with lower absorption (Suprasil 3001/3002) has been chosen for the AdV mirrors. The bulk absorption for this material is three times smaller than that used for Virgo (0.2 ppm/cm at 1064 nm \cite{ref:mir4}), while the other relevant parameters (quality factor, index homogeneity, residual strain, birefringence) are the same or better. Reducing the absorption in the substrates is certainly of interest, as the power absorbed causes thermal lensing in all transmissive optics. However, the thermal effects are still dominated by coating absorption. It is therefore more important to improve the absorption of the coatings than the absorption of the silica.

Hereafter, a detailed list of the AdV substrates is reported together with their main characteristics: 
\begin{itemize}
\item Input Mirrors (IM) - The AdV mirrors will have the same diameter as the Virgo mirrors (35 cm) but will be twice as thick (20 cm) and twice as heavy (42 kg). A high-quality fused silica (Suprasil 3002) with a very low bulk absorption (0.2 ppm/cm) has been used, as these optics transmit a relatively large amount of power (of the order of 2 kW). 
\item End Mirrors (EM) - Suprasil 312, a fused silica grade of lower optical quality (and cost) has been used for the end mirrors, as in this case the mirrors will be reflecting most of the light. The only constraint in this case is the substrate mechanical quality factor, which has to be sufficiently high as to avoid increasing the thermal noise above the level determined by the mechanical losses in the coating \cite{ref:mir5}.
\item Beam Splitter (BS) - The BS will be 55 cm in diameter and 6.5 cm thick. A high-quality fused silica grade (Suprasil 3001) has been chosen. This type of Suprasil is particularly suitable for the BS because it is an optically-isotropic 3D-material. It is highly homogeneous and has no striations in all three directions. 
\item Compensation Plates (CP) and Pick-off Plate (POP) - These components have been machined from the Virgo+ input mirrors, made of Suprasil 312 SV. The absorption measured on these silica substrates was lower than 1 ppm/cm.
\item Power/Signal Recycling mirrors (PR/SR) - The PR/SR are 35 cm in diameter and 10 cm thick and are made of Suprasil 312.
\end{itemize}

\begin{figure}[ht]
\centering
\includegraphics[width = 0.49\textwidth]{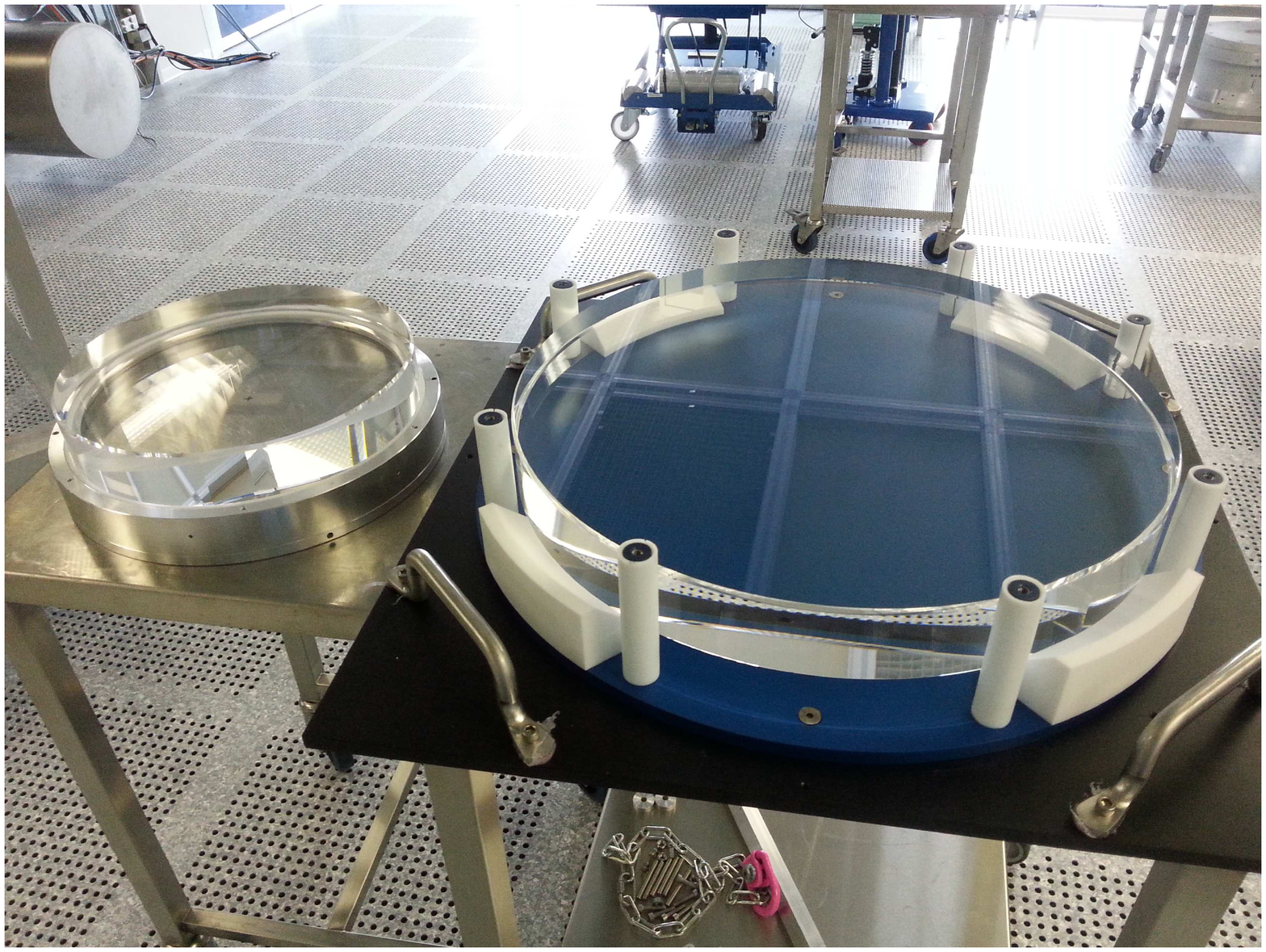}\hfill
\includegraphics[width = 0.49\textwidth]{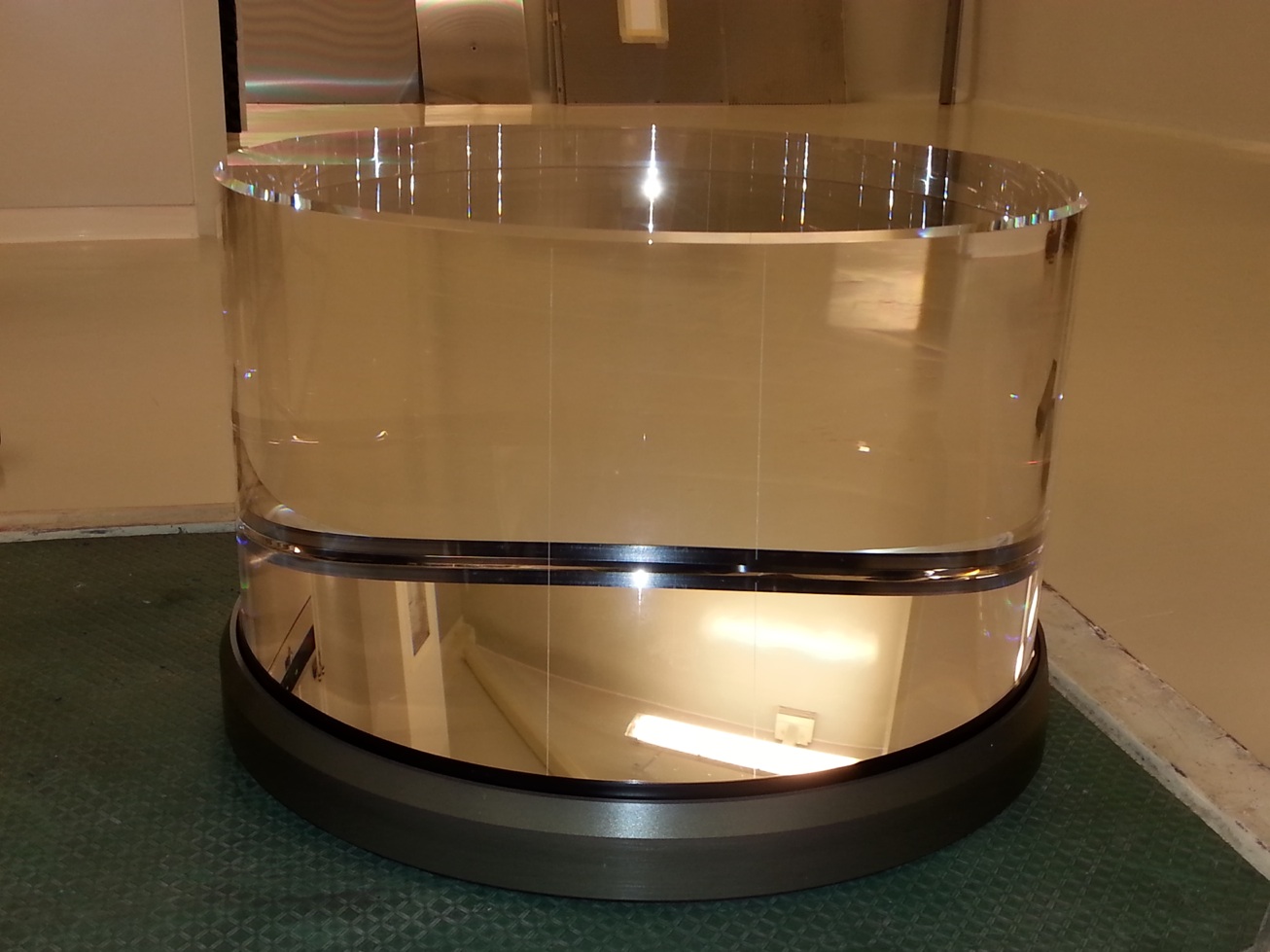}
\caption{\sf LEFT: Power Recycling substrate (35 cm in diameter, left) beside the large Beam Splitter of 55 cm. RIGHT: a test mass.}
\label{MIRphoto1}
\end{figure}

All the substrates needed for the interferometer, as well as the spare parts, were produced and delivered by HERAEUS at the end of 2012 (Figure \ref{MIRphoto1}).

\subsection{Polishing}

The polishing quality is characterized by two different parameters: the flatness and the micro-roughness. The first parameter gives the RMS of the difference between the perfect surface (typically a sphere for spherical mirrors) and the actual surface as measured by a phase map interferometer (for Virgo a flatness of a few nm was achieved \cite{ref:mir2}). The second parameter gives a measurement of the mirror surface roughness at small-scale lengths, from a few microns, up to about 1 mm (of the order of 0.05 nm in Virgo).

Both effects contribute to the scattering of light from the fundamental mode to higher order modes and generate losses and extra noise. Depending on the difference in the losses between the two cavities, these could be the source of finesse asymmetry and contrast defect thus modifying the constraints on other subsystems.

To meet the round-trip losses requirement in the AdV cavities (75 ppm) a flatness requirement of 0.5 nm RMS on 150 mm diameter  for the arm-cavity mirrors (IM, EM) was set. The shape of the Power Spectral Density (PSD) of the surfaces is important too, as different PSD shapes causes different losses in a FP cavity with an equal flatness. Thus, it was required that the RMS in the frequency range 50 m$^{-1}$ - 1000 m$^{-1}$, must not exceed 0.15 nm.

The flatness specifications for all of the other optics were set to be lower than 2 nm RMS on 150 mm diameter.

The first polished substrates were delivered by ZYGO at the beginning of 2014.

In order to characterize the large substrates before and after coating, some upgrades of the existing metrology benches were made (new sample holder for the CASI scatterometer; new, stronger motors for the absorption bench).
To be able to measure the flatness of the AdV substrates and mirrors at the level required (RMS flatness of 0.5 nm), a new interferometer at 1064 nm, coupled to an 18" beam expander was purchased. This interferometer uses a new technique, {\it wavelength shifting}, which makes it possible to characterize substrates with parallel surfaces (such as the IM), eliminating the rear-side interferences.

The first flat substrates (CP, POP, BS) were characterized with this new tool. A flatness of $\sim$0.5 nm RMS on the central 150 mm part was measured, which was much better than the specifications required for these optics ($<2$ nm RMS). At the time of writing, we also know that the polishing of the first Input Mirror (cavity mirror) achieved a flatness of 0.17 nm RMS on 150 mm diameter (power, astigmatism removed), compared to a requirement of 0.5 nm RMS.

\subsection{Coating}

The mirror coatings determine both the total mechanical losses of the mirrors and their optical losses.

\paragraph{Mechanical losses}
At present, the lowest mechanical losses measured for Ta$_2$O$_5$ coating have been those obtained with Ti doped Ta$_2$O$_5$. The losses value is within the AdV requirement of 2.3 10$^{-4}$ \cite{comtet}. One option to further reduce the mechanical losses involves optimizing the thickness of the layers of Ta$_2$O$_5$ and SiO$_2$ (we will refer to this as {\it optimized coating}). Since the Ta$_2$O$_5$ is the more lossy material, it is possible to reduce the mechanical losses of the multi-layer by reducing the amount of Ta$_2$O$_5$ and increasing the amount of SiO$_2$. For a given required reflectivity, it is possible to find an optimum combination. The coating machine is able to produce such optimized multilayers with a reasonable accuracy. The {\it optimized coatings} for the IM and EM will be used for AdV.

\paragraph{Absorption losses}
The high-reflectivity mirrors (EM, IM) currently have an absorption level between 0.3 ppm and 0.4 ppm at 1064 nm, thanks to the use of Ti doped Ta$_2$O$_5$ and {\it optimized coatings}  \cite{ref:mir8}.

\paragraph{Coating and finesse asymmetry}
Test masses of the same kind (IM or EM) are coated together in order to have the same optical performances. In this configuration, the difference between the transmission of the two IM or EM will be lower than 1\% and, consequently, the AdV cavities will be very similar. Otherwise, the asymmetry in finesse and power on the dark fringe might be too large. The finesse asymmetry is dominated by the transmission mis-match between the two IM rather than the losses induced by the flatness of the cavity mirrors (as in AdV, the mirror RMS flatness is very low).

\begin{figure}[ht]
\centering
\includegraphics[width = 0.45\textwidth]{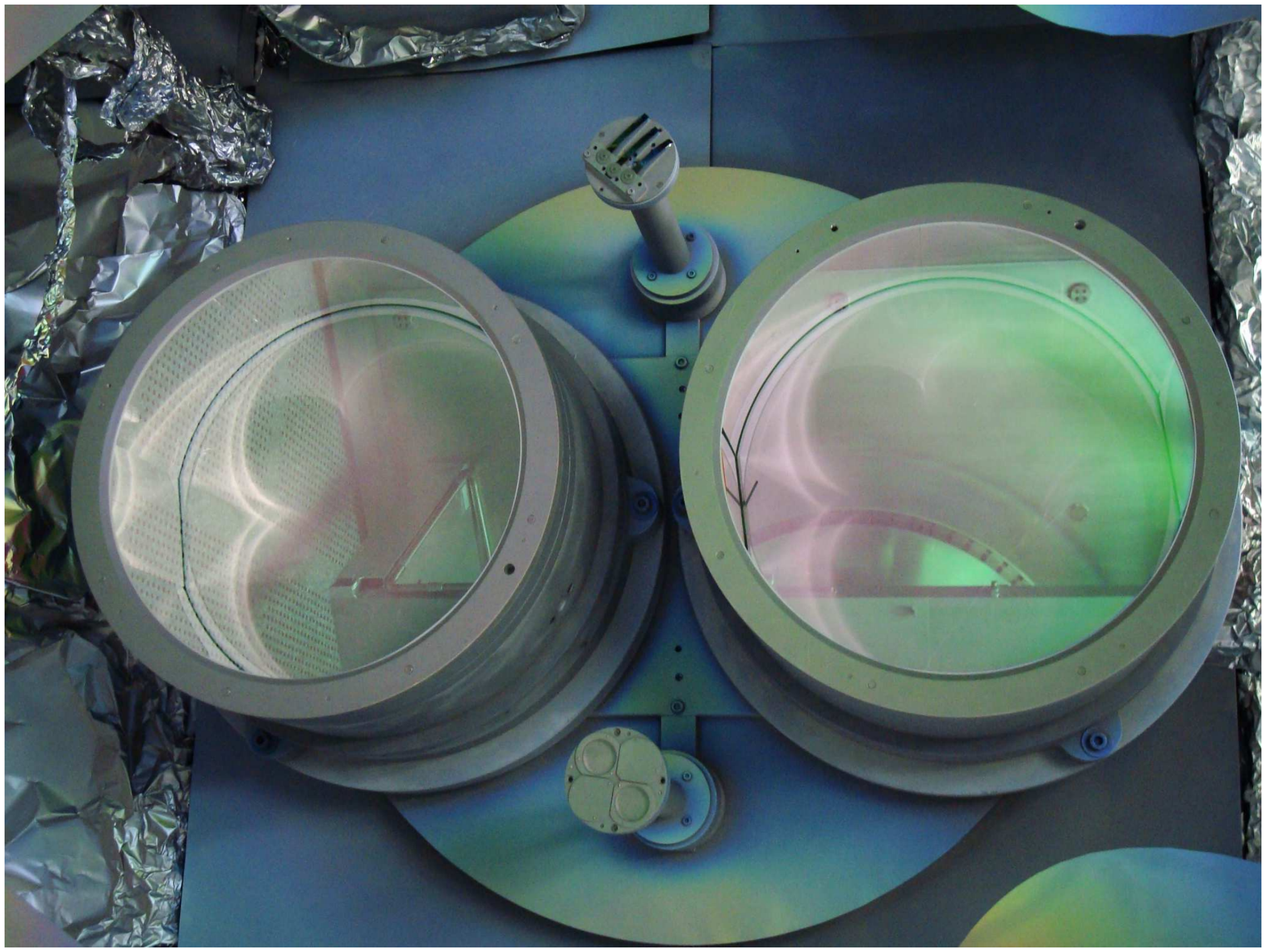}
\caption{\sf Planetary motion of the large substrates inside the coating chamber.}
\label{MIRphoto4}
\end{figure}

\paragraph{Coating uniformity}
The coating must not spoil the flatness requirements set for the polishing. Therefore, a considerable uniformity in the deposition process is needed. The only possible solution to obtain this uniformity on two large substrates at the same time, is to use a planetary motion coupled to a masking technique. A new planetary system was manufactured and installed in the LMA large Virgo coater (Figure \ref{MIRphoto4}). Optical simulations have shown that a flatness of 0.5 nm RMS can be reached when using this technique.

\section{Managing thermal aberrations and optical defects} \label{sec:tcs}
\subsection{Introduction}

Thermal lensing in the optics that are crossed by the probe beam was observed in both Virgo~\cite{Virgo1} and LIGO~\cite{Ligo1} and required the installation of Thermal Compensation Systems~\cite{Mau} (TCS). Advanced detectors will be characterized by a higher circulating power (from 20 kW in the initial interferometers to 700 kW in the second generation detectors) and thermal effects will thus become even more relevant. Besides thermal effects, optical defects in the various substrates of the recycling cavity and figure errors on reflective and transmissive surfaces contribute to the aberrations, as do spatial variations in the index of refraction of the substrates.

Such effects change the cavity mode, thus spoiling the matching between the laser and the power-recycling cavity and leading to a decrease of the recycling cavity gain and therefore in the sideband power. Since sidebands are used to extract the auxiliary control signals, thermal lensing affects the possibility to operate the detector at high input powers. The ultimate consequence is a loss of signal-to-noise ratio at high frequencies due to the increase of shot noise. 

Thermal expansion will change the profile of the high-reflectivity surface, creating a bump in the center of the test mass faces. The optical simulations show that, to maintain the arm-cavity mode structure, it will be necessary to control the radii of curvature of all test masses within $\pm$2 m from the initial RoC \cite{ADV:TDR}.

In Advanced Virgo, TCS will need to compensate for optical aberrations in the power-recycling cavity and to tune the RoC of the test masses, acting on both input and end test masses.

A useful way to picture the optical distortion effect, is to use the fractional power scattered out from the TEM$_{00}$ mode~\cite{Hello, vinet}, termed ''coupling losses'', and the Gaussian-weighted RMS of the optical path length increase.

In Advanced Virgo, the sideband field coupling losses, due to all aberrations, would amount to $\sim $50\%,  corresponding to an RMS of about 125 nm.
The Advanced Virgo TCS needs to reduce the coupling losses by at least a factor of $10^3$ (corresponding, roughly speaking, to a maximum RMS of about 2 nm) to allow the correct operation of the detector at design sensitivity.

\subsection{Thermal compensation actuators}

\begin{figure}[h]
\begin{center}
\includegraphics[width=.95\textwidth]{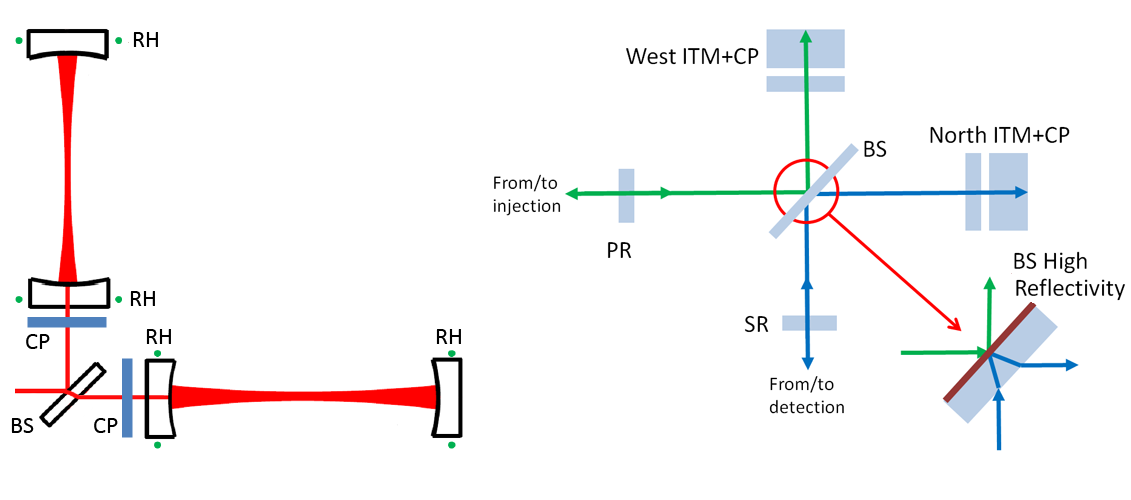}
\caption{Left: Actuation scheme of the Advanced Virgo TCS: blue rectangles represent the CP (heated by the CO$_2$ lasers), while the green dots around the test masses are the ring heaters. Right: Conceptual layout of the Hartmann Wavefront Sensor (HWS) probe beams in the recycling cavity. For the sake of clarity the beams have different colors, but the wavelength is the same.}
\label{fig:advtcs}
\end{center}
\end{figure}

The conceptual actuation scheme of the Advanced Virgo compensation system~\cite{AdVTCS1, AdVTCS2, AdVTCS3} is shown in Figure~\ref{fig:advtcs}.
The wavefront distortions in the recycling cavities will be corrected with an appropriate heating pattern generated by a CO$_2$ laser, the wavelength of which is almost completely absorbed by fused silica. For the control of the radii of curvature of all the test masses, ring-shaped resistive heaters (RH) will be used.

In Advanced Virgo, due to the sensitivity improvement, it will no longer be possible to illuminate the input test masses directly with the CO$_2$ laser, as was done with the initial detectors~\cite{Mau}. The displacement noise introduced by the intensity fluctuations of the CO$_2$ laser would spoil the detector sensitivity in the 50 Hz - 100 Hz frequency band. To make TCS compliant with Advanced Virgo noise requirement, the relative intensity noise of the CO$_2$ laser should be reduced to the level of $10^{-8}/\sqrt{\mathrm{Hz}}$ at 50 Hz, one order of magnitude below what it is possible to achieve with the present technology. This implies the need of an additional transmissive optic, a Compensation Plate (CP), upon which the compensating beam can act. The CP are placed in the recycling cavity, where the noise requirements are by a factor of $\pi/2F$ less stringent than in the Fabry-Perot cavity.

This scheme also makes it possible to reduce the coupling between the two degrees of freedom (lensing and RoC), and therefore to have a control matrix that is as diagonal as possible.

The optimum thickness of the CP is a trade-off, which has been reached by minimizing the heat that escapes from it laterally and by taking into account the need to accumulate enough optical path length. The distance between CP and IM is 20 cm, which makes it possible to minimize the radiative coupling between the two optics. In fact, the heated CP radiates heat towards the test mass. The heating of the IM is uniform, but since the side of the input mirror radiates a part of the heat, a radial temperature gradient is established. This gives rise to an increase in optical path length, which adds to the thermal lensing.

The position of the RH along the barrel of the IM is defined so as to maximize its efficiency.

In order to optimize the heating pattern to be applied to the CP, the aberrations have been classified according to their symmetry properties, regardless of their origin:
\begin{itemize}
  \item aberrations with cylindrical symmetry;
  \item non-symmetric defects.
\end{itemize}

Studies relating to the optimization of the heating pattern have been carried out with a Finite Element Model (FEM). For those aberrations with cylindrical symmetry, the modeling has shown that the optimum heating pattern~\cite{AdVTCS3} would reduce the residual coupling losses to about 6 ppm (about 0.1 nm RMS), thus leading to a reduction factor of about $10^5$, with about 18 W of CO$_2$ power falling on the compensation plate.

For the non-symmetric optical defects, full-3D modelling is required, making this kind of simulation rather computationally expensive. The results of the optimization procedure show that, by depositing heat in the right CP positions, the residual optical path length RMS can be reduced by a factor of 20 for spatial frequencies below 40 m$^{-1}$~\cite{AdVTCS3} and amounts to 0.35 nm, well within the requirements. The method selected to generate this heating pattern is based on a CO$_2$ laser scanning system. This technique, developed at MIT~\cite{law}, comprises a pair of galvanometer mirrors, to move the laser beam on the surface of the CP, and an acousto-optic modulator to modify the power content of the beam.

\subsection{Sensing for thermal compensation}

Aberrations in the recycling cavity optics will be sensed by several complementary techniques. The amplitude of the optical path length increase will appear in some interferometer channels, as will the power stored in the radio frequency sidebands. These are scalar quantities that can only give a measurement of the amount of power scattered into higher order modes. Furthermore, phase cameras~\cite{phase1,daypc} will sense the intensity distribution and phase of the fields in the recycling cavity (carrier and sidebands). 

Each optic with a significant thermal load will be independently monitored.  The HR face of each test mass will be monitored in off-axis reflection for deformation. The input test mass/compensation plate phase profile will be monitored on an on-axis reflection from the recycling-cavity side. The TCS control loop will then use a blend of all of the signals from the different channels.

The TCS sensors, dedicated to the measurement of thermally induced distortions, consist of a Hartmann Wavefront Sensor (HWS), and a probe beam (at a different wavelength to the interferometer beam), the wavefront of which contains the thermal aberration information to be sensed.

The Hartmann sensor selected for Advanced Virgo has already been developed and characterized on test bench experiments and in the Gingin High Optical Power Test Facility for the measurement of wavefront distortion~\cite{brooks_hws}. This sensor has been demonstrated to have a shot-to-shot reproducibility of $\lambda$/1450 at 820 nm,  which can be improved to $\lambda$/15500 with averaging, and with an  overall accuracy of $\lambda$/6800~\cite{munch}.

The conceptual layout of the HWS beams in the recycling cavity is shown in Figure~\ref{fig:advtcs}. In the picture, the beams have different colors for the sake of clarity; the wavelength is the same for both beams. The beams will be injected/extracted from the injection and detection suspended benches and superposed on to the main interferometer beam with a dichroic mirror at the level of the mode-matching telescopes.  This scheme allows for on-axis double pass wavefront measurement (which increases the signal-to-noise ratio by a factor of two) and makes it possible to probe all of the optics in the recycling cavity.

Additional optics are necessary to fulfill the main optical requirements for HWS beams: to image a plane around the IM HR surface on the Hartmann plate, to illuminate the IM with a 10 cm-in-size Gaussian beam and to match the optimal beam size on the sensor.
The two sensing beams are separated by the BS HR coating.
The beam from the detection bench will also sense the BS thermal lensing, thus allowing for its correction on the north arm CP. 

\section{Mirror suspensions} \label{sec:pay}
The mirror is suspended by four wires to a metal body (the {\it marionette}), which is moved by an array of coil-magnet actuators to control the position of the mirror itself. The marionette is suspended by a central maraging steel wire from the last filter in the Super Attenuator chain, named {\it Filter 7}. Control forces can be exerted from the {\it Filter 7} on both the marionette and the mirror. This system, consisting of mirror, marionette, associated suspensions and actuators, is referred to as the {\it payload}. In AdV additional components requiring seismic isolation or control must also be suspended from the payload: baffles, compensation plates and ring heaters. Though the design concept is the same for all of the payloads, the details are different, depending on the components to be suspended and the size of the mirror.

\subsection{Tests on the Beam Ssplitter payload}

\begin{figure}
\centering{\includegraphics[scale=0.35]{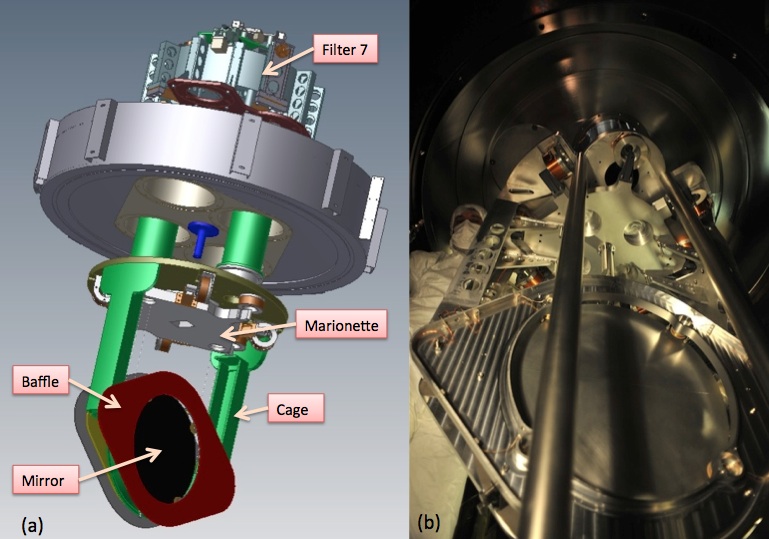}}
\caption{(a) Design of the new Beam Splitter payload. (b) The prototype of the Beam Splitter payload ready to be integrated with a Super Attenuator for testing.}
\label{fig:PAY_BS}       
\end{figure}

The payload for the BS was the first to be produced and has been used as a test bench for the adopted design. Following a first assembly for mechanical tests, the BS payload was suspended from a Super Attenuator, and a series of tests were performed:

\begin{itemize}
\item{\bf Controllability:} using optical levers  we controlled the position of both the mirror and the marionette, checking the recoil on the actuation cage by means of the sensors dedicated to the last seismic filter, where the cage is fixed. Concerning the use of the smaller magnets adopted for AdV, a first test of hierarchical control of the payload was performed  successfully. Namely, using the available actuators and control bandwidths compatible with those foreseen for AdV, the position accuracy of the baffles and that of the mirror were compatible with the operation requirements;
\item{\bf Pendulum Q:} the upper stage of the mirror suspension, the marionette, was optimized in order to minimize related dissipations. Q values of $2 \times 10^4$ have been achieved;
\item{\bf Coupling with electro-magnetic stray fields:} a detailed electromagnetic finite element model was developed and a measurements campaign was carried out on the BS payload, to tune the model and to further develop it for the next payloads. To this purpose, the response of the BS payload to a variable magnetic field, produced by a large coil in its proximity, was measured, with the goal of possibly mitigating the eddy current effects induced in the conductive parts of the payload. In addition, the experimental results have been used to predict the influence of measured environmental magnetic noise on the BS. No relevant contribution to the sensitivity curve is due to magnetic noise on this payload.
\end{itemize}

\subsection{The Input Mirror payload}

\begin{figure}
\centering{\includegraphics[scale=0.5]{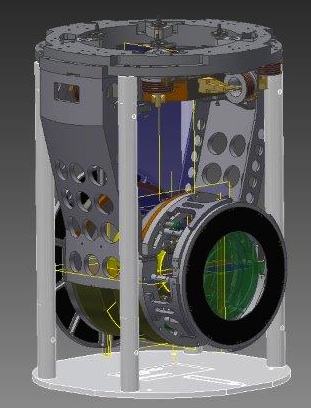}}
\caption{Design  of the AdV payload for Input Mirrors.}
\label{fig:PAY_input}       
\end{figure}

In the following, we describe the most complex payload: the suspension of the mirror at the input of the Fabry-Perot arms (see Figure \ref{fig:PAY_input}). All other payloads (the End Mirror payloads, the Power Recycling payload, the Signal Recycling payload) are composed of a different arrangement of these same parts.

\subsubsection{The Compensation Plate Support}

\begin{figure}
\centering{\includegraphics[scale=0.45]{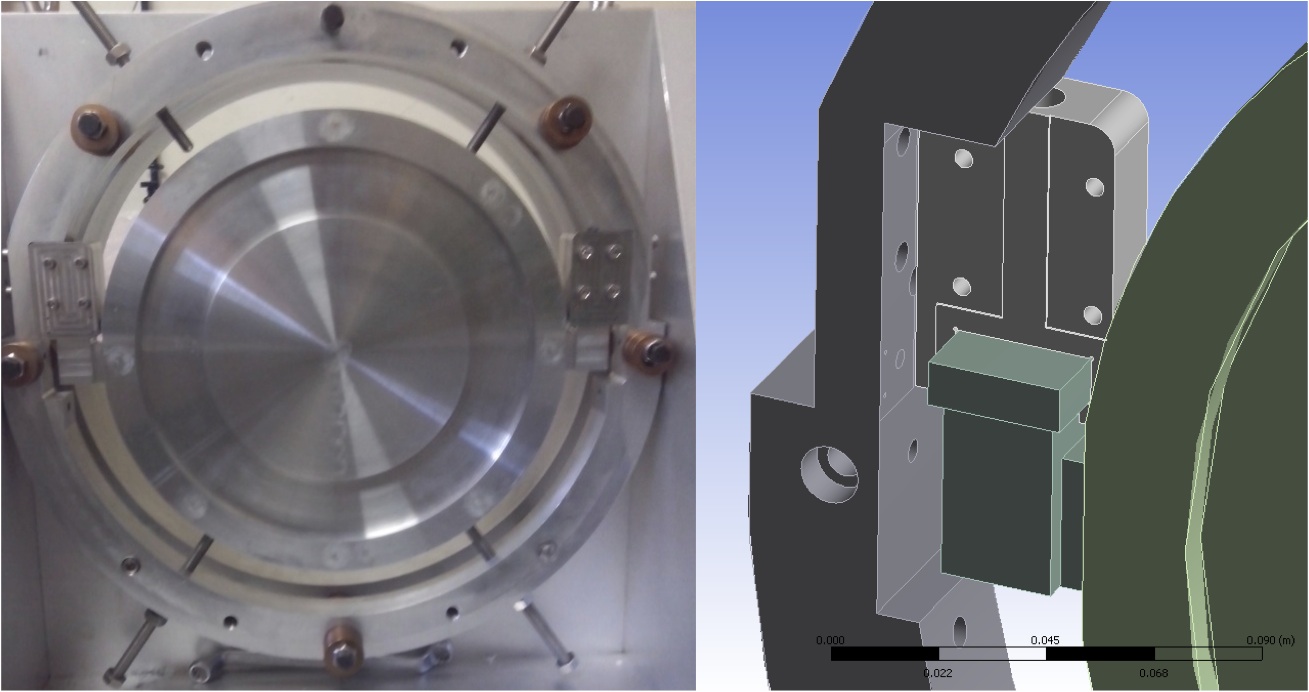}}
\caption{Prototype of the Compensation Plate suspension and detail of the lateral clamps.}
\label{fig:PAY_CPsusp}       
\end{figure}

A Compensation Plate (CP) must be suspended in the recycling cavity in front of the back face of each IM (see Figure \ref{fig:OSD:config}).
The CP is a fused-silica disk with a diameter of 350 mm and a thickness of 35mm.  The plate is placed co-axially with the input test masses at a distance of 20 cm from the Beam Splitter side. It is attached to the cage of the payload by means  of fused-silica {\it ears}, which are silicate bonded to the plate (Figure \ref{fig:PAY_CPsusp}). This kind of suspension is designed to incur a level of stress on the CP comparable to that of a standard mirror, so as to limit birefringence effects. Two stainless-steel clamps hold the other side of the ears and are rigidly connected to an external aluminum ring, which is fixed on to the supporting frame, bolted  to the Filter 7 (the {\it cage}). The CP suspension has been designed to make it possible to use frequencies inside the detection bandwidth, in order to avoid any coupling with the pendulum modes of the mirror. In this way, the CP behaves like any other part of the cage. A detailed study of the possible noise contributions to the sensitivity arising from the thermal motion of the CP and from possible excitation due to the Thermal Compensation System has been carried out. The results are described in \cite{ref:payCP1}.

\subsubsection{The ring heater and large baffle supports}

A Ring Heater (RH), see Section \ref{sec:tcs}, must surround the mirror. The RH is connected by rods to the ring holding the coils on the back of the mirror. Centering of the RH with respect to the mirror could be critical, so, as a consequence, a system enabling remote adjustment of the RH position along the vertical and horizontal axes has been implemented. 
The large baffles are clamped at the end of the payload (see Section \ref{sec:slc}). The supports are designed to provide clamped baffle resonance frequencies that are as high as possible, also taking into account the necessity to limit the total weight of the structure. For this reason, the baffle holders usually also have other functions: for instance, in the BS payload, they are also used as mirror coil holders, while in the input payload they also act as a counterweight, to balance the load of the CP support.

\subsection{The Monolithic Suspensions}

One of the main noise sources limiting the sensitivity of gravitational wave interferometers is the thermal motion of the mirror pendulum (the mirror and its suspension) and of the bulk of the mirror itself (both substrate and coating)~\cite{ref:payTherm2}.  A significant contribution also comes from the friction on the clamping points of the suspension \cite{ref:payTherm5,ref:payTherm6}.  The best way devised so far to reduce these sources of noise, has been to use fused-silica wires, attached to the mirror by welding or using silica bonding, which can reproduce the connection between materials at the molecular level \cite{ref:payTherm7,ref:payTherm8,ref:payTherm9}. We refer to this design as  {\it monolithic suspensions}.

\begin{figure}
\begin{center}
\includegraphics[scale=0.35]{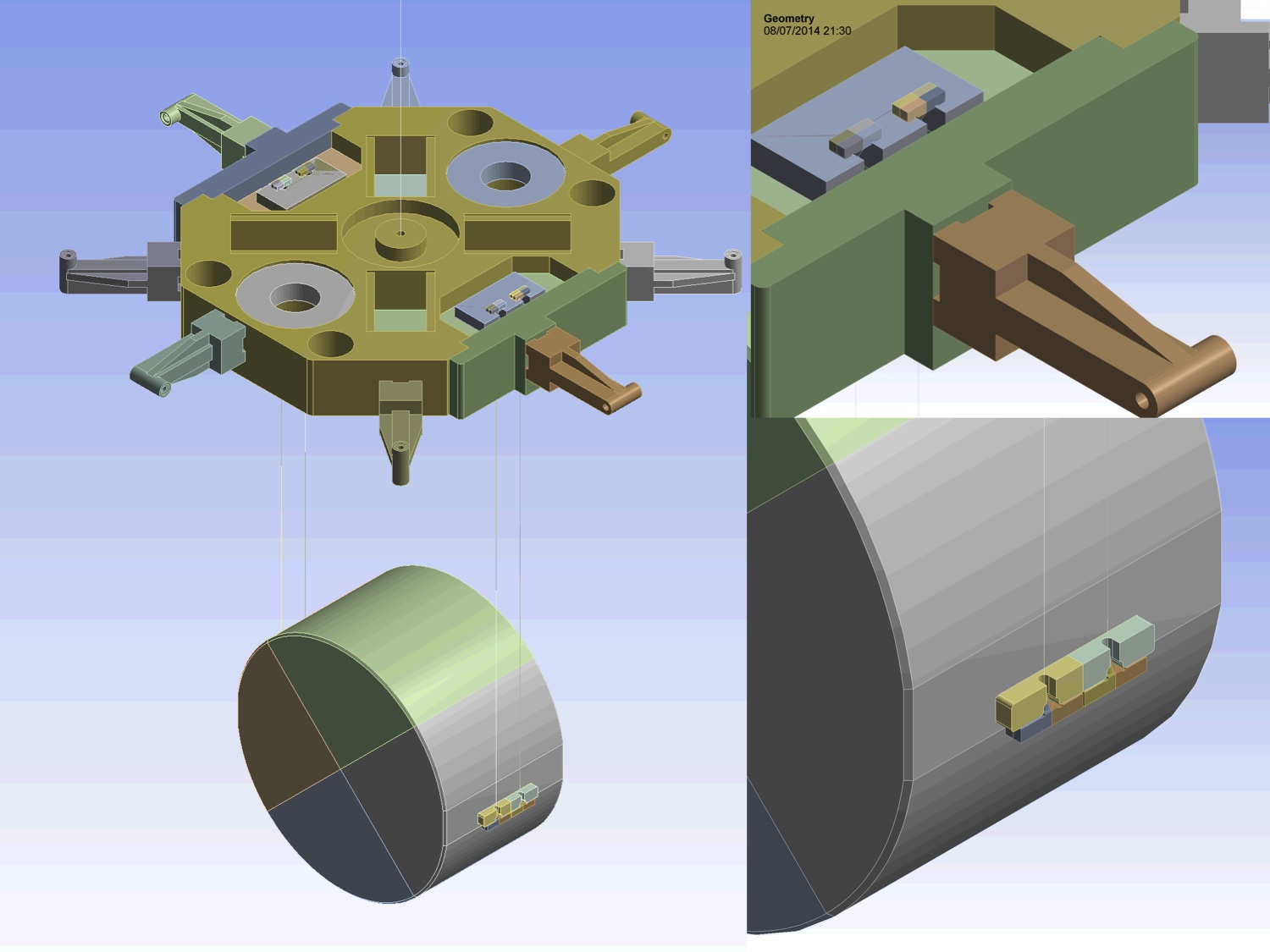}
\caption{Design of the monolithic suspension for input payloads. Two T-shaped fused silica blocks connect the fiber to the ears glued to the mirror and to a steel plate coated with fused silica.}
\label{fig:PAY_MonoScheme}
\end{center}       
\end{figure}

In the suspension design developed for AdV (see Figure \ref{fig:PAY_MonoScheme}) the ends of each fiber are welded into two T-shaped fused silica blocks ({\it the anchors}), which are then connected to the marionette on one side and to the mirror on the other side. On the mirror side, the anchor is glued with a silica bonding technique to a section of fused silica protruding from the mirror ({\it the ear}). Machining the ear out of the mirror is rather difficult, so the ear itself is silica bonded to the mirror before the fiber is assembled.  If silica bonding is correctly applied, fiber, anchor, ear and mirror are a continuous body, and we have a monolithic suspension. The other end of the fiber, welded to another anchor, is connected to the marionette with a stainless steel interface (the upper clamp assembly).

\begin{figure}
\begin{center}
\includegraphics[scale=0.35]{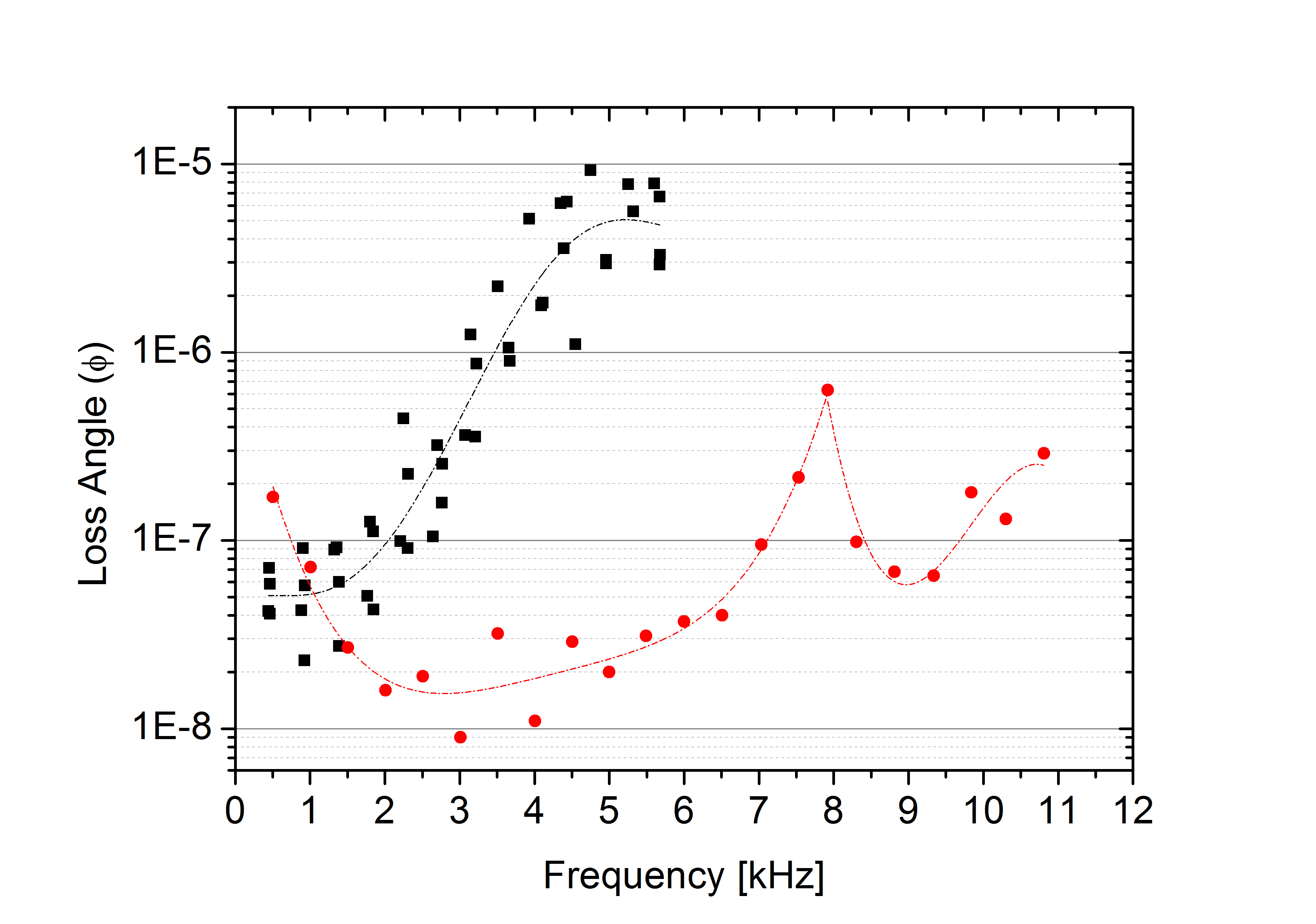}
\caption{Comparison of the loss angle values ($\phi$) of the violin modes measured in Virgo+ (white circles) and in the Perugia facility where the new configuration for AdV has been tested (black squares). At low frequency (below 2 kHz) and at 8 kHz there are two dissipation peaks, due to recoils of the structure supporting the fiber. Relative error on $\phi$ values for each point is about 1\%.}
\label{fig:PAY_01}  
\end{center}     
\end{figure}

In 2009, a solution similar to that described above, was implemented in the {\it Virgo+} configuration \cite{ref:payVirgo+}. The four mirrors of the 3 km Fabry-Perot arms were in operation for about two years, suspended to 0.3 mm diameter glass fibers, before being disassembled to start the upgrade of Virgo to the Advanced state. While this new set up was a success as a technical achievement, the  actual values of Q measured for violin modes were, in general, a factor of 10 lower than expected. We identified the probable cause of these extra dissipations as being in the design of the upper clamp of the fibers \cite{ref:payVirgo+TN}. The design of the upper clamps was revised in order to be able to use monolithic suspensions in Advanced Virgo. The revision was guided by FEM simulation models which were used to optimize the dissipation paths of the new suspension, and by a series of tests performed in a dedicated facility. In the new design, we tried both to implement a reliable way of connecting the fused silica parts to the metal ones, and to limit as much as possible the losses due to the friction between metal and fused silica surfaces. The solution we have selected consists of a $\simeq 1\ kg$ fused silica block, which is  silica bonded to a stainless-steel plate coated with fused silica. The tests revealed an improved behavior of the dissipations of the violin modes of the fibers, with respect to that observed in Virgo+, at all frequencies, except for two intervals around 500 Hz and 8 kHz, where dissipation peaks due to the recoil of the structure could be observed (Figure \ref{fig:PAY_01}). Measuring the $Q$ of the violin modes \cite{ref:payViolinTN} makes it possible to infer the intrinsic dissipations of the suspension and hence the thermal-noise contribution to the sensitivity of the mirror-pendulum motion. Currently, the last details of this configuration are being tested, in order to be implemented in the AdV payloads.

\subsection{Fiber production and test}

Fiber length, fiber stiffness and the position of bending points are constrained by the available space and the choice of resonant frequency of the various modes of the suspension, which are driven strongly by control issues. For instance, the vertical bouncing frequency of the last stage represents the lower limit of the detection band because, although the vertical to horizontal coupling is small (a  value of  $10^{-3}$ is assumed), the vertical oscillation does not have any dilution factor \cite{ref:payDilFac}. So, in order not to spoil the sensitivity, this frequency must be kept below the low frequency limit of the detection bandwidth (i.e. 10 Hz). This can be achieved with a careful choice of the dimensions and shape of the fiber. Also, in order to have the smallest possible number of resonant modes in the detection band, the first violin mode should be as high as possible. On the other hand,  the frequency of the first violin mode and of its harmonics, depends on the fiber cross-section and then ultimately on its breaking stress. As a further constraint, we must also take into account thermoelastic dissipations \cite{ref:payTherEl}, the contribution of which to thermal noise can be made negligible with a suitable choice of fiber diameter.

\begin{figure}
\begin{center}
\includegraphics[scale=0.6]{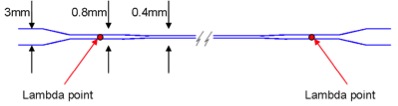}
\caption{Sketch of the fused silica fiber shape in AdV.}
\label{fig:PAY_02}     
\end{center}  
\end{figure}

The AdV mirrors will be suspended to a $400\ \mu m$ diameter fiber, having a load stress of 780 MPa, which is about the same as that of the fibers used for Virgo+ and well below 4-5 GPa, the breaking stress estimated by the tests carried out before Virgo+ fiber production. With this diameter, the bouncing frequency is about 6 Hz, and the first violin mode frequency is at about 430 Hz. However, this is not the optimal value to limit thermoelastic dissipations. So, the fiber is also Òdumbbell-shapedÓ, as shown in Figure \ref{fig:PAY_02}. In this scheme, most of the fiber length has a diameter of $400\ \mu m$, while two short heads, with a diameter of $800\ \mu m$, are in place at both fiber ends. In this way, since most of the bending energy is stored close to the bending points, which are inside the $800\ \mu m$ regions (heads), the cancellation mechanism can minimize the thermal noise due to the thermoelastic effect.
Moreover, the two 3 mm-thick ends (bars), can be used to weld the fiber to the rest of the suspension and, also, to set the bending point on the right position with respect to the mirror and the marionette. In any case, these regions must not be longer than a few millimeters, to limit the bending energy stored there and to make the bending point position independent of the welding shape. A finite element model, implemented by a specifically developed code \cite{ref:payFiberSim}, was used to simulate the elastic behavior and to estimate the thermal noise.

Each fused silica  fiber is produced starting from commercially available, 10 cm-long and 3 mm-thick, high-purity fused-silica cylindrical bars  (suitable materials are Herasil or Suprasil). 
This small cylindrical bar is clamped at both ends and heated in the central region using a 200 W $CO_2$ commercial laser with a $10.6\ \mu m$ wavelength. Subsequently, the two ends are pulled apart, extending the fiber to the desired length and shape. This process is performed by a dedicated machine developed at the University of Glasgow and duplicated in Virgo and further modified to improve the laser focusing.

Following production, the fibers are tested to a load at least double the operation load. Then, if the fiber survives, its bending length is measured. The bending length $\lambda$ of the suspension is the distance of the fiber bending point from the clamped end. Positioning the bending point on the center of mass plane of both the marionette and the mirror allows minimum coupling between the different degrees of freedom.

After this validation, the fibers are then placed in position, clamping the upper part to the marionette and inserting the lower anchor below the lateral supports bonded to the mirror. In the end the anchor and the supports are bonded together with  silicate bonding. 

\subsection{Payload assembly and integration}

As soon as the silica bonding is cured, the monolithic suspensions are integrated with the rest of the mechanical elements described above. The complete payload is inserted into a container with controlled humidity and cleanliness and transported into the clean room under the {\it tower}, which is a vacuum chamber surrounding the Super Attenuator chain. During transport there is a continuous monitor of many physical parameters, such as acceleration, temperature and humidity. The actual suspension of the mirror at the end of the Super Attenuator chain requires several hours of work.

\section{Mirror isolation and control} \label{sec:sat}
The seismic isolation of the AdV mirrors will be achieved by the Super Attenuator (SA), a hybrid (passive-active) attenuation system, 
capable of reducing seismic noise by more than 10 orders of magnitude in all six degrees of freedom (DoF) above
a few Hz. A detailed description of the SA and its performance are given in~\cite{SAT3}.

\subsection{Mechanics} 
Since the performance of the SA measured in Virgo is compliant with the AdV requirements, no major changes in the mechanical design have been introduced. 

\begin{figure}
\begin{centering}
\includegraphics[width=0.4\textwidth]{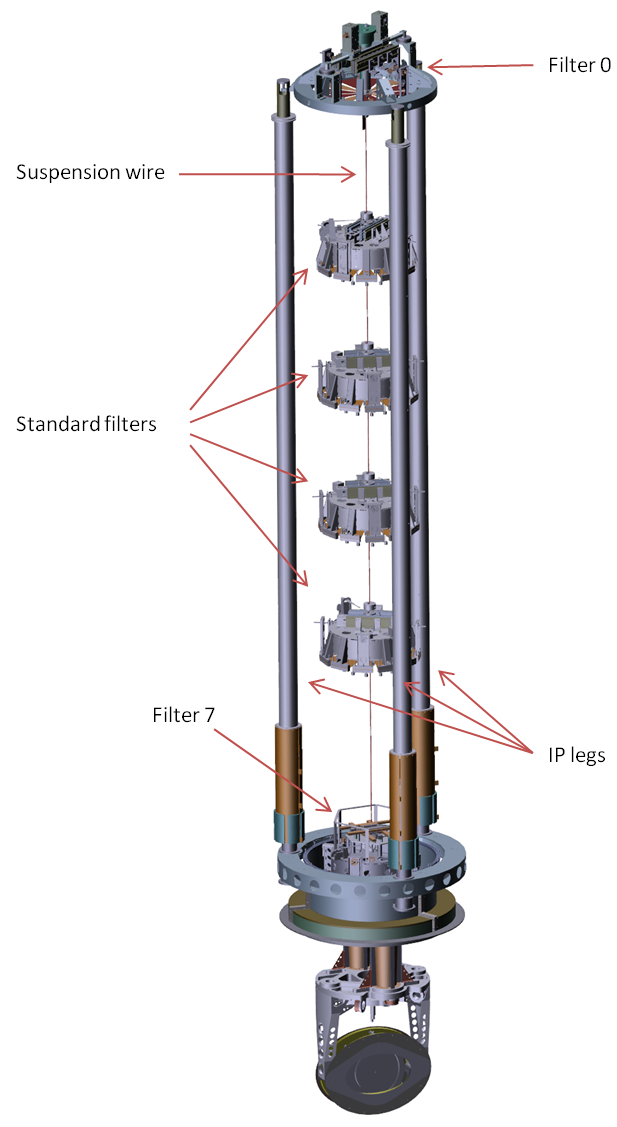}
\par\end{centering}
\caption{\label{fig:SA}The AdV Super Attenuator. We can distinguish, from top to bottom, the three legs of the inverted pendulum, the Filter 0, the top ring, the passive filters 1 to 4, and the mirror suspension. This last stage, composed of the steering filter, the marionette and the actuation cage, is dedicated to the control of the mirror position for frequencies $f>$10 mHz. }
\end{figure}

The AdV SA mechanical structure, shown in Figure \ref{fig:SA},
consists of three fundamental parts: 
\begin{itemize}
\item the inverted pendulum (IP);
\item the chain of seismic filters;
\item the mirror suspension.
\end{itemize}

The IP~\cite{IP} consists of three 6 m-long aluminum monolithic {\it legs}, each connected to ground through a flexible joint and supporting an inter-connecting structure (the top ring) on its top. The top ring, which is a mechanical support for an additional seismic filter, is called {\it Filter 0} and is similar to those used in the chain. It is equipped with a set
of sensors and actuators, which are placed in a pinwheel configuration and which are used to actively damp the IP resonance modes. Two kinds of sensors specifically designed for Virgo/Virgo+, are used: three Linear Variable Differential Transformer (LVDT), with a sensitivity of $10^{-8} \textnormal{m}/\sqrt{\textnormal{Hz}}$ above 100 mHz; and five accelerometers, three horizontal and two vertical, with a sensitivity of $3\cdot10^{-10} \textnormal{m}/\textnormal{s}^{2}/\sqrt{\textnormal{Hz}}$ below a few Hz~\cite{acc}. Two sets of actuators, coil-magnets and motor-springs, are also installed on the top ring. 

The chain of seismic filters is suspended from the Filter 0 and is composed of an 8 m-long set of five cylindrical passive filters,
each one designed to reduce the seismic noise by 40 dB both in the horizontal and vertical degrees of freedom, starting from
a few Hz. 

The payload is suspended from the last seismic filter of the chain, called {\it Filter 7} or the steering filter. 
Given that in AdV, as described in Section \ref{sec:pay}, the last stage of the SA has been completely redesigned, several changes have been introduced in the mechanics of the Filter 7. In order to keep the same total weight for the payload-Filter 7 assembly, the steering filters supporting the Fabry-Perot and Signal-Recycling test masses are now 55 kg lighter than in Virgo/Virgo+.
The mass reduction has been obtained by using a shorter drum steel structure for the filter body and a lighter crossbar, which is
the mechanical structure designed to support the permanent magnets used to lower the vertical resonance frequency of the standard filters. Furthermore, six LVDT and six coil-magnet actuators are now installed in a pinwheel configuration on the ring surrounding the Filter 7, in order to control its motion in all DoF.

\subsection{SA control system}

As, between 200 mHz and 2 Hz, seismic noise is amplified by the resonance modes of the isolation stages, 
an active control of the SA is needed. As in Virgo/Virgo+, the system has been designed using a hierarchical strategy, regulated by the dynamic range of the actuators.
In the ultra-low frequency band ($f<$10 mHz), actuation will be performed on the IP top stage along the $z$, $x$, and $\theta_{y}$
DoF (as is customary, we consider $z$ to be aligned with the suspended mirror optical axis) using three coils, placed in pinwheel configuration on the top ring.
In the 10 mHz$<f<$1 Hz band, the control will act both on the Filter 7, along six DoF using six actuators, and on the \emph{marionette} 
along four DoF ($z$, $\theta_{x}$, $\theta_{y}$, $\theta_{z}$) using eight coils. For frequencies higher than a few Hz, the force will be applied directly to the mirror along $z$, 
using four coils mounted on the actuation cage, which also allow for tiny corrections along $\theta_{x}$ and $\theta_{y}$. 

In Virgo/Virgo+, the controllers were designed using classical Nyquist-like techniques, diagonalizing the sensor-actuator space with static matrices in order to obtain a set of single-input single-output
(SISO) systems~\cite{ID}. In AdV, a multivariable design approach, based on optimal predictive regulators~\cite{Kalman}, will be used. This new approach will have the advantage of being a user-independent and completely automatic design process, with the possibility  to optimize the feedback performance for both the mixed and diagonal term elements of the sensor/actuator transfer function matrix.

As in Virgo/Virgo+, the regulators will be implemented in a digital hard real-time control system based on Digital Signal Processors (DSP). 
The hardware has been completely redesigned for AdV and is constituted by MicroTCA boards, using the RapidIO bus, which integrates front-end electronics, data conversion and data processing in a single unit.
The processor chosen is the Texas Instruments TMS320C6678, a high-performance multicore fixed and floating point DSP with a total computing power of 160 GFLOPs in single precision at 1.25 GHz.
Data conversion is based on the Texas Instruments ADS1675, a 24-bit $\Sigma-\Delta$ analog-to-digital converter with a maximum throughput of four MSPS, and, on the AD1955 Analog Devices, a 24-bit $\Sigma-\Delta$ digital-to-analog converter (DAC), which can process both PCM and DSD data formats. The maximum sampling frequency that can be implemented on the control system is 640 kHz.
A series of multi-channel low-noise power amplifiers, known simply as {\it coil drivers}, are used to drive the coil-magnet pairs installed on the SA. Every coil driver has an on-board DSP and two distinct sections, each driven by an independent DAC channel: one high-power section used for the lock acquisition of the ITF optical cavities, and one low-noise section for linear regime. In this latter mode, the coil drivers can supply up to 0.5 A with a few of pA/$\sqrt\textnormal{Hz}$ of noise. Each suspension will be typically connected to twenty boards, each hosting a TMS320C6678 DSP. The total computing power, in single precision, available for the control of the SA and the processing of its signals will be 3.2 TFLOPs.

\subsection{Tilt Control}

We know experimentally that, during earthquakes or poor weather conditions, seismic noise grows by up to 2 or 3 orders of magnitude in 100 mHz - 1 Hz band, with a maximum (the micro-seismic peak) between 400 and 500 mHz. Since the ground tilt is transmitted without any attenuation at the SA top stage, we estimate that, in these conditions, the angular component of the seismic noise can reach levels high enough to compromise the duty cycle of the interferometer. 
A tilt control of the SA will therefore be implemented in AdV.

To this purpose, a set of piezo-electric actuators (Physik Instrumente P-239.30), capable of providing a force of 4500 N with a dynamic range of 40 $\mu$m, are installed within the feet supporting the IP bottom ring. 
Three LVDT, which monitor the vertical displacement of the ring, are used in a closed loop with the piezos in order to increase the linearity of the actuators.

At the same time, a sensor capable of providing tilt measurements that are uncontaminated by translational components, is required to implement a tilt control of the suspension, as the SA accelerometers produce a signal proportional to a linear combination of horizontal acceleration and tilt. While several studies on mechanical gyroscopes with high sensitivity at low frequencies have been done~\cite{Tilt1, Tilt2}, 
the most promising angular sensor candidate in terms of angle/acceleration cross-coupling is the Hemispherical Resonator Gyroscope (HRG)~\cite{Tilt3, Tilt4}. 
An HRG, custom-made by the Russian firm MEDICON for AdV, is currently being tested.

\section{Laser} \label{sec:psl}

\subsection{Overview}
The AdV laser source is a high power (HP) continuous-wave laser, which is stabilized in frequency, in intensity and in beam pointing. The high-power requirement is necessary to overcome the shot-noise limit, while the stabilization brings the technical noises of the laser down to a level where they no longer mask the tiny gravitational wave signal. 
A laser power of at least 175 W in TEM$_{00}$ mode is required to meet the sensitivity goal.
In absence of a perfect symmetry in the cavities, the interference between the two arms of the Michelson reflects all of the frequency and power fluctuations of the laser. In order to detect such a small GW signal as $10^{-23}$ in relative strain, the relative frequency and power fluctuations of the laser have to adhere to the same order of magnitude. Then, for a laser light emitting at a 1 $\mu$m wavelength, we end up with a few $\mu \mathrm{Hz} / \sqrt{\mathrm{Hz}}$ for the laser frequency fluctuations, while in free-running conditions a quiet laser produces a noise of a few $\mathrm{kHz} / \sqrt{\mathrm{Hz}}$. This establishes a stabilization factor of 9 orders of magnitude, which is quite hard to obtain with a single stage of servo-loop. Therefore, a multi-stage servo-loop is then used for the frequency stabilization, with different references, ranging from a rigid Fabry-Perot cavity to the arms differential of the Michelson itself. 
To minimize laser power fluctuations, the Michelson operates on the dark fringe, but the DC readout method used in advanced detectors requires a small deviation from the dark fringe. Combined to the fluctuating radiation-pressure noise suffered by the mirrors, the laser-power fluctuations must be in the range of $10^{-9}/ \sqrt{\mathrm{Hz}}$, while the beam pointing has to be below $10^{-11}$ $\mathrm{rad} / \sqrt{\mathrm{Hz}}$.

\subsection{HP Laser for AdV}

Achieving such high-power output with a highly-stable laser is a serious challenge, which is managed by separating the two functions of HP output and stabilization: a low-power stable laser is used to transfer its stability to a HP oscillator by injection-locking and/or amplifying it through an internal laser process, helped by a slow servo-loop to prevent drifts. While, in the first case, the stable output of the low-power laser is automatically transferred to the HP laser, the amplifier case contains no filtering cavity and transfer of stability is achieved only in saturation regimes. 
The HP oscillators used so far in Virgo/Virgo+ have been based on Nd:YV04 (Neodymium-doped Yttrium Orthovanadate) crystals, which are the best choice for the 100W-class lasers. The first phase of the AdV project requires a medium-power laser of around 35-40W at the input of the mode cleaner, while the final phase will need about 200W output, such that at least 175W in TEM$_{00}$ mode can be delivered as requested. 
The medium-power laser is today an Nd:YVO4 oscillator amplifying a 20W injection-locked laser to deliver around 60W. The free-running frequency noise of this amplifier copies that of the master laser, which is a monolithic, commercial non-planar ring oscillator of 1W (Innolight NPRO). Though intrinsically stable in the short term, it will nevertheless be frequency stabilized to the Michelson arms.

To reach the ultimate power we plan to use amplifiers based on fiber technology. An R\&D project is in progress with the goal of delivering the final laser for installation in 2018. 

\subsection{Stabilization of the HP laser }

\subsubsection{Frequency stabilization}
The frequency stabilization acts on the piezo-transducer of the master laser. The error signal is a combination derived from a multi-stage servo-loop, using as references: the rigid Reference Cavity (RC), the Input Mode Cleaner (IMC) and the differential arms of the Michelson. 

\subsubsection{Power stabilization of the laser }
Comparing the light power fluctuations to a stable voltage gives the error signal, which is fed back to the HP amplifier pumping diode current  for the power stabilization. The principle sounds simple, but a voltage fluctuation measured out of a photodiode can contain, not only the light fluctuations, but also any other fluctuation due to surface inhomogeneity (e.g. air pressure fluctuations, dust crossing the beam). Therefore, a careful selection of photodiodes has to be made, and these must be used in a quiet environment, such as under vacuum. Furthermore, we know that resonant cavities act as low-pass first-order filters for laser amplitude and beam-jitter noise. For amplitude noise, the corner frequency of this filtering is equal to half the line-width of the cavity.
Therefore, placing the power-stabilization photodiode under vacuum after the IMC relaxes the constraints of the servo-loop in the MHz range and, in our case, amplitude and beam-jitter noises begin to be attenuated by the IMC above $800-900$ Hz. Nevertheless, a power stabilization servo-loop is still necessary inside the detection range, mostly below 500 Hz. The photodiodes have been designed to fulfill the specifications, i.e. a relative noise intensity (RIN) of $1.2 \times10^{-9} /\sqrt{\mathrm{Hz}}$ at the frequency of 30 Hz, which is the most stringent requirement. To reach this level of stabilization, the required shot-noise limit on the photodiodes has to be lower than $10^{-9} /\sqrt{\mathrm{Hz}}$, which requires an equivalent of about 400 mA. Due to the limited standing power of fast photodiodes, the light sensor is composed of two sets of two photodiodes sharing a total amount of 400 mA of photocurrent.  Two photodiodes are coherently combined and provide a measured noise floor limit of $1.3 \times10^{-9} /\sqrt{\mathrm{Hz}}$ with a 400 mA photocurrent. The two others will be used for out-of-loop verification. 

\subsubsection{Laser beam shape and jitter control}
The last control of the laser beam concerns the beam shape and the beam pointing noise. To lower the beam noise at the entrance of the Injection Bench and to filter out the high order modes inherent to HP lasers, a Pre-Mode-Cleaner (PMC) is positioned on the Laser Bench. A compromise has been found for its specifications between the finesse and the high level of stored light in a small beam waist. To avoid direct feedback to the laser, the PMC is a triangular non-monolithic cavity, which carries a piezo transducer to control its length relative to the laser frequency. This PMC has already been used for the Virgo+ phase and was able to work with a 50W laser, with a finesse of 500 in a beam of 500 $\mu$m. It is useful to remember that the PMC lies on a horizontal plane and therefore reduces the horizontal beam jitter.

\section{Light injection}  \label{sec:inj}
\subsection{Overview}

\begin{figure}[htb]
\centering
\includegraphics[viewport=0 150 850 520,width=1\textwidth,clip]{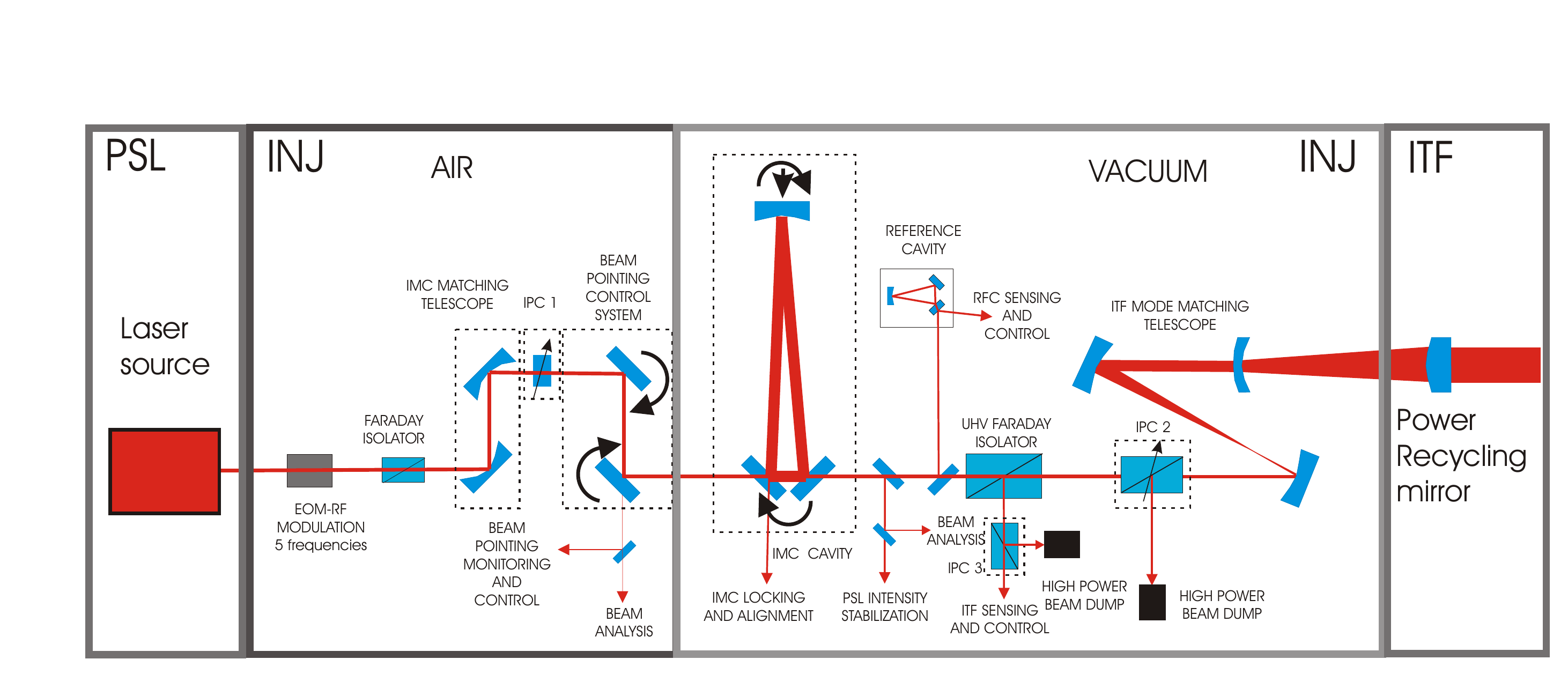}
\caption{\sf General input optics scheme.}
\label{INJ_baseline}
\end{figure}

Figure \ref{INJ_baseline} shows an overview of the {\it input optics} found between the laser and the interferomer and designed to provide a beam with the required power, geometrical shape, frequency and angular stability. The main requirements for the input optics are: 
\begin{itemize}
\item transmission to the ITF $>70 \%$ TEM$_{00}$;
\item  non-TEM$_{00}$ power $<5 \%$ ;
\item intensity noise     $2\times 10^{-9}$/$\sqrt{\rm Hz}$ at 10 Hz;
\item beam jitter         $ <10^{-10}$ rad/$\sqrt{\rm Hz}$  ($f>$10 Hz);   
\item frequency noise (for lock acquisition)         $<$1 Hz rms; 
\end{itemize}
 
An Electro-Optic Modulation (EOM) system provides the RF phase modulations needed for the interferometer (ITF) cavity control. A system based on piezo actuators and DC quadrant photodiodes, the Beam Pointing Control (BPC) system, has been developed to reduce, as much as possible, low-frequency beam jitter before it enters the vacuum system. Another system (EIB--SAS) reduces the beam jitter induced by the structural resonances of the optical-bench legs. A power-adjustment system (Input Power Control), consisting of a half waveplate and a few polarizers, is then used to tune the ITF input power. Some matching and steering of in-air optics is required to properly couple the beam with the in-vacuum suspended Input Mode Cleaner cavity (IMC). The IMC geometrically cleans the beam and reduces its amplitude and beam-pointing fluctuations before the ITF. The resonant IMC, the length of which is locked on the reference cavity (RFC), serves as a first stage of frequency stabilization for the main laser. After the IMC, an intensity-stabilization section provides the signal for stabilizing the laser Relative Intensity Noise (RIN) and thus to reach the requirements. Then, an in-vacuum Faraday isolator prevents interaction between the ITF rejected light and the IMC and laser system. Finally, a mode-matching telescope (the ITF mode-matching telescope) is used to match the laser beam to the interferometer. 

\subsection{Electro-optic modulator}

\begin{figure}[htp]
  \centering
  \includegraphics[scale =0.25]{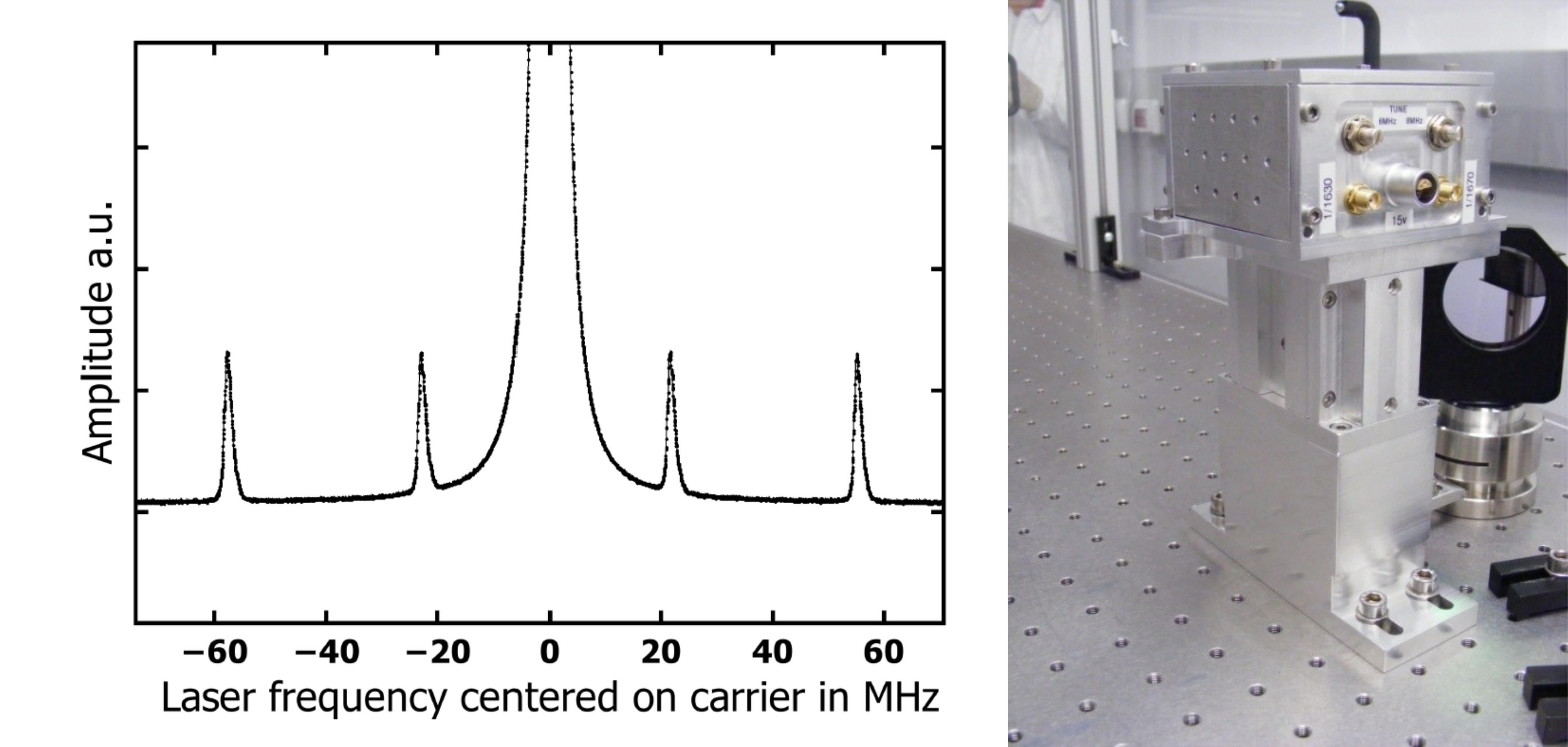}
  \caption{22 and 56MHz sidebands, measured with a scanning Fabry-Perot cavity (left); AdV Electro-optic modulator and the External Injection Bench (EIB).}
  \label{fig:EOM}
\end{figure}

In Advanced Virgo, five different modulation frequencies are required. To create them, we use three custom modulators, located between the laser system and the vacuum vessel on the external injection bench (EIB). Because of the high input laser power, we have selected a low absorption electro-optic material: the RTP(Rubidium Tantanyle Phosphate RbTiOPO$_{4}$) from Raicol Crystals Ltd. This material has a very low absorption level ($<$50ppm/cm @ 1064nm), thus reducing the thermal-lensing effects. In fact, a thermal focal length of $\sim 10$m for 200W of input optical power has been measured for each modulator. Each crystal surface is wedged and has a trapezoidal shape. A horizontal wedge of one degree on both sides allows the separation of the linear polarizations by walk-off. We can also avoid the cavity effect and reduce the residual amplitude modulation (RAM) \cite{Ram}, which could be a source of noise (RAM $<10^{-6}$ required). These modulators have been tested in the laboratory and a modulation depth higher than 0.1 has been obtained.  They are currently being installed on the detector. Figure \ref{fig:EOM} shows a view of one modulator and the spectrum of the laser beam after it has passed through the modulator. 

\subsection{Beam Pointing Control system}

Beam-pointing fluctuations can be a relevant source of noise. Indeed, in case of geometrical asymmetries between the arms of the ITF, created by spurious misalignments of the ITF optics, the input-beam jitter creates a phase noise directly affecting detector sensitivity~\cite{misalgnVirgo, misalgnLigo}. \\
The Advanced Virgo `Beam Pointing Control system' (BPC)~\cite{bpc} is designed to monitor and mitigate below 10 Hz, the beam-pointing noise at the input of the IMC. It is composed of two quadrant photodiodes, which sense the input-beam displacement, and two tip/tilt piezoelectric actuators, which compensate it. The sensing design places the two quadrants in the focal and image planes of the input of the IMC to respectively sense the tilt and the shift of the beam. \\
The BPC system achieves a control accuracy  of $\sim 10^{-8}$~rad for the tilt and $\sim 10^{-7}$~m for the shift, along with a sensing noise of less than 1~$n$rad/$\sqrt{\rm Hz}$, making it compliant with the requirements.

\subsection{EIB--SAS}
\begin{figure}[htbp!]
\centering
\includegraphics[width=11cm]{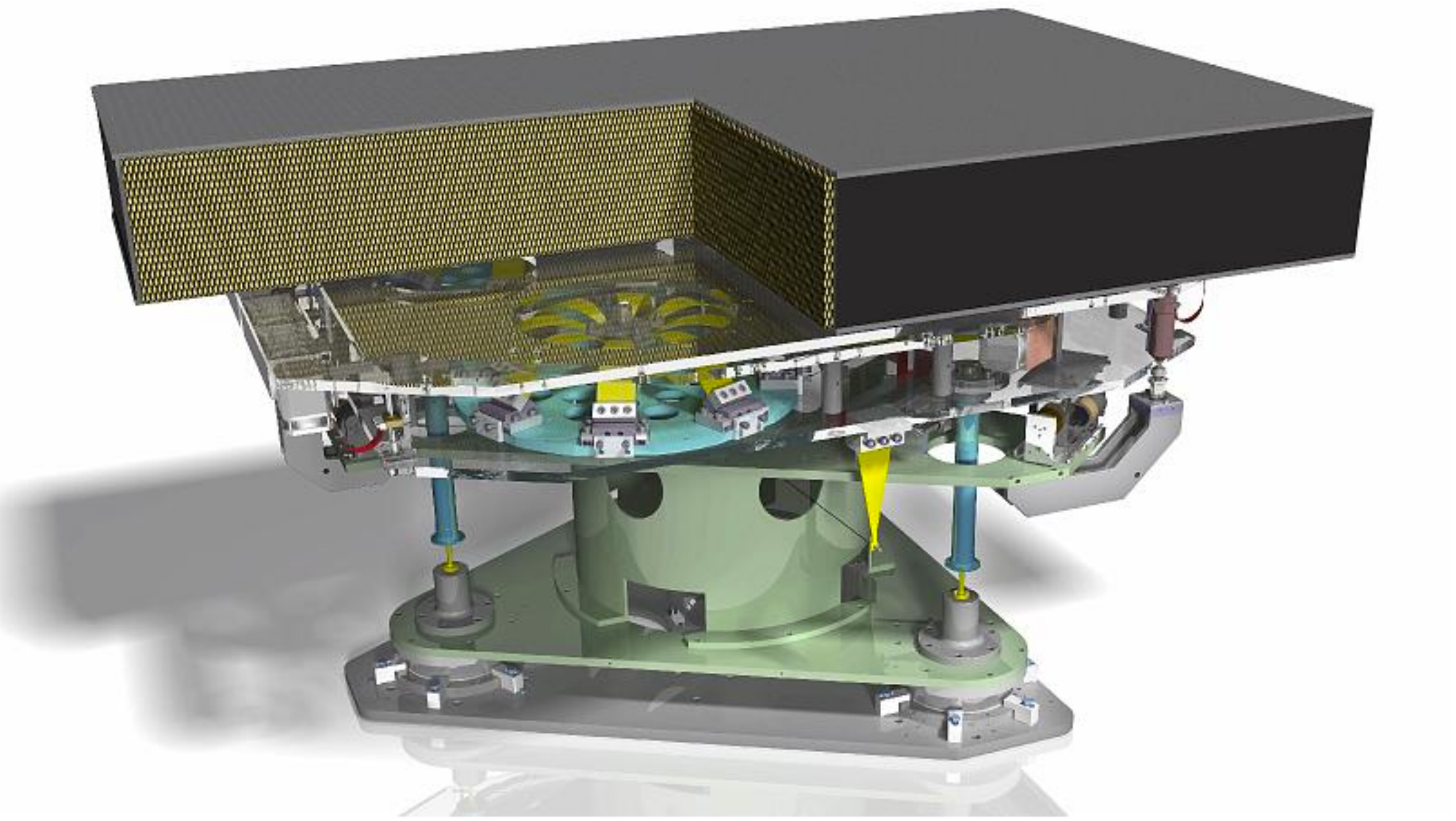}
\caption{\it Impression of the External Injection Bench supported by the EIB--SAS. The EIB is a standard 2400$\times$1500-mm optical bench with a mass of 840~kg. The inverted pendulum legs allow the intermediate platform ({\it spring-box}) to move horizontally.
The spring-box hosts the three GAS springs, which allow the optical bench to move along vertical, pitch and roll degrees of freedom. }
\label{fig:eibsas}
\end{figure}

The excess motion of the External Injection Bench (EIB), due to the seismic excitation of its rigid body modes, can be a major source of angular and lateral beam jitter in the injection system and thus limit the detector sensitivity. A new actively-controlled bench support system, called EIB--SAS (EIB Seismic Attenuation System), has been created to meet the requirement of having a neutral behavior with respect to the ground motion in a broad frequency range (DC to several hundreds of Hz). The EIB-SAS (see Figure~\ref{fig:eibsas}) is a single-stage six-degree-of-freedom vibration isolator, the design of which is derived from the Advanced LIGO HAM--SAS prototype~\cite{hamsas}, which has been more recently upgraded and adapted at the AEI~Hannover, in order to be incorporated into their 10 m prototype interferometer~\cite{aeisas}.

Passive seismic attenuation, from 20~to~70~dB in vertical and from 30~to~70~dB in horizontal between 2~and~400~Hz~\cite{blom}, is achieved by means of a combination of a short (0.5~m long legs) inverted pendulum (IP) and three sets of cantilever blades in {\it geometric anti--spring} (GAS) configuration ~\cite{vsanni}. All six fundamental rigid body modes have natural frequencies tuned between 0.2 and 0.5~Hz. 
The EIB--SAS is equipped with position (LVDT) and inertial (geophones) sensors and voice-coil actuators to actively damp the low-frequency eigenmodes and to stabilize the position and orientation of the supported bench. Higher frequency modes, originating from the GAS spring lateral compliance and from the vertical compliance of the inverted pendulum structure, are treated with tuned dampers placed on the spring-box. The internal modes of the IP legs are cured with eddy current dampers. Stepper motor-driven correction springs are used for initial (and periodical over the long term) coarse alignment of the bench.

\subsection{IMC cavity}

The input mode cleaner (IMC) cavity is an in-vacuum, triangular cavity comprising suspended optics, which has a length of 143.424~m and a finesse of 1200 (corresponding to an input/output coupler reflectivity of 2500 ppm, and total round-trip losses lower than 100 ppm). The finesse and length of the cavity are unchanged with respect to Virgo, as they are still thought to be a good compromise when considering the high spatial-filtering effect of the cavity and issues linked to the high finesse and length, such as radiation pressure, thermal effects and the small angle back-scattering from the end mirror. On the other hand, stricter polishing requirements have been set to improve the cavity performance in terms of throughput losses and back-scattering. At the time of writing, the mirrors have been polished, coated and installed. Simulation using measured mirror maps yielded an estimated 2~\% throughput loss and 300~ppm effective ``reflectivity''.   

\subsection{Faraday isolator}

Light back-reflected from the ITF may scatter inside the IMC and interfere with the input beam \cite{Bondu}. This issue can be tackled by adding an in-vacuum Faraday isolator on the Injection Bench suspended between the PR mirror and the IMC. Given the AdV high input power, a standard Faraday Isolator (FI) would exhibit loss of optical isolation, due to thermally-induced birefringence \cite{khazanov1999, mosca2010}  and modification of mean-angle rotation, due to heating of the TGG crystal under laser radiation. This last effect is particularly problematic under vacuum, where thermal dissipation is mainly insured by radiation \cite{FaradayAO}. Indeed, a standard device would also exhibit very high thermal lensing, which it would be necessary to correct inside the Faraday itself, in order to avoid large power-dependent mis-matching inside the system.

\begin{figure}[htp]
\centering
\includegraphics[width=14cm]{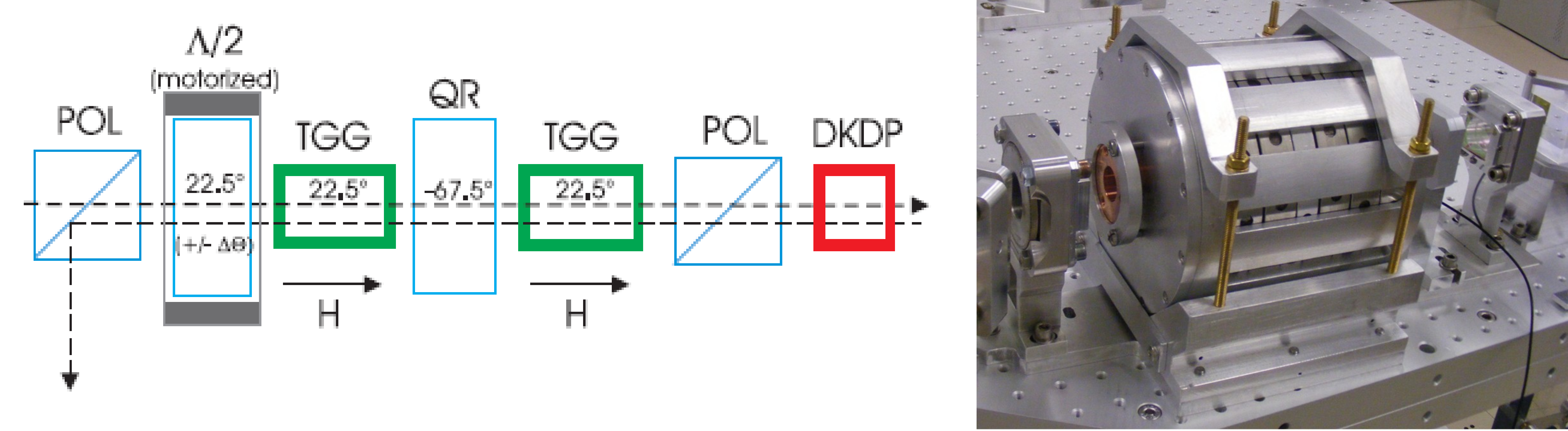}
\caption{Left: Scheme of the high power-compatible Faraday isolator. Right: View of the assembled device.}
\label{Faraday}
\end{figure}

A dedicated FI, allowing for compensation of all of these effects, was developed for AdV \cite{FaradayJOSAB} in collaboration with the Institute of Applied Physics (Russia) and is described in Figure~\ref{Faraday}. 

The FI consists of an input polarizer, a half waveplate, two Terbium Gallium Garnet(TGG) crystals, rotating the polarization by 22.5$^{\circ}$ and separated by a -67.5$^{\circ}$ quartz rotator (QR), an end polarizer and a DKDP plate.\\
By using two TGG crystals, separated by a quartz rotator, it is possible to limit the effects of thermal depolarization by compensating with the second crystal the depolarization created in the first \cite{Khazanov3}. The variation of mean angle rotation of the two TGG crystals can be compensated by slightly turning the half wave plate, which can be done remotely \cite{FaradayAO2}. This also makes it possible to compensate for the modification of isolation observed with the whole setup when it is tuned in air and then subsequently placed in vacuum, where the thermal dissipation process is different. Finally, a DKDP (Deuterated Potassium Dihydrogen Phosphate, KD$_2$PO$_4$) plate is used to compensate for thermal lensing created inside the TGG crystals, because of its large negative thermo-optic coefficient \cite{Zelenogorsky}.
Using this device we can insure an isolation ratio of 40dB at 150W, a residual thermal lens as low as 100m at full power and a total transmission of about 98\%.

\subsection{Mode-matching telescope}

The design of the mode-matching telescope (MMT) has been driven by the space constraints on the Suspended Injection Bench (SIB) and the need to limit, as much as possible, the aberrations of the beam \cite{ref1,ref2}. The telescope is made by an afocal two-mirror, off-axis parabolic telescope, which increases the beam size from 2.6 mm (size of the beam inside the Faraday), to 22 mm. The beam is then expanded  by a meniscus lens, and is finally re-collimated by the Power-Recycling mirror, the anti-reflecting (AR) coated surface of which is curved, to the Fabry-Perot cavities. The meniscus lens is needed to compensate for the spherical aberrations of the Power-Recycling mirror. This design allows a  theoretical matching of the input beam with the ITF that is higher than 99.9\%. 

In order to limit the scattered light from the optics, requirements on the surface quality of the parabolic mirrors (roughness $<$ 1 nm) and on the AR coating of the meniscus lens ($< $100 ppm) have been set. In addition, the optical mounts have been designed in order to limit the presence of low-frequency resonances. Moreover, the telescope is suspended from a Super Attenuator, limiting the modulation of the scattered light in the AdV detection band by seismic noise. 

The telescope has been installed on the Suspended Injection Bench 1 (SIB1) and was pre-aligned at the beginning of 2014. 

\begin{figure}[h!]
	\centering
	\includegraphics[width=1\textwidth,clip]{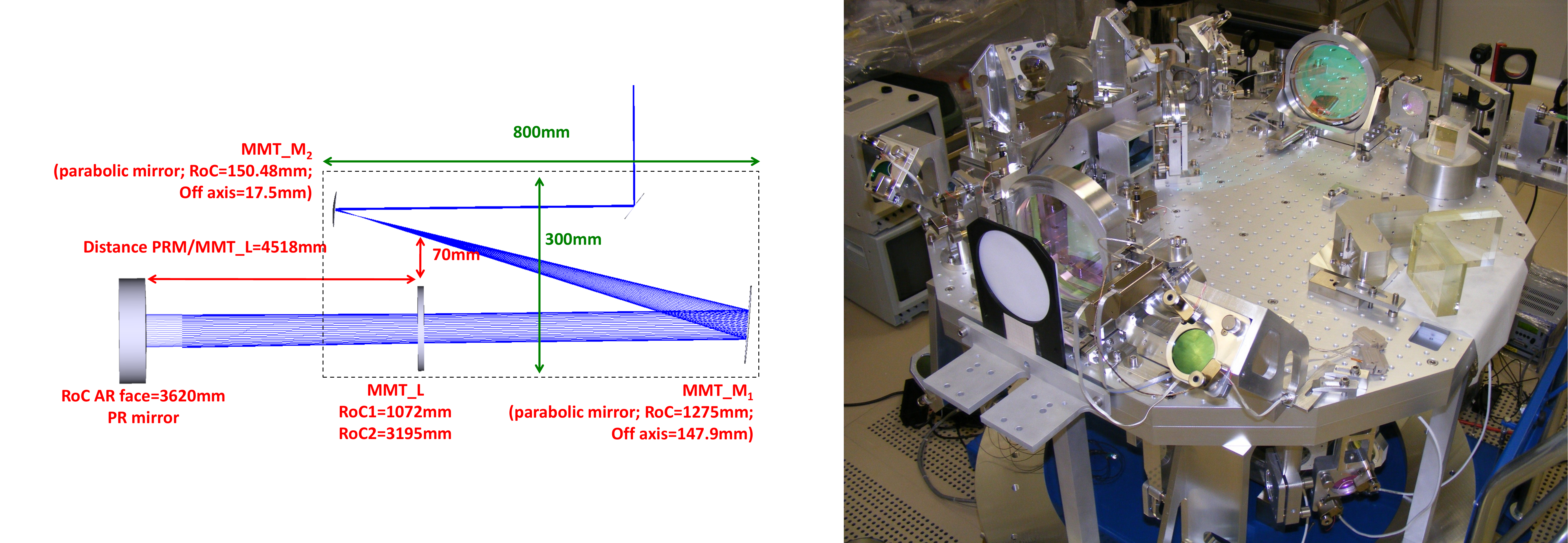} 
	\caption{\label{MMT} Catadioptric telescope of the Suspended Injection Bench 1 (SIB1) layout (left); MMT installed on SIB1 (right) in February 2014.}
\end{figure}

\section{Light detection} \label{sec:det}
\subsection{Overview}\label{section_overview}

\begin{figure}[!hbp]
\mbox{
      \subfigure[]{\includegraphics[width=6.0cm]{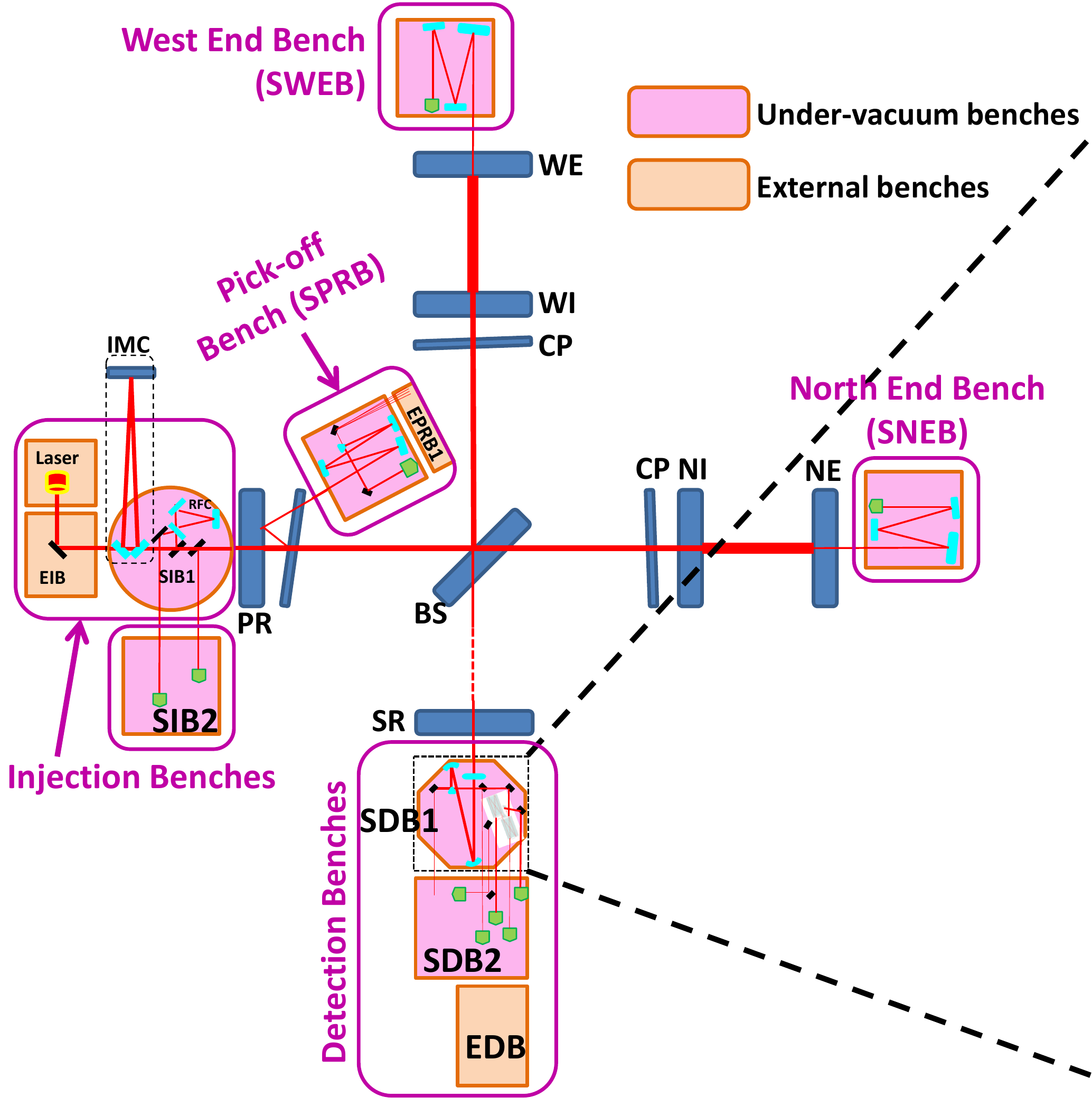}} \quad
      \subfigure[]{\includegraphics[width=6.0cm]{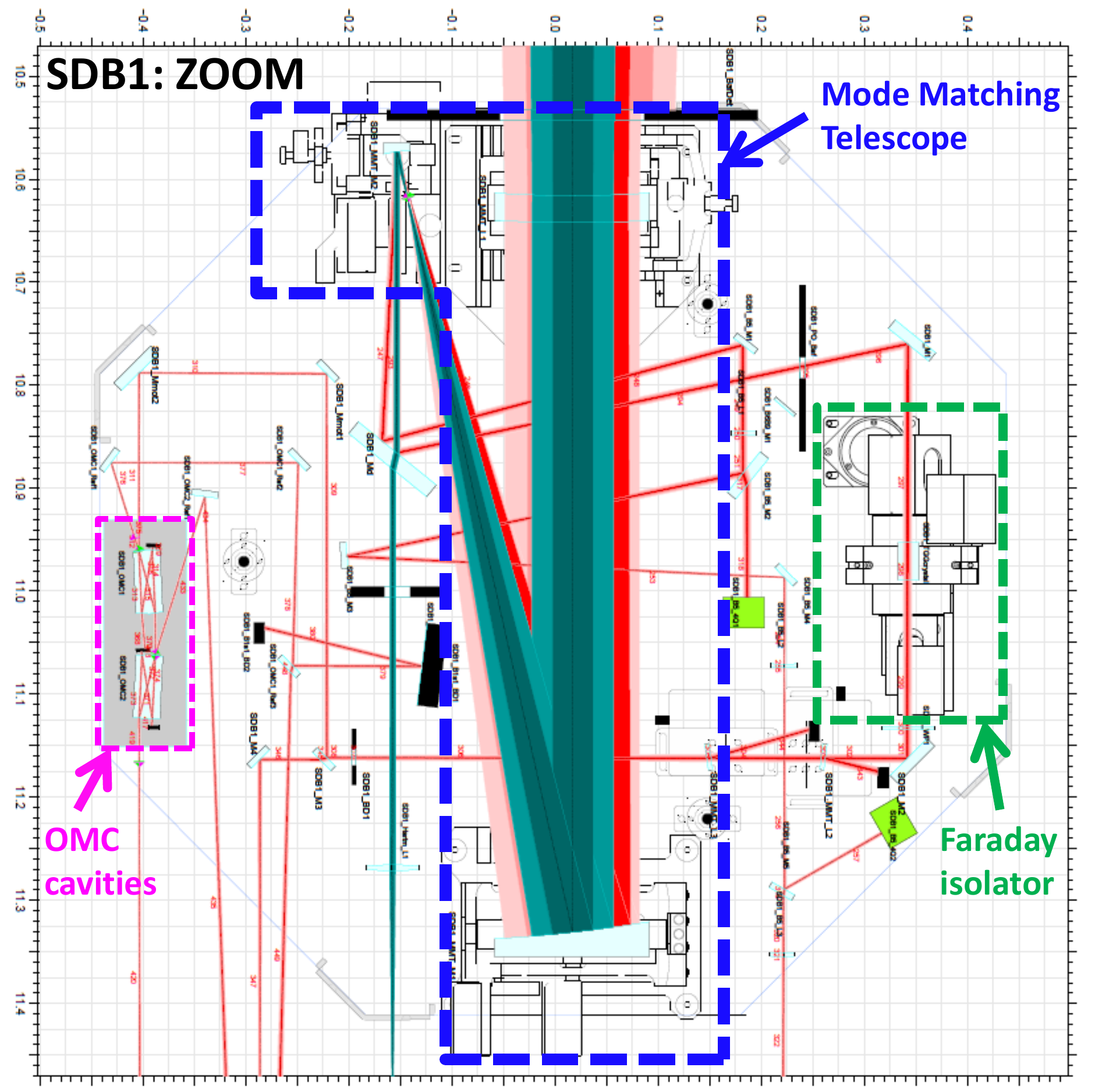}}
     }
\caption{(a) Scheme of the optical benches foreseen for AdV. (b) Optical layout of the first suspended detection bench SDB1. YAG beams ($\lambda$~=~1064~nm) are displayed in red. A Hartmann beam ($\lambda$~=~800~nm) is displayed in green.}
\label{fig:det1}
\end{figure}
The detection system extracts the ITF signals from the anti-symmetric port and from the auxiliary beams. It consists of several optical benches, hosting photodetectors, as well as motorized alignment optics and mode-matching telescopes. The beam from the anti-symmetric port, referred to as the {\it dark-fringe beam}, carries information about a change in the differential arm-length (DARM) and is thus sensitive to gravitational waves (GW). The detection system must comply with the DC detection scheme. Auxiliary beams are sensitive to other ITF degrees of freedom, such as cavity lengths and laser frequency variations, providing signals needed for the ITF control systems. The optical benches designed for the extraction of the main ITF beams are sketched in Figure~\ref{fig:det1}.a and described below.\\
\\
A first detection bench (SDB1) is placed along the path of the dark-fringe beam. An optical layout of SDB1 is presented in Figure~\ref{fig:det1}.b. This bench hosts an Output Mode Cleaner (OMC) the goal of which is to improve the contrast defect by filtering out spurious components of the light, which are induced by interferometer imperfections. In the DC detection scheme the OMC must also filter out the RF sidebands. The SDB1 bench is followed by a second Suspended Detection Bench (SDB2) which hosts sensors for the detection of the dark fringe beam and secondary beams involved in the ITF control. Both SDB1 and SDB2 benches are suspended and placed under vacuum, in order to isolate sensitive optics and photodetectors. An External Detection Bench (EDB) hosts less-sensitive equipment, such as a phase camera~\cite{phase1} and a Hartmann wavefront sensor which, as explained in Section 7, will be used to monitor the thermal aberrations and optical defects in the ITF.\\
A suspended bench denoted as SPRB is used to extract a pick-off beam from the power-recycling cavity. It hosts a telescope to adapt the beam size and Gouy phase to the photodetectors. Similarly a suspended bench equipped with a telescope at the end of each ITF arm (named SNEB for the North Arm and SWEB for the West Arm) allows the extraction of the beam transmitted by the Fabry-Perot cavity. Another suspended bench, named SIB2, extracts the beam reflected by the ITF.\\
\\
One of the design criteria of the detection system has been the minimization of the impact of stray and scattered light, which was one of the noise sources limiting the sensitivity of the first-generation interferometers. In order to limit the amount of back-scattered light, stringent requirements have been set on the micro-roughness (typically $\le$~0.3~nm~RMS for spherical and flat optics, and $\le$~1~nm~RMS for the parabolic mirrors of the telescope) and the anti-reflective coating ($\approx$~100~ppm targeted) of the optics placed along the path of the dark-fringe beam. Moreover, scattered-light noise couples through the displacement of the optics, which is amplified at the resonance frequencies of the mounts. Therefore, the main optical benches are seismically and acoustically isolated.

\subsection{Suspended detection benches}\label{section_bench}
\begin{figure}[!hbp]
\begin{center}
\includegraphics[width=12cm]{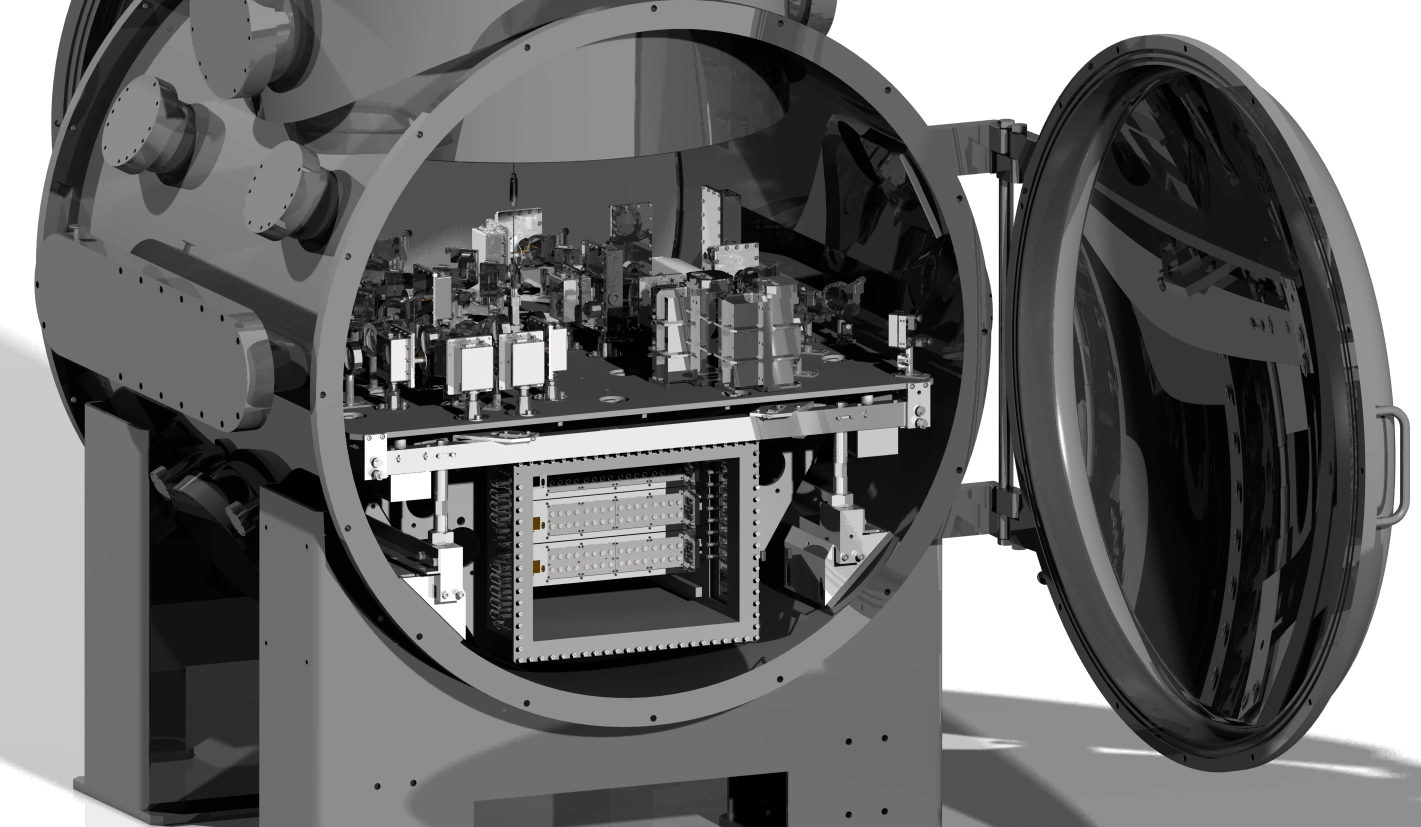}
\caption{\label{fig:det2}3D drawing of the new suspended bench inside its vacuum chamber.}
\end{center}
\end{figure}

In addition to SDB1, which is suspended to a short version of the Virgo Super Attenuator (see Section 6), five new benches will be installed in new vacuum chambers and suspended from complex isolation systems. As shown in Section~13, the designed vacuum chambers are very compact, which is necessary to comply with the space constraints of the existing infrastructure. Accordingly a compact isolation system (see Section~13) and a new suspended bench body~\cite{DET_bench}, which fit in these vacuum chambers, have been designed. A 3D drawing of the bench inside the vacuum chamber is shown in Figure~\ref{fig:det2}. The bench body is made of aluminium and incorporates a sealed air container hosting, in particular, electronic boards used for the readout and control of the sensors located on the bench. This strategy minimizes the number of cables running along the suspension, which helps to keep the isolation system compact, while avoiding mechanical short-circuiting. As a consequence of having the sensing electronics integrated within the bench body, a significant amount of heat has to be dissipated into the vacuum chambers. The heat will be evacuated from the electronic components to the bench body through conduction, using heat links, and from the bench body to the vacuum chamber walls, through radiation. To this purpose, the suspended bench will be anodized in order to maximize its thermal emissivity. For the inner part of the stainless-steel chamber, a sand-blasting treatment is enough to provide a sufficiently high thermal transfer, since its surface is larger than the bench external surface.

\subsection{Output Mode Cleaner}\label{section_omc}

The AdV OMC is made of two identical cavities, placed in series~\cite{DET_omc_design} (see Figure~\ref{fig:det1}.b). Each cavity is made of a monolithic block of fused silica with four polished surfaces, which together form a four-mirror bow-tie cavity. The choice of a compact monolithic cavity has been made to minimize the cavity length noise and to keep mechanical resonances above the detector bandwidth, namely above 10~kHz. The round-trip length ($L_{rt}$~=~248~mm) and the nominal finesse (F~=~143) of the cavity have been chosen to meet the AdV requirements on the filtering of the RF sidebands for DC detection. Having two cavities in series offers the possibility of keeping a moderate finesse value, which eases the lock of the cavity and, for a given locking accuracy, limits the coupling of cavity-length noises, proportional to the square of the finesse. The finesse has also been chosen sufficiently low to keep the internal losses per cavity, expected to be dominated by scattering losses, lower than 1$\%$. One of the cavity surfaces is spherical, which makes the cavity stable. The radius of curvature, equal to 1.7~m, has been chosen to avoid degeneracies between the carrier fundamental mode and the high order modes of the carrier and sidebands, as well as to minimize the cavity thermo-refractive noise, which is inversely proportional to the square of the cavity waist~\cite{DET_omc_roc} (the nominal waist is 321~$\mu$m). The carrier fundamental mode must be kept resonant in the two OMC cavities, which is achieved by controlling each cavity optical-length with two types of actuators: Peltier cells and a PZT. The PZT actuator is also used to modulate the cavity length. The error signal used to control the cavity length is extracted by demodulating the signal provided by the photodetectors placed in transmission of the OMC at the PZT modulation frequency. This control system is designed to satisfy the requirements of a locking accuracy lower than 1.2$\times 10^{-12}$~m.

\subsection{Telescopes}\label{section_telescope}

Two types of telescopes are needed: an Output Mode Cleaner Mode-Matching Telescope (OMC$\_$MMT) on the SDB1 bench (see Figure \ref{fig:det1}.b), and a telescope used to adapt the beam parameters to the photodetectors placed on the end benches (SNEB and SWEB) and pick-off bench (SPRB).\\
\\
The design of the OMC$\_$MMT is very similar to the injection mode matching telescope described in Section 9.7. This design has been driven by the space constraints on the SDB1 bench and the requirements on the beam quality to limit aberrations. This telescope consists of a diverging meniscus lens, used in combination with a curved surface of the Signal Recycling mirror (SR), to reduce the size of the beam from 49~mm to 22~mm and compensate the spherical aberrations induced by the SR lens. An afocal off-axis telescope, made of two mirrors, is then used to decrease the beam size by a factor of 17. Finally, two lenses are used to match the beam to the OMC. Thanks to this design, the theoretical matching of the ITF beam to the OMC is higher than 99.9\%.\\
The telescopes of SNEB, SWEB and SPRB are very similar to those used in Virgo, with a doublet as a main focusing element and a second lens used to tune the Gouy phase on two quadrant photodiodes.

\subsection{Sensors and front-end electronics}\label{section_sensors}

Three main types of sensor are used as part of the detection system:
\begin{itemize}
\item ``Longitudinal'' photodiodes provide signals that are sensitive to length variations in the ITF and, in the case of the photodiodes placed at the dark port, to gravitational wave signals.
\item Quadrant photodiodes provide signals sensitive to the alignment of the ITF mirrors.
\item Phase cameras are used to monitor the surface defects of the ITF mirrors. 
\end{itemize}
The main longitudinal and quadrant photodiodes are hosted on the suspended benches. Each photodiode is placed inside a sealed-air box, the analog outputs of which are connected to the suspended bench electronic container through in-vacuum cables and feedthroughs.\\
\\
{\bf Longitudinal photodiodes:}\\
Longitudinal photodiodes are high-quantum-efficiency InGaAs PIN photodiodes, with clear apertures of 2~mm and 3~mm. The photodiode electronics chain is made of three main parts. First, a preamplifier converts the photodiode current to voltage and provides channels for different bandwidths: a large-dynamics DC channel to be used up to a few Hz; a highly-sensitive audio channel, high-pass filtered above 5 Hz; and a RF channel with a high-frequency bandwidth. The RF signal is then digitized by a demodulation board, which, as explained in Section 15, performs the digital demodulation~\cite{DET_demodulation} at the frequencies of the modulation sidebands (ranging from 6~MHz to 132~MHz) and their double frequencies. The audio and DC signals are digitized by a 'service' board, which also provides the bias voltage and monitors the photodiode air box.\\
With the DC detection scheme, the audio signals of the photodiodes placed at the anti-symmetric port are used to search for gravitational waves. Demodulated signals are used as error signals to control the ITF longitudinal degrees of freedom and thus their noise can also have an impact on the ITF sensitivity. Accordingly, both RF and audio channels are designed so that their associated electronics noise is lower than the photodiode shot noise, by up to a factor of 10 for the audio channel of the dark fringe.\\
\\
{\bf Quadrant photodiodes:}\\
Quadrant photodiodes are Si-PIN diodes, with a 5~mm-diameter total aperture, divided into four sections. By design, these sensors are sensitive to beam misalignments. Two types of electronic configurations are foreseen. Quadrants placed at the SNEB and SWEB terminal benches are optimized for DC sensing: their DC channels, which are designed to be shot-noise limited, provide error signals involved in the alignment of the Fabry-Perot arm cavities. Quadrants placed on other suspended benches are equipped with a RF channel, which, after being demodulated at the sideband frequencies, provides error signals for the alignment of the other ITF degrees of freedom.\\
\\
{\bf Phase cameras:}\\
Phase cameras provide wavefront-sensing capabilities to measure the spatial profiles of each frequency component of the laser field (carrier and sidebands). These spatial profiles contain information about mirror-surface defects and can thus be used to determine corrections to be applied to the mirrors. The AdV phase cameras are based on an optical heterodyne measurement: a reference beam, extracted from the AdV laser bench, is frequency shifted at 80~MHz by an acousto-optic modulator, and travels up to the phase camera benches through optical fibers. The reference and test beams are combined onto a beam splitter and co-propagate towards a detection port of the phase camera. The detection port is made of a pinhole photodiode and a scanning system, making it possible to scan the wavefront over the pinhole diode. Information on the wavefront amplitude and phase are then obtained by demodulating the photodiode output at the beat frequency between the reference beam and the frequency components of the tested beam.

\section{Interferometer control}  \label{sec:isc}

\subsection{Steady state}

To be able to detect gravitational waves, the various cavities of the interferometer must all be kept on resonance with picometer accuracy. To stay at this working point in spite of external disturbances, a high-performing feedback system is needed. This is achieved with a real-time, digital control system (see Section~\ref{sec:daq}), which uses the output of various photodiodes (see Section~\ref{sec:det}) as error signals and actuates on the mirror position using voice-coils (see Section~\ref{sec:pay}). This control system acts on the interferometer as a whole, e.g. a signal from the \emph{dark fringe} photodiode in the Central Building is used to actuate on the end test masses, located three km away.


For the control of the interferometer, it is common to define five \emph{longitudinal} degrees of freedom (DoF), which roughly correspond to signals available on the various photodiodes:
\begin{itemize}
\item The differential length of the two 3 km-long arm cavities (DARM). This is the most important DoF, since it contains the gravitational wave signal. The most sensitive photodiode for controlling this DoF is the DC power of the dark fringe (B1). To obtain a proper error signal, the DARM degree of freedom needs to be moved away from the dark fringe position by some ten picometers. This scheme is known as \emph{DC-detection} \cite{DC_detection}. To filter out the demodulation sidebands, this beam is first passed through the OMC. This DoF is controlled by actuating differentially on the two end test masses.

\item The common length of the two arm cavities (CARM). A demodulated signal in reflection of the interferometer is used as an error signal. Instead of actuating mechanically on some mirrors, this DoF is controlled by actuating with high bandwidth on the frequency of the main laser. At very low frequencies, this frequency is stabilized on a monolithic \emph{reference cavity}, by actuating in a common way on the end test masses.

\item The differential length of the short Michelson interferometer, formed between the Beam Splitter and the two input test masses (MICH). As an error signal, a demodulated signal from the pick-off beam taken inside the power recycling cavity is used. Actuation is performed on a combination of various mirrors.

\item The length of the power-recycling cavity (PRCL), which is the cavity between the power-recyling mirror and the two input test masses. As error signal, a demodulated signal in reflection of the interferometer is used, while the actuation is performed on the power-recycling mirror only.

\item The length of the signal-recycling cavity (SRCL), which is the cavity formed between the signal-recycling mirror and the two input test masses. As error signal, a demodulated signal from the pick-off beam is used, while the actuation is performed on the signal-recycling mirror only.
\end{itemize}

For most of these DoF, the error signals are obtained by a heterodyne scheme similar to the Pound-Drever-Hall technique. To this end, modulation sidebands at 6, 8 and 56 Mhz are added to the laser in the injection system (see Section \ref{sec:inj}). The first four of these DoF were also used in Virgo+, in a configuration which is known as a power-recycled interferometer. Initially, Advanced Virgo will be run in the same configuration. Only in a second step, will the signal-recycling mirror be added, which adds the new SRCL degree of freedom. This configuration is known as a dual-recycled interferometer.

Slightly different photodiodes, demodulation sidebands and demodulation phases are used in the power-recycled and double-recycled configuration. To choose the optimal photodiodes for controlling the various DoF, the optical response of the interferometer is first simulated in the frequency domain using Optickle \cite{Optickle}. For each DoF, the photodiode and demodulation phase is chosen, in which a particular DoF is visible with the highest signal-to-noise ratio (SNR). Since most photodiodes are sensitive to multiple DoF, the effect of the control system must be taken into account. This whole design is therefore an iterative procedure, in which the various feedback loops are optimized until the various accuracy requirements are fulfilled \cite{steady_state}.

\subsubsection{Noise subtraction}

The optical response of the interferometer to the various degrees of freedom is not a diagonal system, meaning that the main photodiode is not only sensitive to DARM, but also to a lesser degree to signals from the other degrees of freedom (mainly MICH and SRCL). For this reason, also the sensing noise of these auxiliary degrees of freedom will also end up in the gravitational wave signal. To reduce this effect, a noise subtraction technique is used, in which the appropriately-filtered actuation signal of the auxiliary DoF is added to the actuators used for DARM. A subtraction of a factor of 500 has been achieved in the past \cite{noise_subtract}. A similar subtraction factor will be needed for Advanced Virgo, in order not to spoil the sensitivity at low frequencies.
A similar noise-subtraction technique is used when reconstructing the h(t) signal used in data analysis~\cite{bib:DAQ_VirgoHrec}.

\subsection{Lock acquisition}

When all cavities are on resonance, the optical response of the interferometer acts like a linear system. This allows for the design of a robust control system using standard techniques, which can keep the interferometer \emph{locked} for many hours, failing only in case of large external disturbances, such as earthquakes. Bringing the interferometer to this working point, however, requires a complex procedure, since the uncontrolled mirrors initially swing with a large amplitude, causing complex non-linear transient signals. The procedure to get from this uncontrolled state to the final working point, ready to collect science data, is called the \emph{lock acquisition}. Developing this procedure is done with the help of time-domain simulation tools, such as E2E \cite{E2E}. A typical lock acquisition consists of many steps, in which the control of the various DoF is handed from one photodiode to another.

In the past, with the Virgo+ interferometer, this was achieved with a deterministic procedure called the \emph{variable finesse} technique \cite{variable_finesse}.  In this procedure, the interferometer is initially locked with the MICH degree-of-freedom at \emph{half fringe} and the power-recycling mirror misaligned. After aligning this mirror, MICH is brought gradually towards the \emph{dark fringe}, which allows for a controlled build-up of the powers inside the cavities. Since Advanced Virgo will initially start without a signal-recycling mirror, the optical scheme is very similar, so we will use the same procedure. The main difference is that at the end, the lock of DARM will change from a demodulated to a DC signal.

The signal-recycling mirror will be installed only in a second stage, which adds a fifth DoF to be controlled: SRCL. Developing a lock-acquisition procedure that only uses standard photodiode signals has been shown to be very hard \cite{Ward}. To solve this issue, an auxiliary laser system operating at a wavelength of 532 nm was developed at LIGO, which can measure the arm-cavity lengths independently of any other degrees of freedom \cite{green_lock_40m}. For Advanced Virgo, we plan to implement a similar system. This system is still being developed, but the coatings of the main mirrors have already been designed to take the extra wavelength into account. 

\subsection{Angular control}

Apart from the longitudinal degrees of freedom, there are also the two angular degrees of freedom of each mirror, to be controlled with nano-radian accuracy~\cite{AAadV}. 
For this reason a global angular control system, the Automatic Alignment (AA), which uses the signals coming from the interferometer itself, is required.\\
The implementation of the angular control for Advanced Virgo will have to face a more complex configuration with respect to that of Virgo~\cite{AA_VSR1,AA_VSR2}.
The main differences between the Advanced Virgo and Virgo interferometer configurations are the higher circulating power and the presence of the signal-recycling cavity. These modifications produce an improvement in the interferometer sensitivity, but an increase in complexity in the development of the AA control system.\\
The high circulating power produces strong radiation-pressure effects which modify the opto-mechanical transfer function. This has to be taken into account in the control scheme design, while the presence of the signal-recycling cavity increases the number of degrees of freedom available to control  the off-diagonal couplings in the AA error signals.\\
The control scheme, based on a wavefront-sensing scheme, which uses the modulation-demodulation technique, has been modeled with the frequency domain simulation tools Optickle~\cite{Optickle} and Finesse~\cite{Finesse} and a simulated control noise below the design sensitivity safety factor (a factor of 10 below the design) has been obtained~\cite{AAadV}.

\section{Scattered-light mitigation: baffling}  \label{sec:slc}

\subsection{Introduction}
We define \textit{stray light} as the laser light in the interferometric antenna that does not follow the designed path. This light can recombine with the main beam after probing the position of different structures outside of the intended path, thus causing spurious information to enter the detection port. Possible origins of stray light are, for instance, scattering off surfaces with residual roughness, secondary beams/spurious reflections due to non-ideal anti-reflective coating, scattering from point-like defects (such as dust, troughs or scratches on a mirror), and diffraction due to a limited aperture of optical components.

Stray light will carry phase information which depends on the length of the scattering path before recombination with the fundamental mode. This path length is, in turn, modulated by the longitudinal motion of the scattering element $z_s(t)$ with respect to the test-mass position. The effect of the phase modulation carried by the stray-light field affects both the amplitude and the phase of the total fundamental field. Furthermore, it is inherently non-linear with respect to the scatterer displacement noise $z_s(t)$, and may be described by linearized equations only when $z_s(t)<<\lambda/4\pi$.

\subsection{Advanced Virgo solution}
In order to mitigate the effect of stray light on the strain sensitivity of the interferometer, complementary approaches can be pursued: either the displacement noise $z_s(t)$ of a potential scatterer can be controlled (emphasis on seismic isolation) or the fraction of recombined light $f_{sc}$ can be minimized (emphasis on optical quality of the surface). A trade-off between those two approaches is considered on a case-by-case basis.\\ 
In AdV, baffles and diaphragms are placed in the vacuum enclosure, mainly to obscure rough surfaces and sudden discontinuities from the line-of-sight of the core optics. These are designed to intercept and absorb stray light as much as possible. Furthermore, they must be characterized by low scattering and low reflectivity as well. When this is difficult to achieve, because of requirements on the roughness being too stringent, the baffles are also required to be suspended, so that their effect on the strain sensitivity remains negligible. 

\begin{figure}[htbp]
\begin{centering}
\includegraphics[width=.60\textwidth]{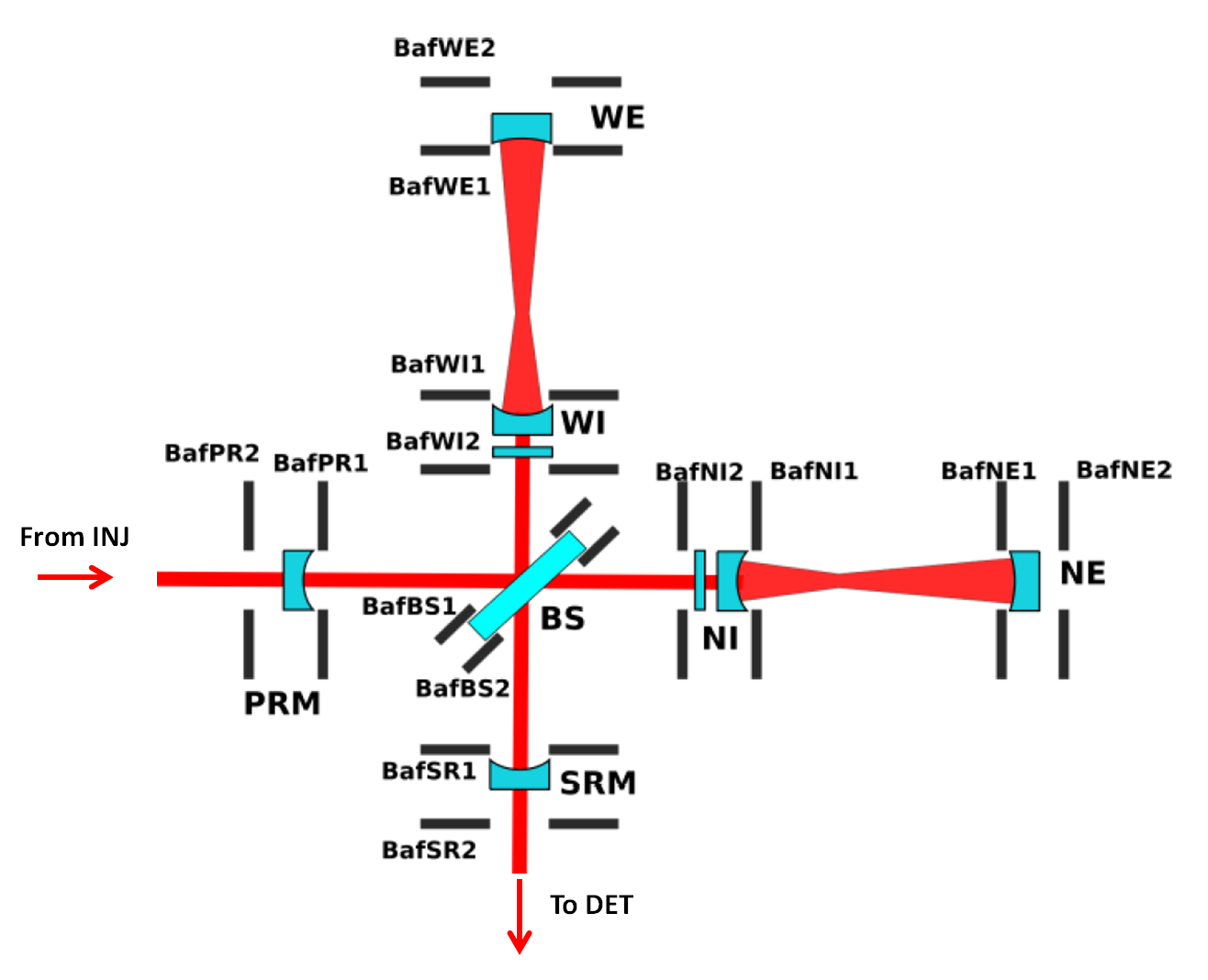}
\par\end{centering}
\caption{\label{fig:Sus_baffles} Schematics of the baffles suspended to the core optics-related Super Attenuators. 320 baffles are welded to the pipes in the long arms, not shown here.}
\end{figure}

\begin{figure}[htbp]
\begin{centering}
\includegraphics[width=.80\textwidth]{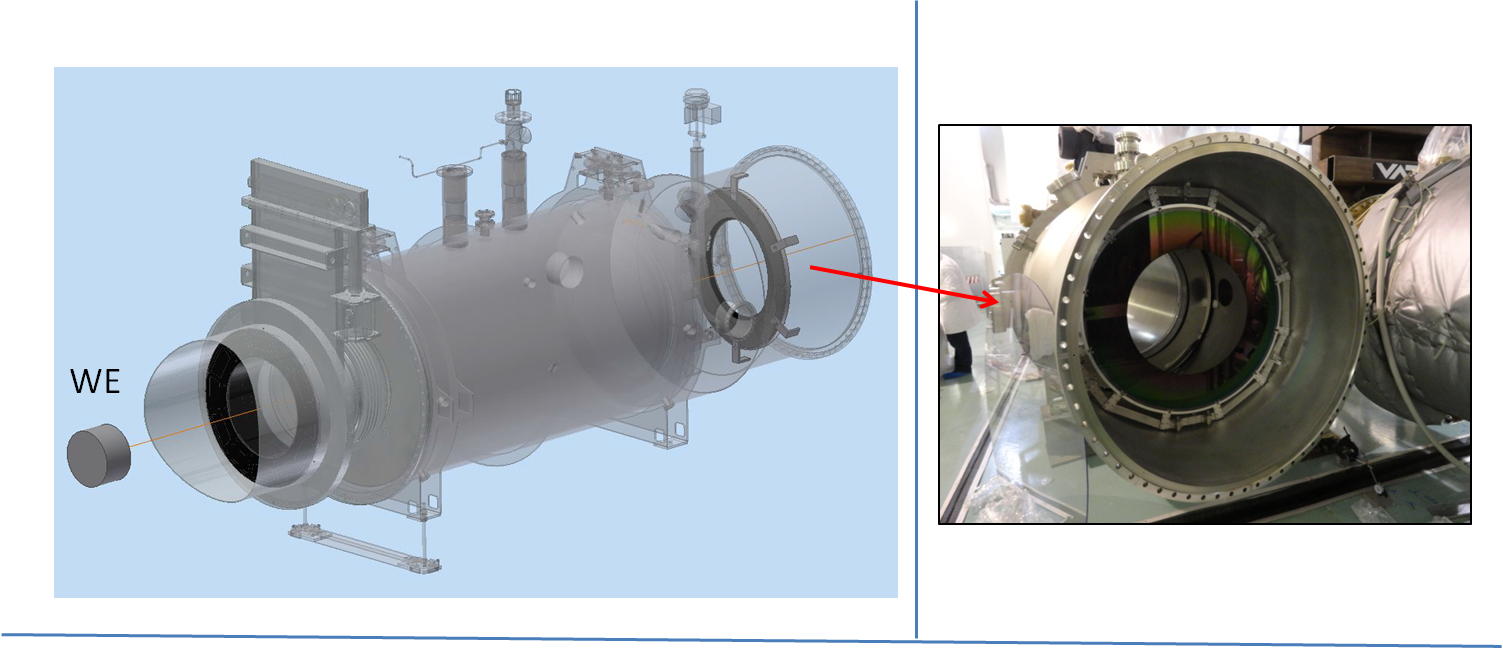}
\par\end{centering}
\caption{\label{fig:BafCryo_WE} The large cryogenic traps separate the tanks housing the test masses from the long vacuum pipes of the arms. Inner structures need to be shielded from the test-mass view, so dedicated baffles are installed. The West End cryotrap was the first to be equipped with such baffles, shown in the right panel.}
\end{figure}

\begin{figure}[htbp]
\begin{centering}
\includegraphics[width=.60\textwidth]{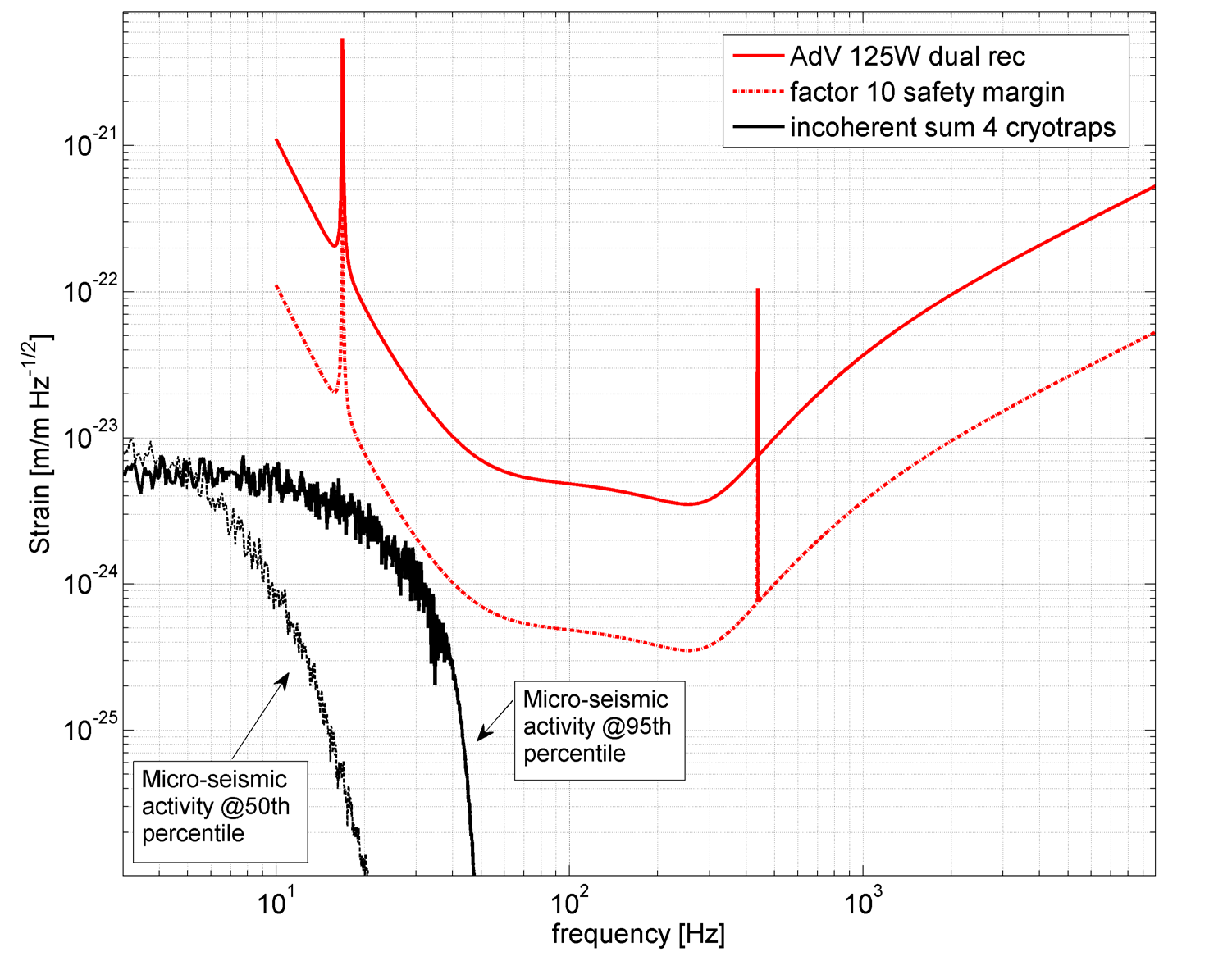}
\par\end{centering}
\caption{\label{fig:cryotrap_noise_all} Projection of the displacement noise of the cryotrap baffles. The overall noise projection is estimated as the incoherent sum of the displacement noise coming from various baffle surfaces, multiplied by a factor of 2 to account for the four cryotraps of the arm cavities (incoherent sum). The solid line is for high microseismic activity (95th percentile upper limit), dashed line for usual microseismic activity (50th percentile upper limit).} 
\end{figure}

\begin{figure}[htbp]
\begin{centering}
\includegraphics[width=.70\textwidth]{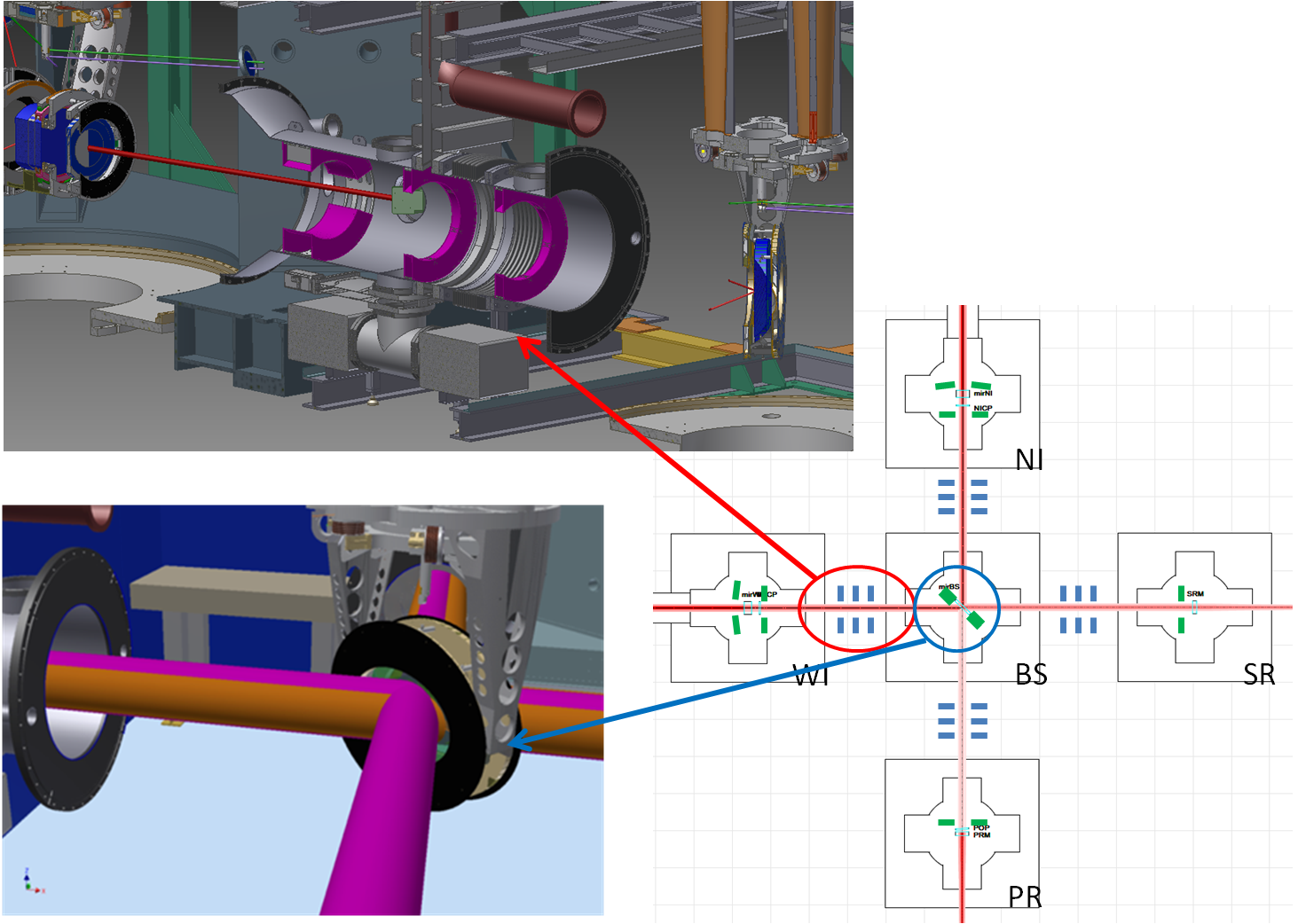}
\par\end{centering}
\caption{\label{fig:CITF_baffles} Schematics of the baffle positions inside the Central Area. Baffles connected to the vacuum links are depicted in blue, while the green baffles are suspended from the core optics-related Super Attenuator. In the top-left panel, a 3D-section view of the baffles foreseen in the link WI-BS is shown, where clearance for the CO$_2$ laser for the TCS is made possible with a special baffle shape. The bottom-left panel shows a 3D rendering of the baffle integrated into the Beam-Splitter payload. Also depicted: beams impinging on the Beam-Splitter mirror travel clearly through the baffles.}
\end{figure}

We have used FOG \cite{FOG} and figure error maps for the mirrors to compute the intensity distribution around the mirror. Indeed, given the optical design of Advanced Virgo \cite{ADV:TDR}, the input power is lost predominantly in the arm cavities. Most of the losses from the scattering from the core optics are accounted for by the low spatial frequency figure errors of the mirrors (below $10^3$ m$^{-1}$). So, a large amount of light will be lost at very small angles, only around the coating diameter of the arm cavity mirrors. Furthermore, spurious phase information, generated inside the arm cavities, is directly comparable, in terms of impact on the dark fringe, to the gravitational signal, so the baffling around the test masses is the most important, while its installation requires considerable delicacy. Because of this, baffles are integrated into the test-mass payloads, thus exploiting the seismic isolation provided by the Super Attenuators. Figure \ref{fig:Sus_baffles} shows a layout of the baffles suspended from the core optics-related Super Attenuators.  Unfortunately, the large cryogenic traps in the arms cannot be shielded by suspended baffles, although they are affected by small-angle scattering from the test masses. Therefore, additional baffles connected to the vacuum pipes are foreseen, to prevent the inner structure of the cryotraps from being seen from the mirrors. Figure \ref{fig:BafCryo_WE} shows the first of these baffles, installed in the West-End cryotrap. In \cite{Chiummo13a} details are given for the expected overall noise due to back-scattering from the cryotrap baffles. The noise-projection curves are reported in Figure \ref{fig:cryotrap_noise_all}: according to the current design, the noise contribution to the sensitivity is negligible even for severe micro-seismic activity.\\
The rest of the light will be scattered off at a larger angle and lost in the 3km vacuum tube, where one hundred and sixty baffles per arm, installed for the first generation of Virgo, have been shown to also be compliant with AdV sensitivity. More baffles will be installed in the central interferometer links (Figure \ref{fig:CITF_baffles}) and in the vacuum tanks housing the suspensions \cite{Chiummo13}.

\subsection{Material characterization}
After a survey of the materials used in Virgo+, in KAGRA and in LIGO/aLIGO, we narrowed the choice for the construction of the baffles down to two candidates, namely Silicon Carbide (SiC), and a coating of Diamond-Like Carbon (DLC) on a stainless-steel substrate. While SiC stood up as the best choice from the point of view of laser-induced damage threshold ($\sim$30 kW/cm$^2$) and Total Integrated Scattering (TIS) which could be reduced to as low as 50ppm \cite{SiC}, the creation of large baffles in SiC resulted as too expensive. DLC, on the other hand, features very good absorption for $\lambda = 1064nm$, and an acceptable damage threshold, although much smaller than that of SiC ($\sim0.5$~kW/cm$^2$)  \cite{DLC}.  The chemical vapor deposition (CVD) technique makes it possible to have conformal coating, i.e. DLC films replicating the roughness of the substrate, so the TIS value is determined by the choice of the substrate. As for SiC, the mis-match of the refractive index requires an AR coating on top of the DLC layer to reduce the direct reflection (reflectivity of about $\sim15\%$ at $0^{\circ}$ angle of incidence).\\ 
DLC on mirror-finish stainless steel, topped with AR coating is so far the reference choice, when expected laser intensity is well below $\sim0.5kW/cm^2$.

\section{Scattered-light mitigation: photodiode seismic and acoustic isolation}  \label{sec:sbe}
\begin{figure}[htbp!]
\centering
\includegraphics[width=5cm]{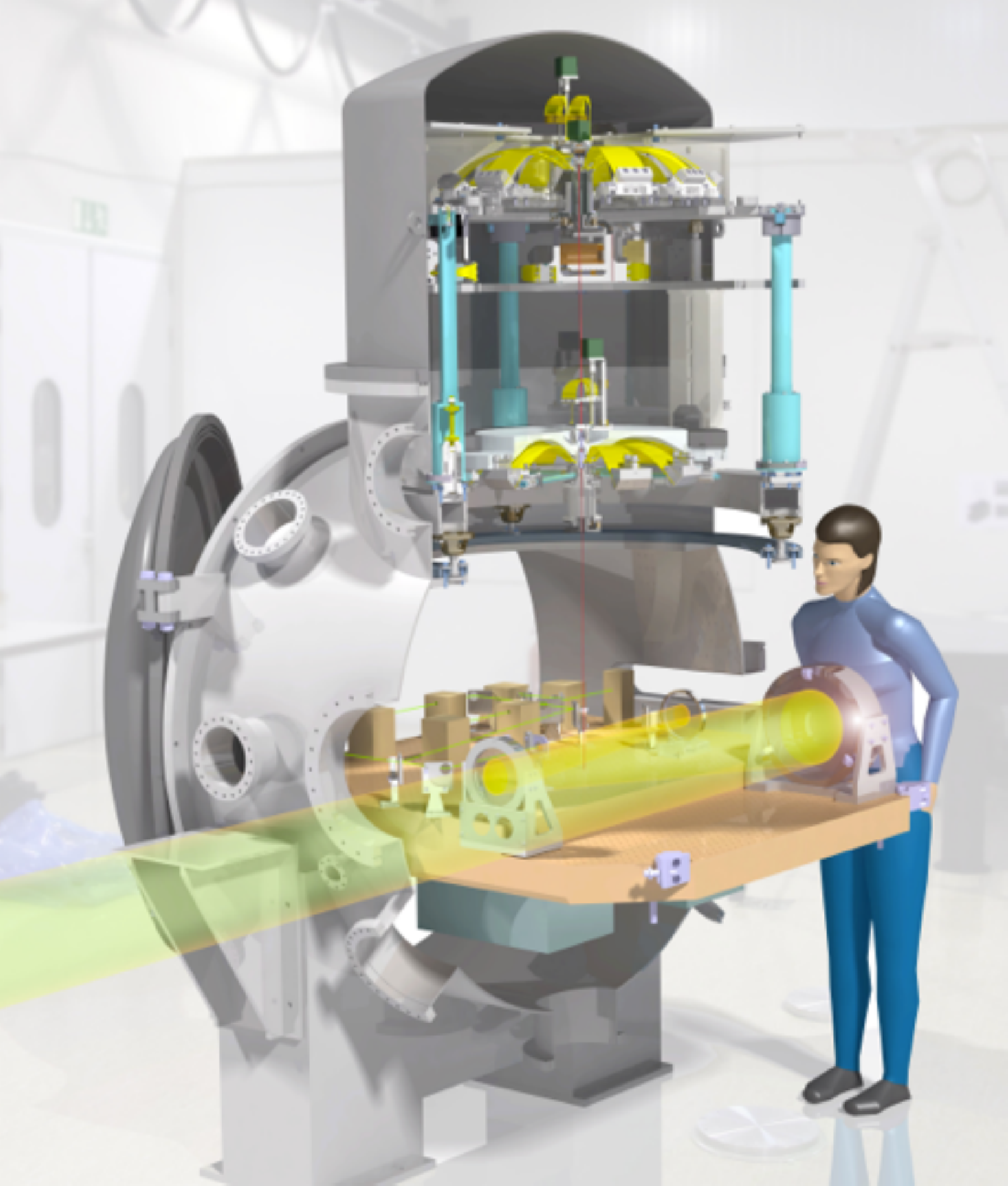}
\caption{\it Impression of the Minitower for the end benches.}
\label{fig:omino}
\end{figure}

The five optical benches, SNEB, SWEB, SIB2, SPRB, SDB2, are suspended in vacuum in order that all of the relevant photodiodes for longitudinal and angular control can be better isolated from in-band seismic and acoustic noise. Moreover, suspending the benches makes it possible to implement global control strategies in which low-frequency relative motion between the interferometer optics and pick-off telescopes (and related QPDs) is suppressed, making the detector less vulnerable to up-conversion of large amplitude motion of scattering sources around the microseismic peak. An integrated system, called the Minitower (see Figure~\ref{fig:omino}), consisting of a multi-stage vibration isolator named MultiSAS and vacuum chamber, is provided for each bench.

The MultiSAS is a compact (1200~mm high, 1070~mm in diameter) multi-stage seismic-attenuation system, designed to provide the best possible vibration isolation for the suspended benches, considering the limited space available in the facility. The MultiSAS design has been driven by the requirements of the end benches, the most demanding in terms of seismic isolation. The end benches carry telescopes and DC  quadrant photodiodes for the interferometer automatic alignment, for which shot noise-limited controls require QPD residual linear ({\it horizontal} and {\it vertical}) and angular ({\it pitch}, {\it roll} and {\it yaw}) motion at 10~Hz of  $2.1\cdot10^{-12}$ m/$\sqrt{\rm{Hz}}$ and $3.3\cdot10^{-15}$ rad/$\sqrt{\rm{Hz}}$ respectively.

\begin{figure}[htbp!]
\centering
\includegraphics[width=.50\textwidth]{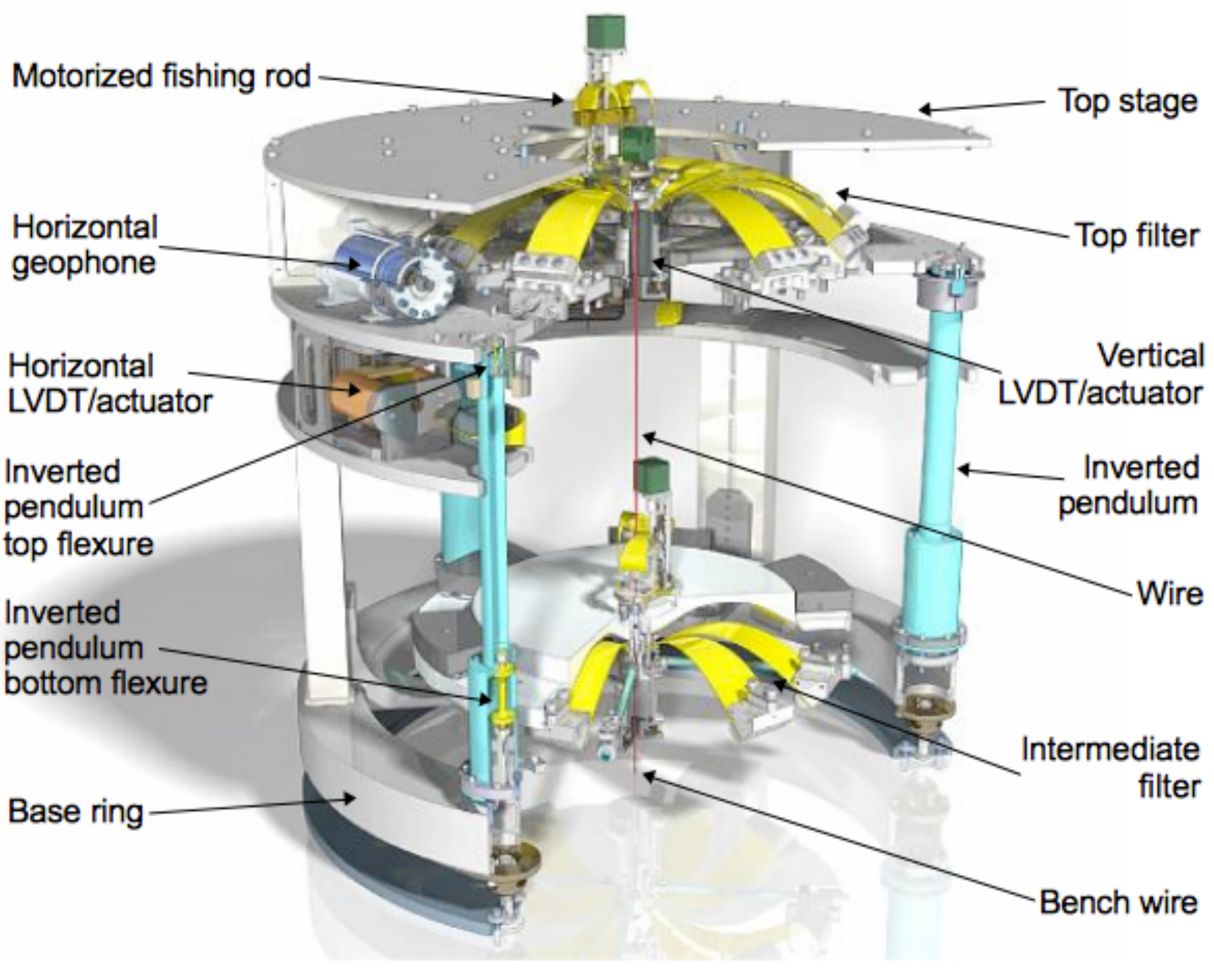} 
\hfill
\includegraphics[width=.35\textwidth]{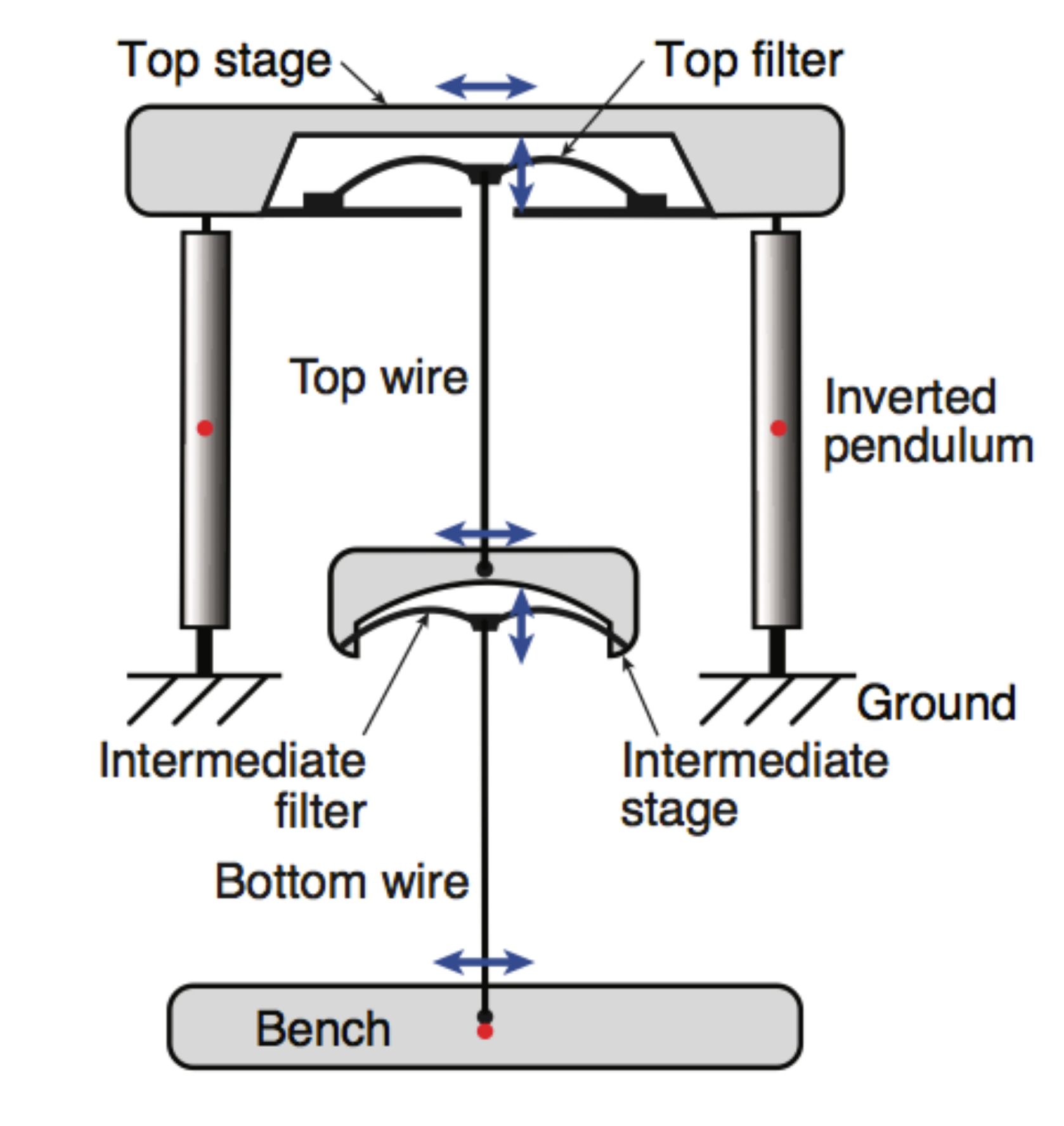} 
\caption{\it Left panel: MultiSAS overview. Right panel: MultiSAS conceptual design; two vertical (GAS springs) and three horizontal (IP and two simple pendulums) isolation stages are provided.}
\label{fig:multi3d}
\end{figure}

MultiSAS, following the philosophy of the Super Attenuator, is a hybrid seismic isolator in which bulk attenuation is provided passively by means of a chain of low-natural-frequency mechanical oscillators ({\it filters}). Filters are created with simple pendulums in horizontal and GAS springs in vertical~\cite{tamasas}. Feedback is used only to damp the rigid body eigenmodes of the system and to maintain, in the long term, the position and orientation of the optical bench.
High-frequency internal modes are handled, if needed, by means of passive resonant dampers.
The main components of the MultiSAS chain are illustrated in Figure~\ref{fig:multi3d}:

\begin{itemize}
\item{}
A horizontal pre-isolation stage, built with a short (about 0.5 m) Inverted Pendulum (IP), tuned to a natural frequency of around 100~mHz, attenuates the bench motion in the microseismic peak band and allows its precise positioning at very low frequencies. The IP also provides an inertial platform on which to detect the recoil and actively damp the rigid body modes of the suspended chain. Passive attenuation between the IP itself and the optical bench is exploited to apply controls without re--injecting noise. The IP is equipped with LVDT position sensors, geophones (L4--C from Sercel) and voice-coil actuators for dynamic controls. Stepper motor-driven correction springs are foreseen for static positioning with micrometer accuracy. 
\item{}
A vertical pre-isolation stage, called the {\it top filter}, consisting of a GAS spring, supported by the IP platform and tuned to 200~mHz. The top filter is equipped with a stepper motor-driven static positioning spring, the so called {\it fishing rod}, and with a collocated LVDT/voice-coil pair for feedback control.
\item{}
A 100~kg-mass intermediate stage, called the {\it intermediate filter}, consisting of a GAS spring, tuned to 400~mHz and suspended from the pre-isolation stage via a 694 mm-long wire. The filter is equipped with a fishing rod and a LVDT/voice-coil actuator for monitoring and test purposes.
\item{}
The 320 kg-mass optical bench, suspended from the intermediate filter, via a 760 mm-long wire. By adjusting the position of the bending point of the suspension wire with respect to the bench center of mass, the tilt modes of the payload will be tuned around 200~mHz, for better decoupling from the horizontal pendulum modes of the chain. The bench is equipped with two motorized counterweights for remote fine balancing and with a set of collocated LVDT coil-magnet actuators, referred to ground, for the local control of the angular degrees of freedom. 
\end{itemize}

\begin{figure}[htbp!]
\centering
\includegraphics[trim= 30 210 65 230, clip, width=.4\textwidth]{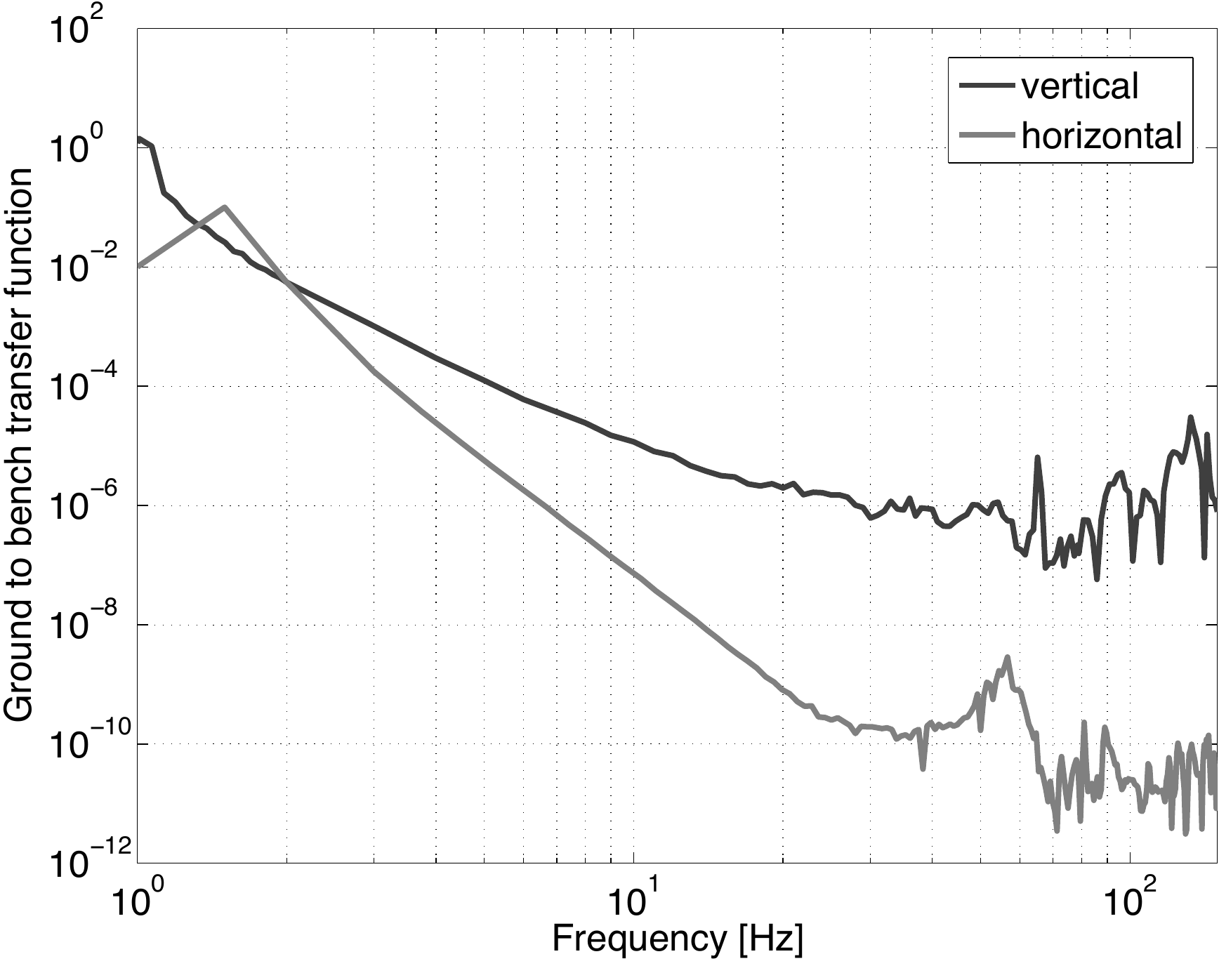} 
\hfill
\includegraphics[trim= 30 210 65 230, clip, width=.4\textwidth]{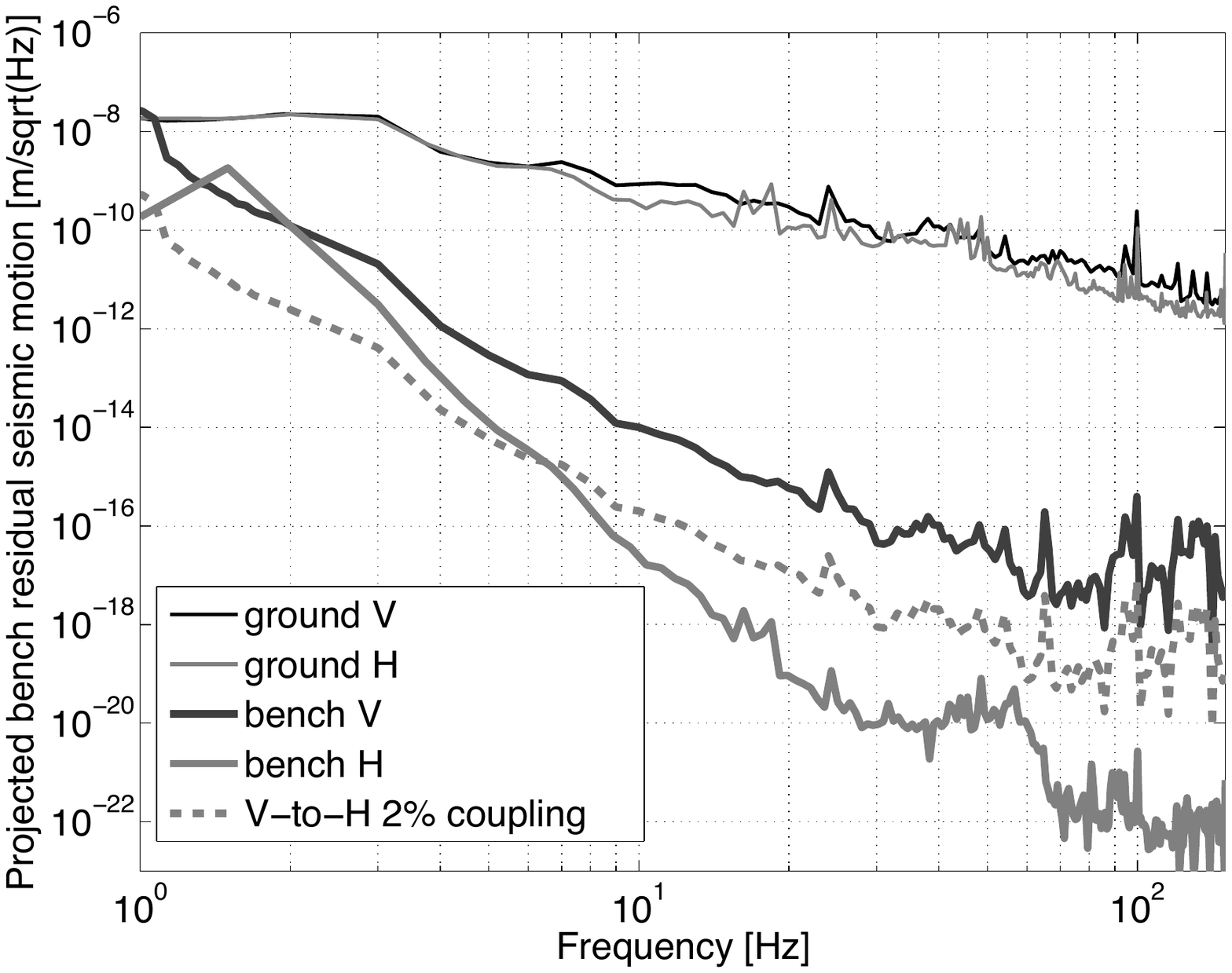} 
\caption{\it Left panel: MultiSAS measured vertical and horizontal transfer functions. A residual transmission around $\rm 10^{-6}$ is reached in vertical. In horizontal  an attenuation limit around $\rm 10^{9}$ is achieved above 20~Hz. The bump between 50 and 60~Hz is due to the residual transmission around the GAS filter keystone modes. Right panel: Estimated bench seismic motion in Advanced Virgo. The residual horizontal displacement is expected to be dominated, above a few Hz, by the vertical-to-horizontal coupling, measured to be about 2$\%$.}
\label{fig:tfunc}
\end{figure}

\noindent More details of the MultiSAS control system are presented in \cite{beker}. 
The MultiSAS seismic attenuation performance has been characterized by measuring its transfer function (see Figure~\ref{fig:tfunc}--left~panel). The vertical
isolation ratio exceeds 100~dB at all frequencies above 10~Hz. In horizontal, thanks to the three-stage design, a residual transmissibility around $-{\rm 180~dB}$ has been measured. Transfer functions have been convoluted with the ground motion spectra measured at the Virgo site to make estimates of the seismic motion of the benches (fig.~\ref{fig:tfunc}--right~panel).

\section{Data acquisition}  \label{sec:daq}

\subsection{Real-time interferometer synchronization and control}

The interferometer is a complex apparatus composed of a lot of sub-systems, which have
to work together in order to reach and keep the detector working point, with optimal sensitivity to GW. 
Of the order of one hundred active feedback systems are necessary to tune and control the whole detector.
As an example, the positions of the suspended mirrors and benches have to be controlled
both individually, with local sensors when the interferometer is not locked, 
and as a whole, depending mostly on the signals seen by photodiodes monitoring different beams in the ITF
during the lock acquisition and when the detector is kept at its working point.

The AdV detector-control scheme moves towards using fewer analog and more digital feedback loops than in Virgo.
Having more digital loops adds more flexibility to the design and configuration of the loop filters,
and improves the loop stability.
It also helps to monitor the detector by making it easier to extract intermediate information 
from digital processes, rather than adding probes into analog electronics boards.

\subsubsection{Centralized timing distribution}

In order to perform real-time control loops, all the detector devices must be synchronized:
the ADC channels that digitize the sensor outputs, 
the digital cameras that acquire images used for the controls, the DAC channels that generate actuator signals 
and the real-time processing units (PC or DSP) where the digital filtering of the loops are processed.
The centralized timing distribution system is based on the GPS timestamp,
in order to be also synchronized with other experiments.
The IRIG-B signal generated by a GPS receiver is distributed through 
the whole interferometer with optical fibers and finally to the front-end digital electronics.
It is used to lock the local oscillators of the devices and to timestamp the sampled data with a GPS time. 
The propagation delays in the distribution system are fine-tuned  such that the timing signal is the same down to all digital devices, 
within better than 50~ns. In this way, all the data-sampling clocks are synchronous.
The timing jitter between the clocks of different devices in the setup is lower than $0.1\,\mathrm{ns/\sqrt{Hz}}$ above 1~Hz~\cite{bib:DAQ_noteTimingJitter}.

\subsubsection{Fast online data network and real-time processing}

A custom fast-data network has been built in order to exchange and process in real-time the data from the sensing electronics to the driving electronics, and to provide these data to the data-collection system.

The global longitudinal and angular controls of the interferometer are performed through synchronous fast digital loops,
running at $\sim 20\,\mathrm{kHz}$ and $\sim 2\,\mathrm{kHz}$ respectively. 
The signals are collected by the sensing electronics in different locations of the detector. 
They are sent through optical fibers to real-time processing units that run the digital servo-loops
and send the data to other processing units or to the driving channels of the different interferometer suspensions. 
A set of multiplexer/demultiplexer boards is used to route the data packets in the network. 
The data can be delivered from any ITF device to any other other ITF device with latencies usually dominated by the
optical fiber propagation time.  
New devices, called DAQ-boxes, will be installed in AdV: these are digital mother boards that can host specific mezzanines
(for ADC, DAC, digital demodulation channels, camera triggers, etc.) and manage the interface with the timing and fast-data networks. 
The same interface is set in the electronics used to control the suspensions.
Some controls are done locally: these are implemented by sending the data directly, on-board, from the sensing part to the driving part,
where the needed filtering is performed by the associated DSP.

Most of the sensor signals are digitized by ADC at $\sim 1\,\mathrm{MHz}$ and digitally low-pass filtered and decimated
on-board, down to frequencies between a few~kHz and a few tens of~kHz. 
The data are then sent to the fast-data network. 
Some are then used as error signals for feedback loops and others are used to monitor the different sub-systems and the detector environment.

\subsubsection{Digital demodulation}

Demodulation will be used for many applications where error signals are needed.
For example, LVDT will be used to locally control the positions of some suspended benches.
Instead of acquiring the error signal in the low-frequency band (below a few Hz), where the $1/f$ noise of the ADC might not be negligible, the probe signal, which is modulated at typically $10\,\mathrm{kHz}$, is acquired at $\sim40\,\mathrm{kHz}$, 
and then demodulated in a real-time PC or in the on-board DSP in order to extract the error signal in a band 
in which the ADC noise is lower than at low frequencies.
Implementing such demodulation digitally adds flexibility to the choice and number of modulation frequencies,
as well as the possibility to modify them online to study, for example, cross-talk between LVDT.

The signals measured by the photodiodes are special cases. As explained in Sections~\ref{sec:det} and~\ref{sec:isc}, 
most of the photodiode signals are demodulated at frequencies between 6~MHz and 132~MHz. 
In Virgo, the output signals of analog demodulation boards were digitized by standard ADC. 
For AdV, digital demodulation will be used instead. The RF signals of the photodiode
are digitized in dedicated fast ADC at $\sim500\,\mathrm{MHz}$ and the demodulation processing is computed on-board by a FPGA.
The in-phase and in-quadrature output channels are then low-pass filtered and decimated in the FPGA and in DSP, such that the 
frequencies of the final output channels, sent to the fast-data network, are $\sim 20\,\mathrm{kHz}$.
In this digital scheme, even in the case of demodulation at multiple frequencies, the RF output of a photodiode is acquired only once 
in a fast ADC and then different digital computations are processed for the different frequencies.
Compared to the analog solution, this reduces the number of cables and electronics boards.
Another important advantage is that it is not needed to distribute the local oscillator clocks from the laser modulator to
all the demodulation devices. The reference is the distributed timing signal used to lock the local clocks of the boards.
The on-board timing distribution adds some low frequency phase noise that can be subtracted online since it is common to all demodulation channels:
the phase noise can be monitored using either the demodulated photodiode signal at twice the main frequency, which also
provides some sideband amplitude noise, or a $10\,\mathrm{MHz}$ clock, which is distributed through the timing network.

\subsubsection{Expected digital channels}
The estimated numbers of front-end channels to be acquired or driven through the fast-data network of AdV
are of the order of 2200~ADC with sampling frequencies higher than 1 kHz, 
1400~DAC, 100~digital demodulation inputs and 40~cameras synchronized with the timing system.

\subsection{Data collection}

The goal of the data collection is to build different data streams and provide them both for long-term storage on disk
and online for commissioning, noise studies, data-quality definitions and $h(t)$ reconstruction before data analysis.
Data are generated all over the detector: they include interferometer sensing and control signals, monitoring signals
from all the detector sub-systems and environmental-monitoring signals.
They are acquired through the fast-data network ($\sim 400\,\mathrm{MBytes/s}$) 
and merged in different streams at the front-end of the data collection pipeline.
Data are then reduced (decimation, compression, image processing, etc.) in order to limit the flow on the Ethernet network and
to require a reasonable storage amount of disk space. Data are then provided to several online processes 
(detector automation and monitoring, data-quality tools, $h(t)$ reconstruction) which enrich the streams 
with additional computed channels that can be shared with other online processes.
The data can be reached by a data-display tool to study the detector behavior online,
with a latency of the order of a second for the front-end data.

At the end of the data-collection tree, different data streams are stored on disk using the gravitational wave experiment data format~\cite{bib:DAQ_FrameFormat}.
The raw-data stream (2~TB/day) contains all of the channels at their nominal sampling frequency, including
the channels generated by the front-end electronics and the channels generated by the online processes along the data-collection tree.
Other streams are generated: 
a fast stream, with the channels stored at their maximum frequency, before any decimation of the data collection pipeline,
in order to monitor and debug the front-end electronics;
a Reduced Data Stream (RDS), 30~GB/day, with the channels decimated at lower frequencies than nominal (typically $\sim100\,\mathrm{Hz}$) for commissioning and noise studies at low frequencies; 
trend-data streams, with statistical information computed over one second (4~GB/day) and a few tenths of seconds over all of the channels; 
the $h(t)$ data stream (7~GB/day), which contains the reconstructed GW strain channel and data-quality flags, 
and that can be directly used by the data analysis pipelines.
The data are kept on circular buffers at the Virgo site, with a depth of $\sim6\,\mathrm{months}$ for the raw data.
During science runs all the data are transferred to computing centers for permanent storage and offline data analysis.

\subsection{Calibration and $h(t)$ reconstruction}
The main output of the calibration and reconstruction procedures is the $h(t)$ time series, which describes the reconstructed GW strain signal. This is the main input for GW searches. 
For AdV, the requirements from the different data-analysis groups, searching for CBC, bursts, continuous waves and stochastic sources,
call for uncertainties below  $10\%$ in amplitude, $100\,\mathrm{mrad}$ in phase, $25\,\mathrm{\mu s}$ in timing. 
These are $1\,\sigma$ uncertainties given in the range 10\,Hz to 2\,kHz.

During Virgo~\cite{bib:DAQ_VirgoHrec}, the systematic uncertainties of the reconstructed $h(t)$ time series 
already met these requirements. These were estimated to be 8\% in amplitude, while $h(t)$ was 50~mrad in phase at 10~Hz, with a frequency dependence following a delay of $8\,\mathrm{\mu s}$ at high frequency.

Consequently, the same calibration methods~\cite{bib:DAQ_VirgoCalib} will still be used for AdV.
Similarly, the same method will be used for the $h(t)$ reconstruction, in particular at the start of AdV, in the power-recycled configuration.
Developments will be needed to adapt the reconstruction when the SR mirror is added.

In parallel to the standard calibration methods, a photon calibrator has been used in Virgo~\cite{bib:DAQ_VirgoHrec} to validate,
with independent measurements, the reconstructed $h(t)$ time series. An improved design of the photon calibrator is being defined for AdV.
Two such devices will be installed, one on each end test mass.

\section{Vacuum upgrades}  \label{sec:vac}

The main vacuum enclosure is one single volume of about $7000$\,m$^{3}$, consisting of several parts  \cite{VirgoP,ADV:TDR}:
\begin{itemize}
\item two $3$\,km-long straight tubes, $1.2$\,m in diameter, hosting the interferometer arms;
\item eleven vertical cylinders, called {\it towers}, containing the Super Attenuators. These chambers have a diameter of $2$\,m. Three are 4.5 m high and host the {\it short} Super Attenuators suspending the Input Mode Cleaner mirror and the Injection/Detection benches; seven are $11$\,m high and contain the {\it long} Super Attenuators suspending all of the other optics;
\item one $142$\,m-long tube, $30$\,cm in diameter, containing the Mode Cleaner optical cavity;
\item several valves, with up to $1$\,m aperture;
\item several pumping groups, including many types of pumps and vacuum gauges;
\item a few large cryostats.
\end{itemize}

To meet the sensitivity goal, a residual gas pressure of about $10^{-9}$\,mbar inside the $3$\,km-long arm tubes is required, in order to suppress several sources of noise, the dominant source being the fluctuations of the refractive index due to the statistical fluctuations of the gas molecule density on the 3 km path of the laser beams.

In order to cope with the vacuum and cleanliness prescriptions, a careful material selection has been performed for all of the in-vacuum components, and suitable vacuum/thermal treatments have been developed and performed prior to installation:

1)  \emph{Bake-out} Since it is mandatory to remove the layers of water molecules that stick to the inner wall 
of the enclosure, a bake-out of the vacuum pipes may be performed at a later stage to reach the final residual pressure.
The chosen process consists of heating the pipes, once under vacuum, to $150^{\circ}$C for about one week. The baking equipment was installed and tested before the operation of Virgo. However, Virgo has been operated in unbaked conditions, since a pressure of $10^{-7}$ mbar was not limiting the sensitivity. The tubes will be baked only when necessary to improve the sensitivity of AdV and convenient for the schedule.
The towers will not be baked to avoid the risk of damaging the suspended payloads. The migration of water molecules from the unbaked towers to the $3$\,km pipes will be prevented by the installation of cryogenic traps (cryotraps) at each end of the pipes.

2)  \emph{Firing} With the Advanced Virgo operating pressure already in mind, all the constituents of the Ultra-High Vacuum (UHV) enclosure (pipe modules and tower lower chambers) have been fired in air at $400^{\circ}$C in order to reduce by at least a factor of $100$ the outgassing of hydrogen from the stainless-steel walls. This makes it possible to meet the base pressure requirements with only one pumping station every $600$\,m of pipe.

3)  \emph{Cleanliness} The cleanliness of the optical surfaces is one of the most critical aspects of operating
the detector. For this reason, the residual gas must be free of condensable organic molecules (hydrocarbons), 
and a conservative hydrocarbon partial pressure of  $10^{-13}$\,mbar is required for the mirror vacuum chambers. 
The four large cryotraps at the ends of the 3 km pipes will contribute to the fulfillment of the requirement by condensing hydrocarbon vapors. 
Each of the seven long towers is divided into two compartments: a lower vacuum chamber, hosting the mirror; and an upper vacuum chamber, containing mechanical elements and the Super Attenuator cables. 
The two volumes are separated by  a mechanical structure (the so called {\it Separating Roof}) with a low-conductance pipe in the center, for the passage of the suspension wire connecting the payload to the upper stage. This low conductance (1.6\,l/s or 0.4\,l/s when differentially pumped) makes it possible to tolerate, in the tower upper part, a pressure up to  $10^{-6}$\,mbar dominated by water. 
Similar conditions are also obtained in the three short towers, which are separated from 
the mirror towers by two additional cryotraps. The common goal for all the evacuated volumes is to keep them free 
of dust, avoiding optics contamination during any maintenance operations. 

\subsection{Pumping system}

Most of the Virgo pumps are still used in Advanced Virgo.
The upper and lower chambers of each tower are separately pumped by a turbo-molecular group, consisting of a $1500$\,l/s turbo, backed by a dry scroll pump, also used for roughing. Turbo pumps of the hybrid type with magnetic suspensions are used to reduce noise and maintenance. To further reduce vibrations, the scroll pumps have been moved to a distant and acoustically-isolated room, while the turbo pumps are anchored to ground and connected to the towers with soft bellows.
Over the next few years, and only if really needed, the turbo pumps will be progressively replaced by quieter ion pumps (which have lower seismic, acoustic and magnetic emissions). 

In the arm tubes, roughing and intermediate phases are achieved with a large dry mechanical pumping group. Evacuation continues with a turbo-molecular group similar to those of the towers every $600$\,m. Final vacuum will be reached after baking and maintained by titanium sublimation pumps, which are most suited to pump hydrogen ($3000$\,l/s) and smaller ion pumps (70\,l/s) to handle inert gases. Periodical regeneration of a Ti evaporated layer will slightly lower the pumping duty cycle; normally five out of seven groups will be in operation along each arm.

\begin{figure}
\begin{center}
\includegraphics[width=0.65\textwidth]{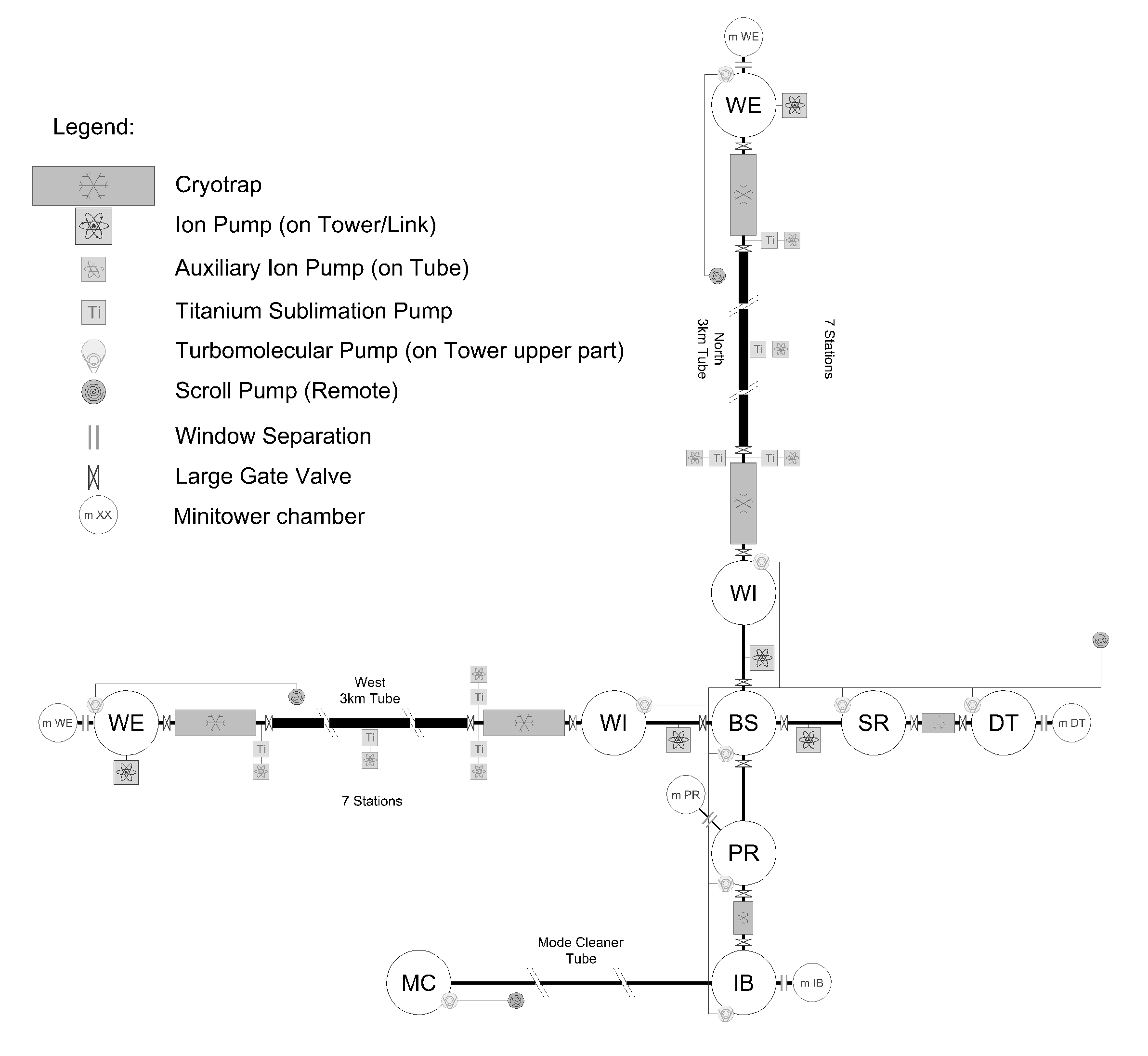}
\caption{Schematic of the AdV pumping system.}
\label{vac1}
\end{center}
\end{figure}

A layout of the pumping system is shown in Figure \ref{vac1}.

\subsection{Cryogenics}

\begin{figure}
\begin{center}
\includegraphics[width=0.6\textwidth]{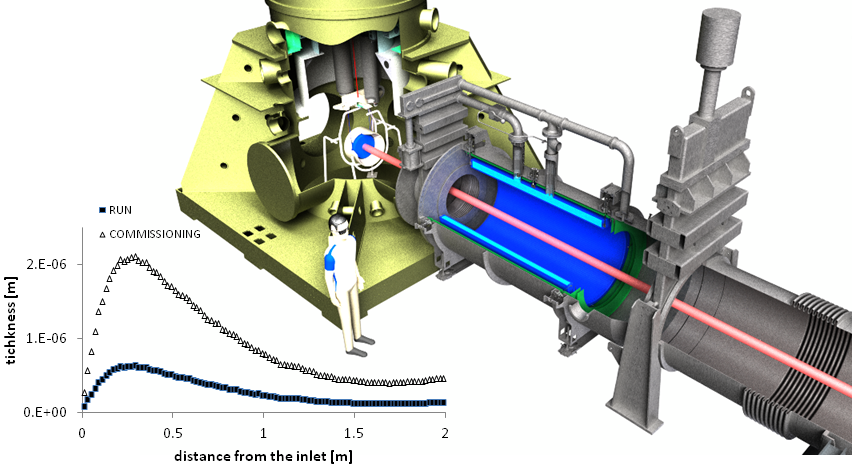}
\caption{View of the arm cryotrap: the inner cryogenic vessel is in light gray. On the left: evaluation of the ice layer build up inside a trap after one year of service in 'run' or 'commissioning' mode.}
\label{vac3}
\end{center}
\end{figure}

The water migrating from the towers to the 3 km pipes will be stopped by four large cryotraps, one at each end of the arm pipes, and by two smaller ones, between the bench towers and the recycling towers. 
The large cryotraps are installed between the input/end towers and the corresponding large valve. They consist (Figure \ref{vac3}) of a vacuum pipe, which is about 3\,m long and 1.4\,m in diameter, containing a $2$\,m-long tank with the shape of a sleeve and filled with about 300\,l of liquid nitrogen at 77K. The sleeve has been built in aluminum for a more uniform temperature distribution; with an inner diameter of 0.95\,m, it has a 6\,m$^{2}$ cold surface to capture water and other condensable gases. The outer wall of the cold tank is centered away from the beam axis maximizing the liquid/gaseous nitrogen open surface, to allow a smooth evaporation and reducing the bubbling noise. The cold tank is thermally shielded by a two-fold aluminum foil screen, while the free aperture for the laser beam is limited to a 600\,mm diameter by suitable thermal and optical baffles.

The escaping fraction of water molecules is estimated to be about $3$\%; the thermal loss, verified on the first built trap, is $250$\,W, corresponding to a consumption of  $5.6$\,l/h of liquid nitrogen and  $1$\,l/s of gas at $80$\,K. Consumption depends also on the thickness of the ice deposit, which grows progressively by about 1 micron per year (Figure \ref{vac3}) and causes a slow increase of infrared emissivity. Regenerations are not envisaged to take place more than once per year; the process will take about one week or slightly more.

The cryogenic liquid is continuously supplied by a distribution plant equipped with three standard reservoirs (1 x $30,000$\,l for the Central Building + 2 x $15,000$\,l for the End Buildings) which lie horizontally, in order to reduce the hydraulic head and the wind grip. The overall site consumption is estimated as $1000$\,l/d (including the reservoir auto-consumption) and the refilling of each tank is planned once per month, from a truck coming on site during maintenance breaks. Transfer lines have been installed with slopes, which have been designed to avoid sections subject to gas accumulation and risk of noisy intermittent flow.

The two smaller cryotraps, 0.9\,m long and with an aperture of 350 mm, play the same role. These can also be operated in batch mode (without the seismic noise due to the moving dosing valve) and feature a relatively large volume of liquid nitrogen ($200$\,l) to reach four days of autonomy.

\subsection{Other components}

The introduction of a signal-recycling mirror also required a new vacuum tank for its Super Attenuator. Moreover,
all the link pipes connecting the towers in the Central Building are being replaced with pipes with larger apertures, to cope with the larger laser spot size. These have been designed to have mechanical resonant frequencies above 30 Hz, to avoid strong coupling with seismic noise.
Each new link pipe is made of three to four parts, including expansion bellows and gate valves with diameters up to $650$\,mm. They also house the extraction or injection ports for different auxiliary beams (optical levers, pick-off beams, thermal compensation beams).

The new vacuum system is completed by a complex control system, based on thirty-three main stations. Each station makes it possible to operate the equipment (pumps, gauges, valves) in local or remote mode, assuring inter-locks and loops that actively protect the system against incorrect actions or failures.

\section{In conclusion}
In this paper we have reviewed the main aspects of the Advanced Virgo project, in the context of the forthcoming network of second-generation interferometric GW detectors. Despite the absence of a detection, the first-generation detectors were a great success: they established the infrastructure which will also be used for the second generation; the interferometric detector technology has been demonstrated, since the design sensitivities has been reached (and even surpassed) and very good robustness has been achieved; a world-wide network of GW detectors has been built and the first steps towards a multi-messenger approach have been walked; concrete methods for the analysis of the data have been established; and many papers providing results that have had an impact on our understanding of the universe have been published. The technology in our hands and the experience achieved with the first-generation detectors, make it possible to increase by a factor of one thousand the volume of the universe we can survey with the advanced detectors and, consequently, the detection probability. This makes the forthcoming years very exciting.

\end{document}